\documentclass[preprint,superscriptaddress,nofootinbib,11pt]{revtex4-1}
\usepackage{graphicx, epsfig}
\usepackage{xspace}
\usepackage{amsmath,amssymb,mathtools}
\usepackage{hyperref}
\usepackage[utf8]{inputenc}
\usepackage{slashed}

\numberwithin{equation}{section}

\newcommand{\exclude}[1]{}
\def\nn{\nonumber}
\newcommand{\vtree}{V^{(0)}}

\newcommand{\veff}{V_{\text{eff}}}
\newcommand{\vone}{V^{(1)}}
\newcommand{\vtwo}{V^{(2)}}

\newcommand{\eps}{\epsilon}
\newcommand{\hc}{\text{h.c.}}

\newcommand{\vphys}{v_\text{OS}}

\newcommand{\aol}{A^{(1)}}
\newcommand{\bol}{B^{(1)}}
\newcommand{\atl}{A^{(2)}}
\newcommand{\btl}{B^{(2)}}
\newcommand{\ctl}{C^{(2)}}
\newcommand{\pr}{\mathcal{P}_R}
\newcommand{\gev}{\text{ GeV}}
\newcommand{\lagr}{\mathcal{L}}

\DeclareMathOperator{\llog}{\overline{\text{log}}}
\DeclareMathOperator{\tr}{\text{tr}}
\def\msbar{{\ensuremath{\overline{\rm MS}}}\xspace}

\begin{document}
\preprint{OU-HET-1077}
\preprint{DESY 20-192}

\title{
 Two-loop analysis of classically scale-invariant models with extended Higgs sectors
}

\author{Johannes~Braathen}
\email{johannes.braathen@desy.de}
\affiliation{
Department of Physics,
Osaka University,
Toyonaka, Osaka 560-0043, Japan
}
\affiliation{
Deutsches Elektronen-Synchrotron DESY, Notkestra\ss{}e 85, D-22607 Hamburg, Germany
}

\author{Shinya~Kanemura}
\email{kanemu@het.phys.sci.osaka-u.ac.jp}
\affiliation{
Department of Physics,
Osaka University,
Toyonaka, Osaka 560-0043, Japan
}

\author{Makoto~Shimoda}
\email{m\_shimoda@het.phys.sci.osaka-u.ac.jp}
\affiliation{
Department of Physics,
Osaka University,
Toyonaka, Osaka 560-0043, Japan
}

\begin{abstract}
We present the first explicit calculation of leading two-loop corrections to the Higgs trilinear coupling $\lambda_{hhh}$ in models with classical scale invariance (CSI), using the effective-potential approximation. Furthermore, we also study -- for the first time at two loops -- the relation that appears between the masses of all states in CSI theories, due to the requirement of reproducing correctly the 125-GeV Higgs-boson  mass. In addition to obtaining analytic results for general CSI models, we consider two particular examples of Beyond-the-Standard-Model theories with extended Higgs sectors, namely an $N$-scalar model (endowed with a global $O(N)$ symmetry) and a CSI version of the Two-Higgs-Doublet Model, and we perform detailed numerical studies of these scenarios. While at one loop the value of the Higgs trilinear coupling is identical in all CSI models, and deviates by approximately $82\%$ from the (one-loop) SM prediction, we find that the inclusion of two-loop corrections lifts this universality and allows distinguishing different BSM scenarios with CSI. Taking into account constraints from perturbative unitarity and the relation among masses, we find for both types of scenarios we consider that at two loops $\lambda_{hhh}$ deviates from its SM prediction by $100\pm10\%$ -- $i.e.$ a quite significant further deviation with respect to the one-loop result of $\sim 82\%$. 
\end{abstract}

\maketitle

\section{Introduction}
The discovery of a Higgs boson of mass 125 GeV at the CERN Large Hadron Collider~\cite{Chatrchyan:2012xdj,Aad:2012tfa} has completed the particle spectrum of the Standard Model (SM) of particle physics, and has established the Higgs sector as the origin of electroweak symmetry breaking (EWSB). Nevertheless, numerous open questions remain unsolved by the SM, and point to the need for Beyond-the-Standard-Model (BSM) physics. Among these issues, one can mention, on the theory side, for instance the hierarchy problem, the lack of a description of quantum gravity, or inflation. Additionally, a number of experimental results have contradicted prior assumptions -- upon which the SM was built -- such as for example (to cite only a few) the need to realise baryogenesis, the evidence for dark matter and dark energy, or the discovery that neutrinos have tiny masses. 

There is no doubt that the Higgs boson plays a central role in new physics, as well as in searches for it. Indeed, many (if not most) problems of the SM can be related to Higgs physics. In order to address these, BSM theories commonly feature a Higgs sector with extended particle content or modified dynamics -- as there is no compelling argument for it to be minimal. Therefore, shedding light on the structure of the Higgs sector is one of the most pressing tasks for particle physicists. Unfortunately, for the time being, all measurements of Higgs properties seem to be in excellent agreement with SM predictions~\cite{Aad:2019mbh,Sirunyan:2018koj}, and this forces us to investigate how new physics can appear to solve problems of the SM while simultaneously evading detection. A first, but unexciting, possibility is decoupling -- $i.e.$ that all new states are well beyond our experimental reach and their contributions to SM(-like) observables are vanishingly small due to the decoupling theorem~\cite{Appelquist:1974tg}. Another more interesting option is alignment without decoupling~\cite{Gunion:2002zf}, which occurs\footnote{Alignment is usually defined in models with several Higgs doublets, and is then the limit in which one of the CP-even mass eigenstates is colinear in field space with the total electroweak vacuum expectation value. } when the couplings of the 125-GeV Higgs boson to fermions and gauge bosons are SM-like at tree level, possibly because of some mechanism or symmetry of the Lagrangian -- see for instance Refs.~\cite{Dev:2014yca,Deshpande:1977rw,Silveira:1985rk,Barbieri:2006dq,Benakli:2018vqz,Benakli:2018vjk,Coyle:2019exn,Aiko:2020atr}. We will return to this latter case in the following.\footnote{Slightly non-alignment cases will be almost completely surveyed and, if nothing appears, will be excluded by the synergy of the HL-LHC and future precision measurements of the Higgs boson couplings such as at the ILC, FCCee, CEPC, CLIC, etc. \cite{Aiko:2020ksl}. }

Among the issues of the SM, one that has received considerable attention is the (gauge) hierarchy problem~\cite{Veltman:1976rt}, or in other words the need for a mechanism to stabilise the mass of the Higgs boson $M_h$ at the electroweak (EW) scale. Because the Higgs boson is a scalar, its mass is in the SM not protected by any symmetry, meaning that it typically receives huge -- quadratically divergent -- radiative corrections. In turn, this implies that a cancellation of tremendous -- and unrealistic -- accuracy would have to occur between tree-level and loop-level contributions to $M_h$. Solutions to the hierarchy problem usually involve some new symmetry or mechanism that protects the Higgs mass. For example, supersymmetry~\cite{Fayet:1977yc} introduces a new symmetry between bosons and fermions, so that scalar masses become protected by the chiral symmetry. Meanwhile, in composite Higgs models~\cite{Kaplan:1983fs,Kaplan:1983sm}, there are no fundamental scalars and the Higgs boson appears as a composite state of elementary fermions. But because new physics has so far eluded discovery, experimental constraints on most popular BSM scenarios have become more stringent (see for instance Refs.~\cite{ATLAS:2020uqv,ATLAS:2020ujn} and references therein) and the need to explore new avenues of research is increasingly evident. 

Another interesting possibility to address the hierarchy problem, which we will consider in this paper, comes from the concept of classical scale invariance (CSI), first proposed by Bardeen in Ref.~\cite{Bardeen:1995kv}. The idea of CSI is to forbid all mass terms in the Lagrangian, so that the theory is scale invariant at the tree ($i.e.$ classical) level. Of course, as this is only imposed at tree level, scale invariance is explicitly broken by radiative corrections, and EWSB can occur radiatively, via the Coleman-Weinberg mechanism~\cite{Coleman:1973jx} -- which was extended to multi-scalar cases by Gildener and Weinberg in Ref.~\cite{Gildener:1976ih}. The Higgs vacuum expectation value (VEV) $v$ and all particle masses (proportional to $v$) are then generated via dimensional transmutation. If the CSI model is taken to be UV complete, or in other words if scale invariance is assumed at the Planck scale, the absence of a tree-level Higgs mass parameter together with the generation of the Higgs mass entirely at loop level can ensure that there is no hierarchy problem -- see for instance Refs.~\cite{Bardeen:1995kv,Foot:2007as,Foot:2007iy,Meissner:2007xv,Iso:2009ss,Iso:2012jn}. On the other hand, models in which scale invariance is imposed at the EW scale can often not be extended up to the Planck scale, because of Landau poles appearing in the running of dimensionless parameters. However, even if a Landau pole appears at some intermediate scale $\Lambda$ and thus requires the existence of some further new physics beyond $\Lambda$, tree-level scalar masses are prohibited up to this scale thanks to the CSI. Therefore there is no hierarchy problem up to the scale $\Lambda$, and a problem only remains between $\Lambda$ and the Planck scale -- arguably a less severe issue than in the SM. The hierarchy problem is thus not entirely solved in such theories, although it can be somewhat alleviated. 

Nonetheless, motivated by a number of phenomenological considerations, we restrict our attention in this paper on the latter type of CSI models, defined with extended scalar sectors at the EW scale. A first interesting feature of such models is that, in the absence of any (BSM) mass term, new states cannot be decoupled, nor very heavy, and hence should be found in a foreseeable future -- if they exist. Additionally, in CSI models, the 125-GeV Higgs boson, which corresponds in the language of Ref.~\cite{Gildener:1976ih} to the ``scalon'' direction along which EWSB is radiatively realised, is automatically aligned at tree level -- as pointed out in Ref.~\cite{Lane:2019dbc} for a CSI variant of the Two-Higgs-Doublet Model (2HDM) -- thereby explaining current experimental results. 

A further discriminative characteristic of CSI models comes from the Higgs trilinear coupling $\lambda_{hhh}$. This coupling is of great importance as it controls the shape of the Higgs potential, and in turn, the nature and strength of the EW phase transition (EWPT).  In particular, one of the requirements to make the scenario of electroweak baryogenesis work~\cite{Sakharov:1967dj,Kuzmin:1985mm,Cohen:1993nk} is for the EWPT to be of strong first order. In turn, a large deviation of the Higgs trilinear coupling from its SM prediction is one of the main signs of a strong first-order EWPT -- for massive models, a deviation of at least 20\% was found to be necessary in Refs.~\cite{Grojean:2004xa,Kanemura:2004ch} -- together with the special spectrum of gravitational waves that would be produced by bubble collisions during the EWPT~\cite{Grojean:2006bp}.

Couplings of the Higgs boson -- and especially its trilinear coupling -- can potentially differ significantly from their SM values in BSM theories with extended scalar sectors, because of \textit{non-decoupling effects} from loop corrections involving the BSM states -- this was found at one loop, first in the 2HDM in Refs.~\cite{Kanemura:2002vm,Kanemura:2004mg}, and later in various BSM models with extra singlets~\cite{Kanemura:2015fra,Kanemura:2016lkz,He:2016sqr,Kanemura:2017wtm}, doublets~\cite{Kanemura:2015mxa,Arhrib:2015hoa,Kanemura:2016sos,Kanemura:2017wtm}, or triplets~\cite{Aoki:2012jj}. The limit when these BSM corrections are maximal is when the BSM scalars acquire large masses exclusively from the Higgs VEV (rather from some BSM mass parameter). Incidently, this is exactly what happens in CSI models, and as a matter of fact, the Higgs trilinear coupling has been found~\cite{Hill:2014mqa,Hashino:2015nxa,Agrawal:2019bpm} to be at one loop $(\lambda_{hhh}^\text{CSI})^{(1)}=5/3(\lambda_{hhh}^\text{SM})^{(0)}$ in \textit{any} CSI model (without mixing), where $(\lambda_{hhh}^\text{SM})^{(0)}$ is the tree-level SM prediction for $\lambda_{hhh}$ -- in other words $(\lambda_{hhh}^\text{CSI})^{(1)}$ deviates by $\sim 67\%$ from $(\lambda_{hhh}^\text{SM})^{(0)}$. The universality of this prediction is a particularly strong and unique property of CSI models, and follows (as we will see in section~\ref{SEC:GW}) from the especially simple form that the effective potential takes at one-loop order along the ``scalon'' direction. This value of $(\lambda_{hhh}^\text{CSI})^{(1)}$ also means that the EWPT is always of strong first order in CSI theories -- this, together with the synergy between the measurements of gravitational waves and of the Higgs trilinear coupling was investigated in Ref.~\cite{Hashino:2016rvx} (for the example of $O(N)$-symmetric CSI models).

One may then naturally wonder what is the current experimental status for the determination of $\lambda_{hhh}$ as well as future perspectives. It turns out that large deviations of $\lambda_{hhh}$ from its SM prediction are still allowed by experiments: indeed, at the present time the most stringent limits on the trilinear coupling come from the ATLAS collaboration using single-Higgs production searches with LHC Run 2 data and give $-3.2 < \lambda_{hhh}/\lambda_{hhh}^\text{SM} < 11.9$ at 95\% confidence level (CL)~\cite{ATL-PHYS-PUB-2019-009} (see also Refs.~\cite{Ferrari:2018akh,Aaboud:2018ftw} and~\cite{Sirunyan:2018iwt,Sirunyan:2018two} respectively for ATLAS and CMS results using double-Higgs production). However, these limits will be greatly improved at future colliders, thus making $\lambda_{hhh}$ an ideal target to search for BSM effects. We refer the interested reader to Ref.~\cite{deBlas:2019rxi} for a detailed discussion of the current prospects for the determination of $\lambda_{hhh}$ at upcoming experiments, and we only summarise here some of the main results. First of all, the high-luminosity upgrade of the LHC (HL-LHC) is expected to reach an accuracy of about $50-60\%$ with $3\text{ ab}^{-1}$ of data~\cite{Cepeda:2019klc} (see also Ref.~\cite{Chang:2019ncg}) while a high-energy extension (HE-LHC) might -- optimistically\footnote{Note that this result depends on the treatment of backgrounds -- see the discussions in Refs.~\cite{Homiller:2018dgu,Goncalves:2018yva} } -- be able to attain an $\mathcal{O}(15\%)$ precision~\cite{Goncalves:2018yva} at 68\% CL. Turning next to potential lepton colliders, the high-energy stages of the ILC can access the value of $\lambda_{hhh}$ respectively to 27\% accuracy at 500 GeV (with $4\text{ ab}^{-1}$ of data) and 10\% at 1 TeV ($8\text{ ab}^{-1}$ of data)~\cite{Fujii:2015jha} -- both values being once again at 68\% CL. Independently, CLIC could achieve a precision of approximately 10\% (at 68\% CL) by combining the data from all phases (up to 3 TeV)~\cite{Abramowicz:2016zbo,Charles:2018vfv,Roloff:2019crr}. Finally, a possible 100-TeV hadron collider would be able to determine $\lambda_{hhh}$ to a level of accuracy of 5-7\% at 68\% CL using $30\text{ ab}^{-1}$ of data~\cite{Goncalves:2018yva,Cepeda:2019klc,Chang:2018uwu}.

The universality of the value of $\lambda_{hhh}$ in CSI models at one loop also raises the question of the impact of two-loop corrections. Indeed, as the form of the effective potential is changed at two loops -- with new squared-logarithmic terms appearing, see $e.g.$ Refs.~\cite{vanderBij:1983bw,Ford:1992pn,Martin:2001vx} -- one may expect the effective Higgs trilinear coupling derived from it to be modified as well. Moreover, two-loop corrections to $\lambda_{hhh}$ have already been investigated in a number of extension of the SM with enlarged scalar sectors,\footnote{For completeness, we must also mention Refs.~\cite{Brucherseifer:2013qva,Muhlleitner:2015dua} in which the leading two-loop $\mathcal{O}(\alpha_s\alpha_t)$ corrections to $\lambda_{hhh}$ are computed, respectively for the Minimal-Supersymmetric-SM and Next-to-Minimal-Supersymmetric-SM, for the purpose of matching the precision at which the Higgs mass is calculated in these models. } namely in the Inert Doublet Model~\cite{Senaha:2018xek,Braathen:2019pxr,Braathen:2019zoh}, in the 2HDM~\cite{Braathen:2019pxr,Braathen:2019zoh}, and in a real-singlet extension of the SM~\cite{Braathen:2019zoh} (this last model actually corresponds to the $N$-scalar theory we will consider in this work for $N=1$). It is therefore also important to compute two-loop corrections to $\lambda_{hhh}$ in the CSI theories, to match the accuracy achieved in non-CSI scenarios and allow precise comparisons, as the computational method differs between the two types of models.

In this paper, we have performed the first explicit\footnote{ A study including two-loop corrections to the effective potential, and its derivatives, via renormalisation-group improvement were performed in Ref.~\cite{Hill:2014mqa}. } calculation of dominant two-loop corrections to $\lambda_{hhh}$ in theories with CSI: we extend the known one-loop (effective-potential) calculation to two loops, so as to include the same level of corrections as in our previous works on non-CSI models~\cite{Braathen:2019pxr,Braathen:2019zoh}. We investigate in particular two types of BSM scenarios: first, $O(N)$-symmetric CSI models, and second, a CSI variant of the 2HDM~\cite{Lee:2012jn}.
Moreover, we include the renormalisation scheme conversion, from the \msbar scheme (in which the effective potential is obtained) to the OS scheme, in order to express our results in terms of physical quantities -- note that we will be referring to the \msbar and OS results respectively as $\lambda_{hhh}$ and $\hat\lambda_{hhh}$ to avoid confusions. We also take into account the requirement of generating the 125-GeV mass of the Higgs boson correctly, at the two-loop order. This yields a relation between the masses of all the states in the CSI models, which we use to constrain the BSM parameters (masses and couplings) appearing in our calculations. We also ensure that perturbative unitarity is maintained and that the EW vacuum remains the true vacuum of the effective potential for the parameter points we consider. As mentioned already, the Higgs trilinear coupling has a universal prediction at one loop in all CSI models -- $(\lambda_{hhh}^\text{CSI})^{(1)}=5/3(\lambda_{hhh}^\text{SM})^{(0)}$ -- and our main result in this paper is finding that the inclusion of two-loop corrections breaks this universality. Through our detailed numerical studies, we observe that the different theoretical constraints we impose on the considered scenarios limit the allowed range of values for $\lambda_{hhh}$ at two loops -- the strongest constraint coming from the requirement of correctly reproducing the 125-GeV Higgs mass. Nevertheless, the two-loop Higgs trilinear coupling retains dependence on the parameters of the BSM scalar sectors, and may provide an opportunity to distinguish different BSM scenarios with CSI.

This paper is organised as follows: in section~\ref{SEC:GW} we present general results, applicable to a wide range of CSI models, and we illustrate our calculational scheme with the example of the CSI Standard Model. Next, in sections~\ref{SEC:CSIO(N)} and~\ref{SEC:CSI2HDM}, we study the Higgs trilinear coupling and the relation between masses in scenarios of $N$-scalar models with $O(N)$ global symmetries and also in a CSI variant of the 2HDM. We discuss implications of our present in section~\ref{SEC:Discussion}, before concluding in section~\ref{SEC:Conclusion}. A number of appendices provide details about our computations and some results we deemed too long for the main text. Definitions of one- and two-loop functions are given in appendix~\ref{APP:loopfn}, complementary expressions for the $O(N)$-symmetric models are provided in appendix~\ref{APP:O(N)}, and results in the CSI-2HDM for general masses are presented in appendix~\ref{APP:CSI2HDM}.  Finally, we provide in appendix~\ref{APP:generic} generic results for the coefficients in the effective potential that can be applied for any CSI model without mixing. 

\section{Generalities about classically scale-invariant theories}
\label{SEC:GW}
We begin this paper by recalling known features of CSI models, before deriving some new results applicable for all such models.

\subsection{Classically scale-invariant theories}
Classically scale-invariant models arise from the assumption that at some energy scale -- which we will in this paper consider to be at or near the EW scale -- the scalar potential exhibits scale invariance, by which we mean that all mass-dimensionful quantities vanish at the classical ($i.e.$ tree) level. Radiative corrections explicitly violate the CSI and allow the EW gauge symmetry to be broken dynamically, giving rise to a mass scale via dimensional transmutation. This mechanism to realise spontaneous EWSB radiatively was first studied for a model of a single scalar by S. Coleman and E. Weinberg~\cite{Coleman:1973jx}. However, it is known that it is impossible to explain the Higgs boson mass in this simplest single-scalar case: neglecting the top-quark effects, the original paper of Coleman and Weinberg~\cite{Coleman:1973jx} predicted a scalar of mass less than 10 GeV, and including the top quark renders the (one-loop) potential unstable. In order to reproduce the 125-GeV Higgs mass in models with CSI, it is therefore necessary to consider extended scalar sectors. The treatment of this general case of multi-scalar theories requires more care as one must also identify the direction in field space along which the EW gauge symmetry is broken. This is the purpose of the method devised by E. Gildener and S. Weinberg~\cite{Gildener:1976ih}, which we will review briefly now.

\subsection{The Gildener-Weinberg method}
The tree-level scalar potential of a generic CSI theory with scalars $\{\varphi_i\}$ takes the form
\begin{align}
 \vtree(\{\varphi_i\})=\Lambda_{ijkl}\varphi_i\varphi_j\varphi_k\varphi_l\,,
\end{align}
where the $\Lambda_{ijkl}$ are (dimensionless) quartic couplings -- tadpole, mass, and trilinear terms all have a non-zero mass dimension and must therefore vanish. Clearly, this potential does not admit any other minimum than the origin $\{\varphi_i\}=0$ and in order for the EW gauge symmetry to be broken spontaneously using the Coleman-Weinberg mechanism, it is necessary that the tree-level potential vanish along a ray in field space. This particular flat direction corresponds to the SM-like Higgs boson $\varphi_h$ (called \textit{scalon} in Ref.~\cite{Gildener:1976ih}) which carries the EW vacuum expectation value (VEV) $v$ and acquires a mass of 125 GeV at loop level. 

Along the flat direction, the scalar potential is generated at one-loop order, and it can be written as a function of the order parameter ($i.e.$ the Higgs field) $h\equiv\varphi_h-v$ using the supertrace formula~\cite{Jackiw:1974cv}
\begin{align}
\label{EQ:onelooppot_CW}
 \veff(h)=\frac{1}{64\pi^2}\bigg\{&\tr\left[M_S^4(h)\left(\log\frac{M_S^2(h)}{Q^2}-\frac32\right)\right]-4\tr\left[M_f^4(h)\left(\log\frac{M_f^2(h)}{Q^2}-\frac32\right)\right]\nn\\
 &+3\tr\left[M_V^4(h)\left(\log\frac{M_V^2(h)}{Q^2}-\frac56\right)\right]\bigg\}\,,
\end{align}
where $M_{S,f,V}(h)$ denote respectively the scalar, fermion, and gauge-boson field-dependent mass matrices (at tree level), and $Q$ is the renormalisation scale. Note that, this expression is given in the \msbar renormalisation scheme. Additionally, compared to the paper of Gildener and Weinberg~\cite{Gildener:1976ih}, we have kept here all the terms of the supertrace formula, rather than just the logarithmic terms. As we will see in the next section, this has no impact on any physical quantity or discussion.

Due to the absence of any mass term in CSI models, the field-dependent masses of all particles in theory -- scalars, fermions, gauge bosons -- can be related to the field-independent masses as
\begin{equation}
\label{EQ:general_mass_form}
 m_X^2(h)=m_X^2\left(1+\frac{h}{v}\right)^2\,.
\end{equation} 
We can exploit this particularity of CSI models to write the one-loop effective potential under the simple form
\begin{equation}
\label{EQ:form_Veff_1l}
 \veff(h)=\kappa\left[\aol\ (v+h)^4+\bol\ (v+h)^4\log\frac{(v+h)^2}{Q^2}\right]\,,
\end{equation}
where $\kappa$ is the loop factor defined in eq.~(\ref{EQ:loopfactor}) and 
\begin{align}
\label{EQ:def_AB}
 \aol&\equiv \frac{1}{4v^4}\left\{\tr\left[M_S^4\left(\log\frac{M_S^2}{v^2}-\frac32\right)\right]-4\tr\left[M_f^4\left(\log\frac{M_f^2}{v^2}-\frac32\right)\right]+3\tr\left[M_V^4\left(\log\frac{M_V^2}{v^2}-\frac56\right)\right]\right\}\,,\nn\\
 \bol&\equiv \frac{1}{4v^4}\left(\tr\left[M_S^4\right]-4\tr\left[M_f^4\right]+3\tr\left[M_V^4\right]\right)\,.
\end{align}
We will discuss in the following how two-loop corrections modify this form of $\veff$, however, first we will review the one-loop result for the effective Higgs trilinear coupling that is computed with this potential.

\subsection{The Higgs trilinear coupling at one-loop order}
We study in this paper corrections to an effective Higgs trilinear coupling, calculated as the third derivative of the effective potential with respect to the Higgs field, and evaluated at its minimum
\begin{equation}
 \lambda_{hhh}\equiv\frac{\partial^3\veff}{\partial h^3}\bigg|_{h=0}\,.
\end{equation}
In such a computation based on the effective potential, lower derivatives of the potential also play an important role. First of all, from the first derivative of $\veff$ we obtain the tadpole equation
\begin{align}
\label{EQ:1l_tadpole}
 \frac{\partial\veff}{\partial h}\bigg|_{h=0}=2\kappa v^3\bigg(2\aol+\bol+2\bol\log\frac{v^2}{Q^2}\bigg)=0\,,
\end{align}
which allows us -- as $v\neq 0$ -- to eliminate\footnote{This confirms our earlier statement: a finite shift to the quantity $\aol$ plays no role physically, as $\aol$ is anyway eliminated using the tadpole equation. } the quantity $\aol$
\begin{align}
 \aol=-\bol\left(\frac12+\log\frac{v^2}{Q^2}\right)\,.
\end{align}
Next, the second derivative of $\veff$ defines the effective-potential, or curvature, mass of the Higgs boson, which we denote $[M_h^2]_{\veff}$. We have that
\begin{equation}
\label{EQ:1l_Higgscurvmass}
 [M_h^2]_{\veff}\equiv\frac{\partial^2\veff}{\partial h^2}\bigg|_{h=0}=2\kappa v^2\left(6\aol+7\bol+6\bol\log\frac{v^2}{Q^2}\right)=8\kappa v^2\bol\,.
\end{equation}
This equation has a very important consequence, unique to models with classical scale invariance: it relates on the left-hand side the Higgs curvature mass, close to its known pole mass of 125 GeV, to the tree-level masses of all particles in the theory through the quantity $\bol$ on the right-hand side. If the model we consider contains no extra fermions and gauge bosons but only additional scalars, we obtain~\cite{Gildener:1976ih,Lee:2012jn,Hashino:2015nxa,Lane:2018ycs} using eq.~(\ref{EQ:def_AB})
\begin{equation}
 4v^4\bol=8\pi^2v^2[M_h^2]_{\veff}=\tr[M_S^4]-12m_t^4+6m_W^4+3m_Z^4\,,
\end{equation}
or equivalently
\begin{equation}
 \tr[M_S^4]=8\pi^2v^2[M_h^2]_{\veff}+12m_t^4-6m_W^4-3m_Z^4\,,
\end{equation}
having taken into account the $N_c=3$ colour factor for the top quark and with $m_t$, $m_W$, and $m_Z$ the tree-level masses of the top quark and of the $W$ and $Z$ bosons respectively. Once one considers specific scenarios for the particle contents of BSM CSI theories, it becomes possible to obtain constraints on the masses of the new states -- as studied for example in Refs.~\cite{Hashino:2015nxa,Lane:2019dbc,Brooijmans:2020yij}. We will return to this type of relations among masses in the following, when we consider particular BSM models.  

Returning now to the one-loop potential ($c.f.$ eq.~(\ref{EQ:form_Veff_1l})) we find that the only two parameters on which it depends -- $\aol$ and $\bol$ -- are fixed by the tadpole equation and the knowledge of the Higgs mass. This implies that if we take further derivatives of the one-loop potential, the computed quantities are also fixed in terms of the Higgs curvature mass. Indeed, the effective Higgs trilinear coupling is at one-loop order
\begin{align}
 \lambda_{hhh}\equiv\frac{\partial^3\veff}{\partial h^3}\bigg|_{h=0}=4 \kappa v \left(6 \aol + 13 \bol + 6 \bol \log\frac{v^2}{Q^2}\right)\,,
\end{align}
and using equations~(\ref{EQ:1l_tadpole}) and (\ref{EQ:1l_Higgscurvmass}), we can eliminate $\aol$ and $\bol$, and rewrite $\lambda_{hhh}$ in terms of the tree-level result for $\lambda_{hhh}$ in the Standard Model as~\cite{Hashino:2015nxa}
\begin{align}
\label{EQ:1l_res_lambdahhh}
 \lambda_{hhh}=40\kappa v\bol=\frac{5[M_h^2]_{\veff}}{v}=\frac53(\lambda_{hhh}^{(0)})^\text{SM}\,.
\end{align} 
Similarly, we can compute a one-loop (effective) quartic coupling $\lambda_{hhhh}$
\begin{align}
\label{EQ:1l_res_lambdahhhh}
 \lambda_{hhhh}=\frac{\partial^4\veff}{\partial h^4}\bigg|_{h=0}=&\ 4\kappa \left(6 \aol + 25 \bol + 6 \bol \log \frac{v^2}{Q^2}\right)=88\kappa \bol\nn\\
 =&\ \frac{11[M_h^2]_{\veff}}{v^2}=\frac{11}{3}(\lambda_{hhhh}^{(0)})^\text{SM}\,.
\end{align}
Note that for these two last equations, we made use of the known tree-level results for the Standard Model
\begin{align}
 (\lambda_{hhh}^{(0)})^\text{SM}=\frac{3m_h^2}{v}\,,\qquad\text{and}\qquad  (\lambda_{hhhh}^{(0)})^\text{SM}=\frac{3m_h^2}{v^2}\,.
\end{align}
The predictions in equations~(\ref{EQ:1l_res_lambdahhh}) and (\ref{EQ:1l_res_lambdahhhh}) deviate significantly from the usual SM values -- for the case of $\lambda_{hhh}$ this effect is sufficiently large to be observed already at the HL-LHC. A further crucial aspect of these results is that, in CSI models, the Higgs trilinear and quartic coupling are, at one loop, \textit{completely independent} of the particle content of the theory. However, as we will see now, this universality is lost when two-loop corrections are included in the effective potential. 

\subsection{Two-loop corrections to the Higgs trilinear coupling}
\label{SEC:2llambda_hhh}
At two loops, corrections to the effective potential are obtained by computing one-particle irreducible vacuum bubble diagrams. In addition to logarithmic and non-logarithmic terms -- as are present at one loop -- such two-loop diagrams also contain squared logarithms, meaning that the effective potential takes the form 
\begin{align}
\label{EQ:form_Veff_1+2l}
 \veff(h)=A\ (v+h)^4+B\ (v+h)^4\log\frac{(v+h)^2}{Q^2}+C\ (v+h)^4\log^2\frac{(v+h)^2}{Q^2}\,,
\end{align}
where the coefficient $C$ appears only at two-loop order. Expanding this expression in terms of one- and two-loop quantities, it becomes
\begin{align}
 \veff(h)\equiv&\ \kappa\vone(h)+\kappa^2\vtwo(h)+\mathcal{O}(\kappa^3)\nn\\
         =&\ \kappa\bigg[\aol\ (v+h)^4+\bol\ (v+h)^4\log\frac{(v+h)^2}{Q^2}\bigg]\\
          &+ \kappa^2\bigg[\atl\ (v+h)^4+\btl\ (v+h)^4\log\frac{(v+h)^2}{Q^2}+\ctl\ (v+h)^4\log^2\frac{(v+h)^2}{Q^2}\bigg]  +\mathcal{O}(\kappa^3)\nn\,.
\end{align}
While $\aol$ and $\bol$ are exactly the quantities given in eq.~(\ref{EQ:def_AB}), $\atl$, $\btl$, and $\ctl$ must be derived from the two-loop diagrams contributing to the effective potential -- and will depend on the model one is considering.

Before investigating specific models, it is enlighting to repeat the procedure of the previous section, and derive general expressions for the derivatives of $\veff$, using now the two-loop form in eq.~(\ref{EQ:form_Veff_1+2l}). First, the tadpole equation reads
\begin{equation}
 \frac{\partial\veff}{\partial h}\bigg|_{h=0}=2 v^3 \bigg[2 A + B + 2 (B + C) \log\frac{v^2}{Q^2} + 2 C \log^2\frac{v^2}{Q^2}\bigg]=0\,,
\end{equation}
from which we can isolate $A$ as
\begin{align}
\label{EQ:A_2l}
 A=-\frac12B-(B+C)\log\frac{v^2}{Q^2}-C\log^2\frac{v^2}{Q^2}\,.
\end{align}
Next, the Higgs curvature mass is, after eliminating $A$
\begin{align}
\label{EQ:2l_Mh}
 [M_h^2]_{\veff}=\frac{\partial^2\veff}{\partial h^2}\bigg|_{h=0}
 &=8 v^2 \bigg[B + C + 2 C \log\frac{v^2}{Q^2}\bigg]\,,
\end{align}
so that $B$ can be rewritten in terms of $[M_h^2]_{\veff}$ and $C$
\begin{equation}
\label{EQ:B_2l}
 B=\frac{[M_h^2]_{\veff}}{8v^2}-C\bigg(1+2\log\frac{v^2}{Q^2}\bigg)\,.
\end{equation}
Equation~(\ref{EQ:2l_Mh}) is also important as it constitutes the two-loop version of the relation between masses in CSI models -- which we first encountered at one loop in the previous section. Once we consider particular BSM models in sections~\ref{SEC:CSIO(N)} and~\ref{SEC:CSI2HDM}, we will see that this relation places stringent constraints on the allowed BSM mass and parameter ranges. 

Now, if we compute the effective Higgs trilinear coupling, we obtain at two loops
\begin{align}
\label{EQ:res_2l_lambdahhh}
 \lambda_{hhh}=\frac{\partial^3\veff}{\partial h^3}\bigg|_{h=0}
 &=\frac{5 [M_h^2]_{\veff}}{v} + 32 C v\,,
\end{align}
once $A$ and $B$ are replaced using eqs.~(\ref{EQ:A_2l}) and~(\ref{EQ:B_2l}). 
We see here that because of the new term $C$, appearing from two loops in $\veff$, non-universal, model-dependent corrections arise in the expression of $\lambda_{hhh}$. In practice we will be computing $C$ from the two-loop diagrams contributing to $\veff$ and use this result to obtain $\lambda_{hhh}$ at two loops. However, we note that we can formally use the last equation to determine $C$ in terms of $[M_h^2]_{\veff}$ and $\lambda_{hhh}$, as
\begin{align}
 C=\frac{1}{32v}\bigg(\lambda_{hhh}-\frac{5[M_h^2]_{\veff}}{v}\bigg)\,,
\end{align}
We can finally use this expression of $C$ to compute $\lambda_{hhhh}$ at two loops in terms of $[M_h^2]_{\veff}$ and $\lambda_{hhh}$ only
\begin{align}
 \lambda_{hhhh}=\frac{\partial^4\veff}{\partial h^4}\bigg|_{h=0}
 = \frac{11 [M_h^2]_{\veff}}{v^2}+192 C
 =\frac{6\lambda_{hhh}}{v}-\frac{19 [M_h^2]_{\veff}}{v^2}\,.
\end{align}
This equation implies that once $\lambda_{hhh}$ is measured, the Higgs quartic coupling can be predicted in a universal way for all CSI theories, up to two loops. This being said, once three-loop corrections are included, we can expect this universality to be lost because of 
$\log^3$ terms appearing in $\veff$ -- similarly to what happens with $\lambda_{hhh}$ once two-loop effects are taken into account.

\subsection{Results for the classically scale-invariant Standard Model}
\label{SEC:CSISM}

\begin{figure}
 \centering
 \includegraphics[width=.8\textwidth]{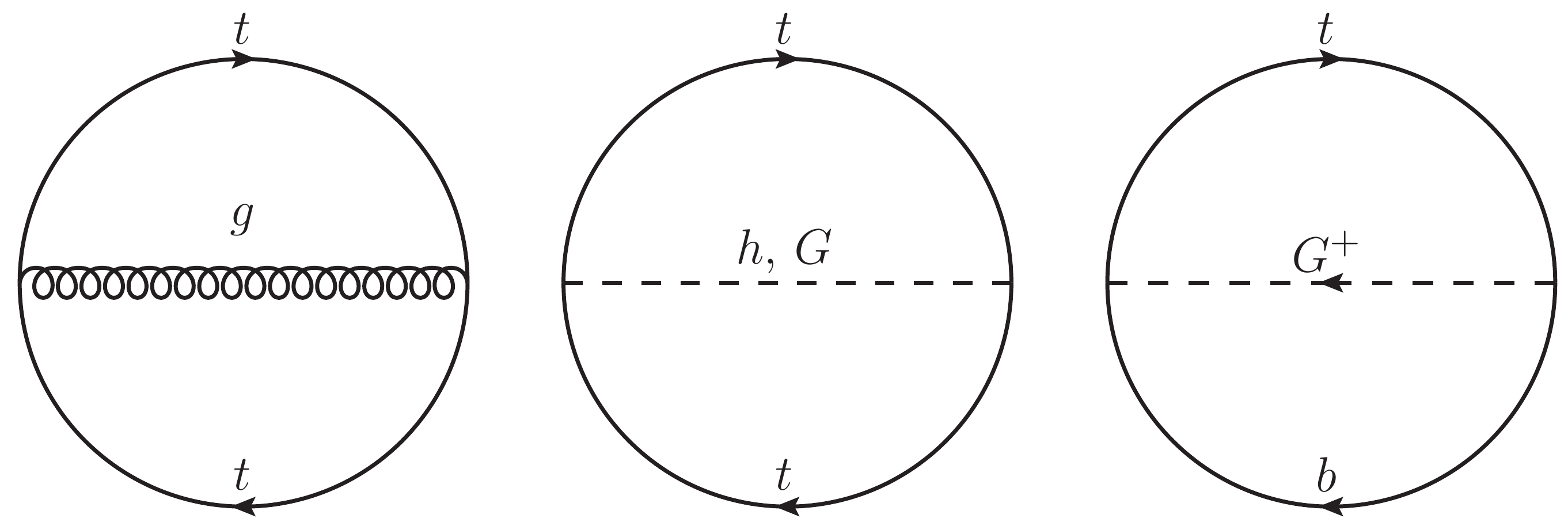}
 \caption{Dominant two-loop diagrams contributing to the SM effective potential. }
 \label{FIG:SM2ldiags}
\end{figure}

We end this section by calculating the leading two-loop contributions to $\veff$ and $\lambda_{hhh}$ for the classically scale-invariant version of the SM, which will also provide an occasion to present the renormalisation scheme conversion that we perform to obtain expressions in terms of physical quantities. We must emphasise that the CSI-SM is already known not to be a valid theory of Nature as it cannot reproduce the Higgs mass correctly~\cite{Coleman:1973jx}, however, its simplicity makes it the ideal setting to explain our calculational setup. 

The tree-level scalar potential of the CSI-SM does not contain a mass term, and reads simply
\begin{align}
 \vtree=\lambda|\Phi|^4\,,
\end{align}
where the Higgs doublet $\Phi$ is defined as
\begin{equation}
\label{EQ:SMdoublet}
 \Phi=\begin{pmatrix}
  G^+\\ \frac{1}{\sqrt{2}}(v+h+iG)
 \end{pmatrix}\,.
\end{equation} 
$G$ and $G^\pm$ are respectively the neutral and charged would-be Goldstone bosons, while $h$ is the 125-GeV Higgs boson. 
At tree level, the field-dependent masses of the Higgs and Goldstone bosons read
\begin{equation}
 m_h^2(h)=3\lambda (v+h)^2\,,\qquad\text{and}\qquad m_G^2(h)=\lambda (v+h)^2\,,
\end{equation}
while the top-quark mass is (like for the usual SM)
\begin{align}
 m_t(h)=\frac{y_t}{\sqrt{2}}(v+h)\,.
\end{align}
The need for a flat direction in the tree-level potential imposes $\lambda=0$, which is also the result of solving the tree-level tadpole equation. 

The dominant contributions to the CSI-SM two-loop effective potential arise from the diagrams shown in figure~\ref{FIG:SM2ldiags} and involving the $SU(3)_c$ gauge coupling $g_3$ and the top Yukawa coupling $y_t$ -- the two EW gauge couplings $g_2$ and $g_Y$, and the light fermion Yukawa couplings are negligible in comparison. These contributions read (in the \msbar scheme)
\begin{align}
\label{EQ:CSISM_Vtwo}
 \vtwo(h)=&-4g_3^2m_t^2(h)\bigg(4J(m_t^2(h))-8m_t^2(h)-\frac{6J(m_t^2(h))^2}{m_t^2(h)}\bigg)\,,\nn\\
 &+3 y_t^2 \bigg[2 m_t^2(h) I (m_t^2(h), m_t^2(h), 0) + m_t^2(h) I (m_t^2(h), 0, 0) + J(m_t^2(h))^2\bigg]\,,
\end{align} 
where $J$ and $I$ are respectively one- and two-loop functions, for which definitions\footnote{Note also that the function we denote as $J$ is the same as the Passarino-Veltman function $A$. We choose the notation $J$ as $e.g.$ in Refs.~\cite{Ford:1992pn,Martin:2001vx} to avoid confusion with the quantity $A$ in the effective potential -- $c.f.$ eq.~(\ref{EQ:form_Veff_1+2l}). } and useful limits are given in appendix~\ref{APP:loopfn}.

In the previous section, we found that to obtain the two-loop corrections to $\lambda_{hhh}$, we need to compute $C$, $i.e.$ the coefficient of the $\log^2$ term in $\veff$. For the CSI-SM, after expanding the expression in equation~(\ref{EQ:CSISM_Vtwo}), we find 
\begin{align}
 \ctl=\frac{24g_3^2m_t^4}{v^4}-\frac{9m_t^6}{v^6}\,.
\end{align}
Using then equation~(\ref{EQ:res_2l_lambdahhh}), we obtain for the effective Higgs trilinear coupling in the CSI-SM, to leading two-loop order,
\begin{align}
\label{EQ:CSISM_lambdahhh_2l}
 \lambda_{hhh}=\frac{5[M_h^2]_{\veff}}{v}+\frac{1}{(16\pi^2)^2}\bigg[\frac{768g_3^2m_t^4}{v^3}-\frac{288m_t^6}{v^5}\bigg]\,.
\end{align}
At this point, we should also discuss the choice of renormalisation scheme for the parameters appearing in the results presented in this work. As the effective potential, from which we derive $\lambda_{hhh}$, is calculated in the \msbar scheme, the expression in equation~(\ref{EQ:CSISM_lambdahhh_2l}) should be understood as being written in terms of \msbar parameters. In the following, we will however choose to convert \msbar expressions into on-shell (OS) ones -- $i.e.$ written in terms of physical quantities and including also finite corrections from wave-function renormalisation (WFR). An OS version of the Higgs trilinear coupling can be obtained as
\begin{align}
 \hat\lambda_{hhh}=\left(\frac{Z_h^\text{OS}}{Z_h^\msbar}\right)^{\frac{3}{2}}\lambda_{hhh}=\left(1+\frac32\kappa\frac{d\Pi_{hh}^{(1)}}{dp^2}\bigg|_{p^2=0}+\mathcal{O}(\kappa^2)\right)\lambda_{hhh}\,,
\end{align}
where $Z_h^\text{OS}$ and $Z_h^\msbar$ are respectively the on-shell and \msbar scheme Higgs WFR constants, $\Pi^{(1)}_{hh}$ is the finite part of the one-loop Higgs boson self-energy, and with $\lambda_{hhh}$ expressed in terms of physical quantities. Note that it suffices to compute the WFR constants to one-loop order, because $\lambda_{hhh}$ is generated at one loop.

In the two-loop result for $\lambda_{hhh}$ in eq.~(\ref{EQ:CSISM_lambdahhh_2l}), $m_t$ and $g_3$ only enter at two loops, and therefore they require no scheme conversion here -- this would only produce corrections at three-loop order. However, it is necessary to include a one-loop scheme conversion for the Higgs \msbar VEV $v$ and curvature mass $[M_h^2]_{\veff}$, both appearing from one loop. First, the \msbar VEV can be related to the Fermi constant at (leading) one loop as
\begin{align}
\label{EQ:VEVconv}
 v^2=\frac{1}{\sqrt{2}G_F}+\frac{3M_t^2}{16\pi^2}\big(2\llog M_t^2-1\big)\,.
\end{align}
For notational convenience, we also define an OS Higgs VEV as $\vphys^2\equiv(\sqrt{2} G_F)^{-1}$. Next, we recall that the Higgs pole ($i.e.$ physical) and curvature masses are related by the following formula
\begin{align}
 M_h^2=[M_h^2]_{\veff}+\Pi_{hh}(p^2=M_h^2)-\Pi_{hh}(p^2=0)\,.
\end{align}
In the CSI-SM, the Higgs mass is generated at loop level, and thus this relation simplifies into
\begin{align}
 M_h^2=[M_h^2]_{\veff}+\frac{M_h^2}{16\pi^2}\bigg[\frac{2M_t^2}{\vphys^2}\big(2+3\llog M_t^2\big)\bigg]\,.
\end{align}
Combining all these intermediate results together, we obtain finally
\begin{align}
 \hat\lambda_{hhh}=\frac{5M_h^2}{\vphys}+\frac{1}{16\pi^2}\frac{35}{2}\frac{M_h^2M_t^2}{\vphys^3}+\frac{1}{(16\pi^2)^2}\bigg[\frac{768g_3^2M_t^4}{\vphys^3}-\frac{288M_t^6}{\vphys^5}\bigg]\,.
\end{align}
It is important to note that, because $M_h^2$ is generated at loop level, the second term, which is proportional to $\kappa M_h^2 M_t^2/\vphys^3$, is formally of \textit{two-loop} order -- only the first term in this expression is of one-loop order. 

Taking from the PDG~\cite{Zyla:2020zbs}, the following values for the physical inputs
\begin{align}
\label{EQ:SMinputs}
 M_h=125.10\gev\,,\quad M_t=172.4\gev\,, \quad G_F=1.1663787\cdot 10^{-5}\gev^{-2}\,, \quad \alpha_s(M_Z)=0.1179\,,
\end{align}
we find numerically
\begin{align}
 \frac{5M_h^2}{\vphys}=&\ 317.8\gev\nn\\
 \frac{5M_h^2}{\vphys}+\frac{1}{16\pi^2}\frac{35}{2}\frac{M_h^2M_t^2}{\vphys^3}+\frac{1}{(16\pi^2)^2}\bigg[\frac{768g_3^2M_t^4}{\vphys^3}-\frac{288M_t^6}{\vphys^5}\bigg]=&\ 323.6\gev\,,
\end{align} 
or in other words, the two-loop corrections to $\lambda_{hhh}$ yield a positive shift of the order of $2\%$.

\section{$N$-scalar models}
\label{SEC:CSIO(N)}
\subsection{Model definitions}
The first class of BSM CSI theories that we consider are extensions of the SM with $N$ additional real singlet scalar states. We will furthermore assume that these models are endowed with a global $O(N)$ symmetry, under which only the singlet scalars are charged while the SM particles do not carry any charge. Although this assumption greatly simplifies the expressions we obtain -- by reducing the number of different masses and coupling constants -- it is important to note that it does not significantly affect our discussion about corrections to the Higgs trilinear coupling. 
Such $O(N)$-symmetric CSI models have been discussed in many references -- see Refs.~\cite{Endo:2015ifa,Endo:2015nba,Endo:2016koi,Helmboldt:2016mpi,Fujitani:2017gma} -- from various points of view. In particular, the exact $O(N)$ symmetry provides a symmetry to stabilise dark matter candidates in these models. It was found in Ref.~\cite{Endo:2015nba} that there is an upper bound on the value of $N$ if one considers direct dark matter search results. However, in this paper, we choose to study $O(N)$ models exclusively from the point of view of Higgs physics and corrections to $\lambda_{hhh}$, and we do not consider potential dark matter aspects of these models. In other words, we consider here $O(N)$ as a phenomenological symmetry -- which greatly simplifies the study of Higgs physics in $N$-scalar theories -- rather than a symmetry to stabilise dark matter. 

We can write the scalar potential of $O(N)$-symmetric models with CSI as
\begin{align}
\label{EQ:O(N)_pot}
 \vtree=&\ \lambda |\Phi|^4+\lambda_{\Phi S}\vec{S}^2|\Phi|^2+\frac{1}{4}\lambda_S(\vec{S}^2)^2\,,
\end{align}
where $\Phi$ is the SM-like Higgs doublet -- same as that shown in eq.~(\ref{EQ:SMdoublet}) -- and  $\vec{S}=(S_1\,, S_2\,, \cdots S_N)$ is a real scalar, singlet under the SM gauge groups and belonging to an $N$-dimensional representation of the global $O(N)$ symmetry group -- another way to see this is to think of $N$ real scalar singlets. Only the neutral component of the Higgs doublet acquires a non-zero VEV -- the singlets do not as this would break the global $O(N)$ symmetry. 

The scalar masses at tree-level are 
\begin{align}
 m_h^2(h)=3\lambda (v+h)^2\,,\qquad m_G^2(h)=\lambda (v+h)^2\,,\qquad m_{S_i}^2(h)=\lambda_{\Phi S} (v+h)^2\equiv m_S^2(h)\,.
\end{align} 
and the tree-level tadpole condition simply gives $\lambda=0$. Finally, it follows immediately from this equation that we can replace $\lambda_{\Phi S}$ by the tree-level BSM scalar mass $m_S$, as $\lambda_{\Phi S}=m_S^2/v^2$.

\subsection{Dominant two-loop corrections to the Higgs trilinear coupling}
\label{SEC:O(N)_lambdahhh}
\begin{figure}
 \centering
 \includegraphics[width=.6\textwidth]{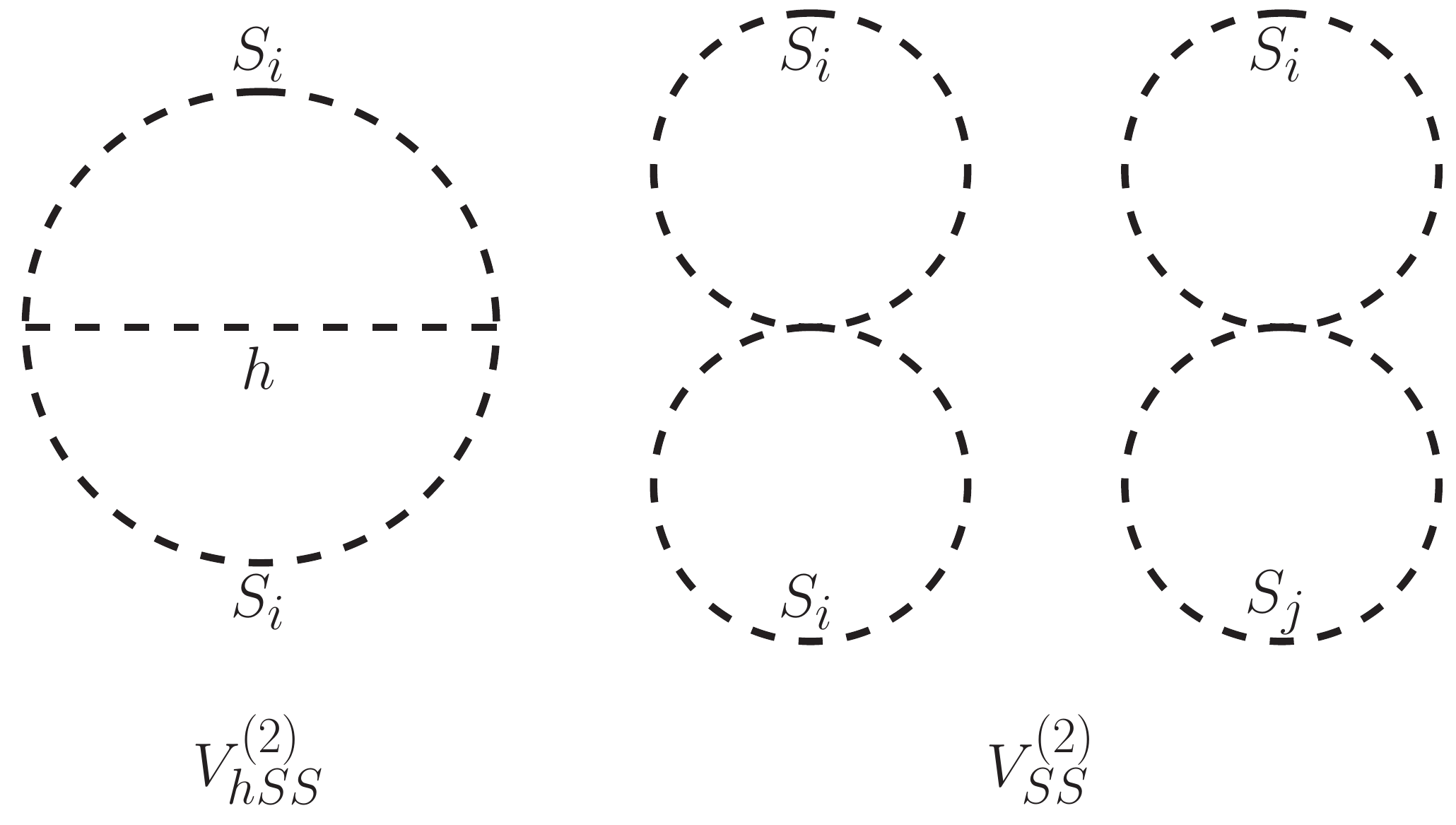}
 \caption{Leading two-loop BSM diagrams contributing to the effective potential of the $O(N)$-symmetric CSI model -- in addition to the leading SM diagrams of figure~\ref{FIG:SM2ldiags}. }
 \label{FIG:O(N)2ldiags}
\end{figure}
We can obtain for the leading BSM contributions to $\vtwo$ by considering the new diagrams involving $\lambda_{\Phi S}$ and $\lambda_S$, shown in figure~\ref{FIG:O(N)2ldiags}. Their expressions are
\begin{align}
 \vtwo=&\ \vtwo_{hSS}+\vtwo_{SS}\,,\nn\\
 \text{with}\qquad\vtwo_{hSS}=&-N\lambda_{\Phi S}^2 (v+h)^2 I(m_S^2(h),m_S^2(h),0)\,,\nn\\
 \vtwo_{SS}=&\ \frac14N(N+2)\lambda_SJ(m_S^2(h))^2\,.
\end{align}
The squared-logarithm terms in these two-loop contributions are, in turn
\begin{align}
 \ctl_{hSS}=&\ \frac{Nm_S^6}{v^6} \,,\nn\\
 \ctl_{SS}=&\ \frac14N(N+2)\frac{\lambda_Sm_S^4}{v^4}\,.
\end{align}
Using once again equation~(\ref{EQ:res_2l_lambdahhh}), we can obtain the leading two-loop order \msbar result for $\lambda_{hhh}$ 
\begin{align}
 \lambda_{hhh}=\frac{5[M_h^2]_{\veff}}{v}+\frac{1}{(16\pi^2)^2}\bigg[\frac{768g_3^2m_t^4}{v^3}-\frac{288m_t^6}{v^5}+\frac{32Nm_S^6}{v^5}+8N(N+2)\frac{\lambda_Sm_S^4}{v^3}\bigg]\,.
\end{align}
Among the terms in the brackets, the first two (those involving $m_t$) are the SM-like contributions, while the last two are the new BSM effects. 
Note that, like in the SM case, $m_S$ and $\lambda_S$ only enter the expression of $\lambda_{hhh}$ at two loops, and hence we do not need to specify their renormalisation scheme. If we want to express the one-loop result in terms of $M_h^2$ and $\vphys$ we must however include scheme conversions for both $[M_h^2]_{\veff}$ and $v$ and take into account WFR effects -- as discussed already in section~\ref{SEC:CSISM}. The only difference we must consider arises from the new momentum-dependent term in the Higgs self-energy in the $O(N)$-symmetric model, which reads
\begin{align}
 \Pi^{(1)}_{hh}(p^2)\supset-\frac{2Nm_S^4}{v^2}B_0(p^2,m_S^2,m_S^2)\,,
\end{align}
where the one-loop Passarino-Veltman function~\cite{Passarino:1978jh} $B_0$ is defined in appendix~\ref{APP:loopfn}. 
Adding this additional term, we find the OS expression for $\hat\lambda_{hhh}$ at leading two loops to be
\begin{align}
 \hat\lambda_{hhh}=&\ \frac{5M_h^2}{\vphys}+\frac{1}{16\pi^2}\frac{5M_h^2}{\vphys^3}\bigg[\frac{7}{2}M_t^2-\frac{1}{6}NM_S^2\bigg]\nn\\
 &+\frac{1}{(16\pi^2)^2}\bigg[\frac{768g_3^2M_t^4}{\vphys^3}-\frac{288M_t^6}{\vphys^5}+\frac{32NM_S^6}{\vphys^5}+8N(N+2)\frac{\lambda_SM_S^4}{\vphys^3}\bigg]\,.
\end{align}
Once again, we emphasise that the terms in the first pair of brackets -- proportional to $\kappa M_h^2/\vphys^3$ -- are formally of two-loop order. 

In principle the above result for $\hat\lambda_{hhh}$ involves 3 BSM parameters, namely $M_S$, $\lambda_S$, and $N$. However, the classical scale invariance produces a further constraint on the BSM parameters, by relating the masses of the particles in the theory, as we will see in the following section.

\subsection{Mass relation}
\label{SEC:O(N)_massrel}
When investigating derivatives of the effective potential in general CSI models in section~\ref{SEC:2llambda_hhh}, we had found that the curvature mass of the Higgs boson is related to the $B$ and $C$ coefficients in $\veff$. In terms of one- and two-loop quantities, this relation reads
\begin{align}
\label{EQ:2l_massrelation}
 8\pi^2[M_h^2]_{\veff}=4v^2\bol+\frac{4v^2}{16\pi^2}\left[\btl+\ctl\left(1+2\log\frac{v^2}{Q^2}\right)\right]\,.
\end{align}
As we have computed the effective potential in $O(N)$-symmetric CSI models, it is a simple matter to derive all the necessary ingredients to apply this relation to this class of models. In addition to $\ctl$, which was given in the previous section, we also need the logarithmic terms at one and two loops -- $\bol$ and $\btl$. Including both SM-like and BSM terms, these are
\begin{align}
 \bol=&\ \frac{1}{4v^4}\left(Nm_S^4-12m_t^4+6m_W^4+3m_Z^4\right)\,,\nn\\
 \btl=&\ \frac{4g_3^2m_t^4}{v^4}\left[12\log\frac{m_t^2}{v^2}-16\right]+\frac{6m_t^6}{v^6}\left[-3\log\frac{m_t^2}{v^2}+8\right]\nn\\
 &+\frac{Nm_S^6}{v^6} \bigg[2\log \frac{m_S^2}{v^2}-4\bigg]+\frac12N(N+2)\frac{\lambda_Sm_S^4}{v^4}\bigg[\log \frac{m_S^2}{v^2}-1\bigg]\,.
\end{align}
Inserting these into equation~(\ref{EQ:2l_massrelation}), we obtain in terms of \msbar-renormalised parameters
\begin{align}
 8\pi^2v^2[M_h^2]_{\veff}=&\ Nm_S^4-12m_t^4+6m_W^4+3m_Z^4\nn\\
 &+\frac{4m_t^4}{16\pi^2}\bigg[g_3^2\left(48\log\frac{m_t^2}{Q^2}-40\right)+\frac{m_t^2}{v^2}\left(-18\log\frac{m_t^2}{Q^2}+39\right)\bigg]\nn\\
 &+\frac{4m_S^4}{16\pi^2}\bigg[\frac{Nm_S^2}{v^2} \bigg(2\log \frac{m_S^2}{Q^2}-3\bigg)+\frac12N(N+2)\lambda_S\bigg(\log \frac{m_S^2}{Q^2}-\frac12\bigg)\bigg]\,.
\end{align}
The first line is simply the one-loop result, while the second and third lines are respectively the SM-like and BSM two-loop corrections to the relation between masses. 

In order to convert this result in terms of physical masses for the SM and BSM states, we require the Higgs and BSM scalar self-energies -- provided in appendix~\ref{APP:CSIO(N)} -- as well as equation~(\ref{EQ:VEVconv}) for the Higgs VEV. We obtain finally
\begin{align}
\label{EQ:O(N)_2l_massrelationOS}
 \frac{4\sqrt{2}\pi^2}{G_F}M_h^2=&\ NM_S^4-12M_t^4+6M_W^4+3M_Z^4\\
 &+M_h^2\bigg[\frac72 M_t^2-\frac16NM_S^2\bigg]+\frac{3M_t^4}{8\pi^2}\bigg[16g_3^2-\frac{6M_t^2}{\vphys^2}\bigg]+\frac{NM_S^4}{16\pi^2}\bigg[(N+2)\lambda_S+\frac{4M_S^2}{\vphys^2}\bigg]\,.\nn
\end{align}
If one considers this relation from the point of view of computing $M_S$ in terms of $\lambda_S$, $N$, and the known SM inputs, the two-loop corrections decrease the extracted value of $M_S$. However, in our numerical investigations, we will instead use this relation to compute $\lambda_S$ as a function of $M_S$ and $N$.

\subsection{Numerical study}
At this point, we have derived all the analytical results necessary to study the possible magnitude of the leading two-loop corrections to the Higgs trilinear coupling in the $O(N)$-symmetric CSI model. Rather than examining the absolute value of $\hat\lambda_{hhh}$, we prefer to investigate by how much it deviates from the prediction in the (non-CSI) Standard Model, at the same leading two-loop order. Therefore, we present in the following results for the BSM deviation $\delta R$, defined as 
\begin{align}
\label{EQ:deltaR_def}
 \delta R&\equiv \frac{\hat\lambda_{hhh}^\text{BSM}-\hat\lambda_{hhh}^\text{SM}}{\hat\lambda_{hhh}^\text{SM}}\,,
\end{align}
where $\hat\lambda_{hhh}$ is the leading two-loop result for the Higgs trilinear coupling in the OS scheme and the SM -- its analytical expression can be found for instance in Ref.~\cite{Braathen:2019zoh} (see also Ref.~\cite{Senaha:2018xek,Braathen:2019pxr}). Taking the same input values as in eq.~(\ref{EQ:SMinputs}), we obtain numerically $\hat\lambda_{hhh}^\text{SM}\simeq 177.4\gev$.

\begin{figure}
 \centering
 \includegraphics[width=.7\textwidth]{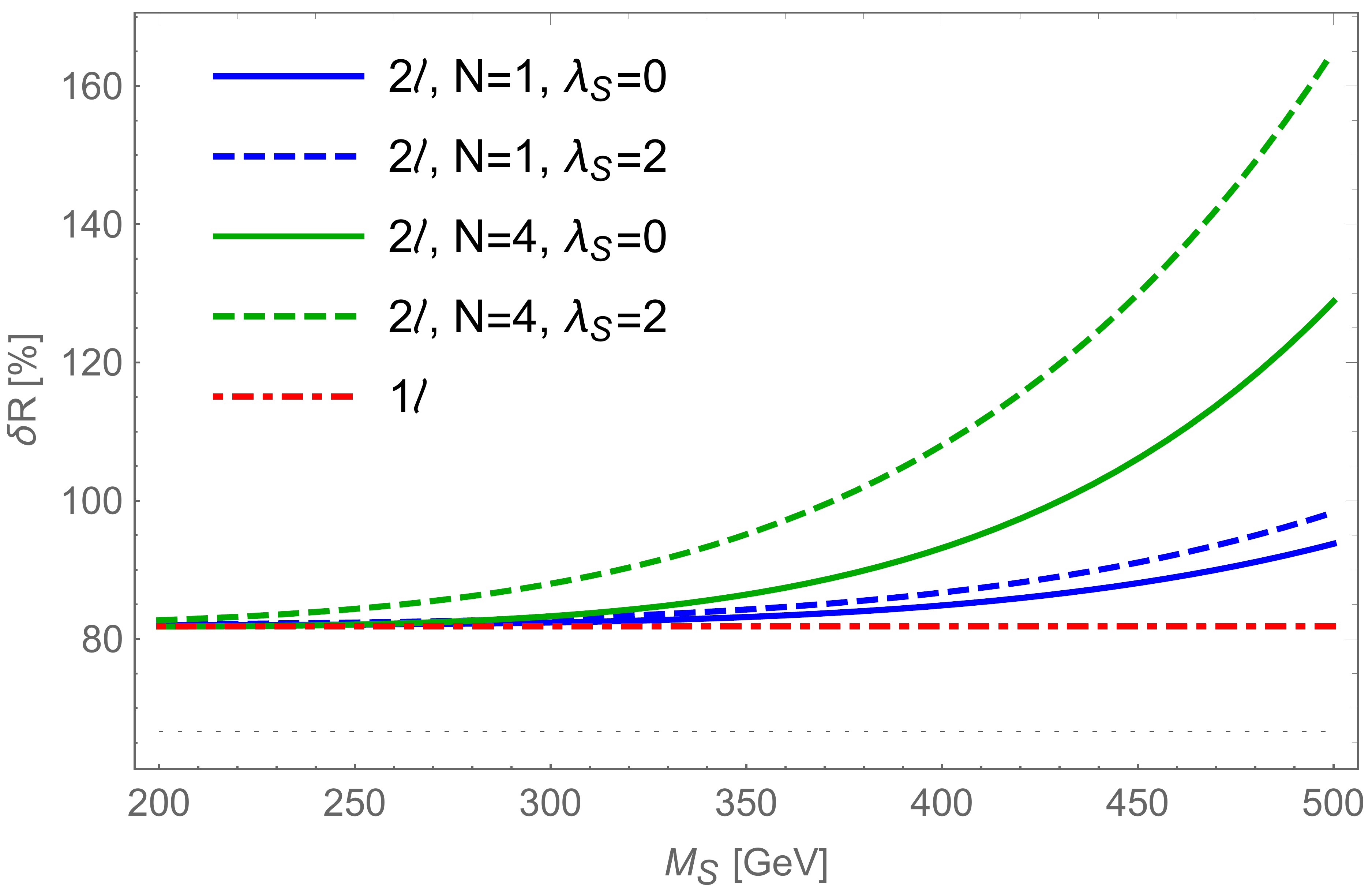}
 \caption{BSM deviation $\delta R$ -- defined in eq.~(\ref{EQ:deltaR_def}) -- of the Higgs trilinear coupling $\hat\lambda_{hhh}$ computed in the $O(N)$-symmetric CSI model with respect to its prediction in the usual ($i.e.$ non-CSI) SM. The blue (green) curves show the results at two loops for $N=1$ ($N=4$). Solid and dashed curves are plotted respectively for $\lambda_S=0$ and $\lambda_S=2$. The red dot-dashed line shows the one-loop result -- $i.e.$ comparing the one-loop CSI result of $5M_h^2/\vphys$ to the one-loop (non-CSI) SM result. The black dotted line corresponds to the comparison of the one-loop CSI result for $\lambda_{hhh}$ to the SM result at tree level -- $i.e.$ the 67\% deviation mentioned in the introduction. }
 \label{FIG:O(N)_deltaRnonCSI1}
\end{figure}

We present in figure~\ref{FIG:O(N)_deltaRnonCSI1} our results for $\delta R$ using only the expressions from section~\ref{SEC:O(N)_lambdahhh} -- in other words, we do not impose the relation between masses from eq.~(\ref{EQ:O(N)_2l_massrelationOS}) yet. The red dot-dashed line shows the one-loop result, comparing the one-loop CSI result $(\hat\lambda_{hhh})^{(1)}=5M_h^2/\vphys$ to the (leading) one-loop expression in the SM -- see for instance Ref.~\cite{Kanemura:2004mg}. We should point out here that this result differs from the $\sim 67\%$ mentioned in the introduction, and shown with the black dotted curve: we have $(\delta R)^{(1)}\simeq 82\%$. This is because the value of $\sim 67\%$ is obtained by comparing the one-loop $(\hat\lambda_{hhh}^\text{CSI})^{(1)}$ to the tree-level SM result for $(\hat\lambda_{hhh}^\text{SM})^{(0)}$, but once we include the one-loop contribution from the top quark in the SM expression we obtain a smaller value for $(\hat\lambda_{hhh}^\text{SM})^{(1)}<(\hat\lambda_{hhh}^\text{SM})^{(0)}$, and hence a larger deviation. We must emphasise that with our definition of $(\delta R)^{(1)}$ we are comparing \textit{one-loop effective-potential results}, which do not include any dependence on external momenta. It should however be noted that, as studied in Ref.~\cite{Kanemura:2004mg}, the momentum dependence in the SM result can be quite significant, and could modify significantly the value of the BSM deviation. In Ref.~\cite{Fujitani:2017gma}, momentum-dependent corrections were also included in the study of the Higgs trilinear coupling in $O(N)$-symmetric CSI models, and were found to be non-negligible.  
Nevertheless, we still choose to define $(\delta R)^{(1)}$ in terms of effective Higgs trilinear couplings, as this allows simpler comparisons with our new two-loop results, calculated with the effective potential. The result we find for $(\delta R)^{(1)}=(\hat\lambda_{hhh}^\text{CSI})^{(1)}/(\hat\lambda_{hhh}^\text{SM})^{(1)}-1$ remains totally independent of the particle content of the CSI model. 

Next, the blue and green curves in figure~\ref{FIG:O(N)_deltaRnonCSI1} are the values of $\delta R$ at two loops, for different choices of $N$ and $\lambda_S$: blue and green curves correspond respectively to $N=1$ and $N=4$; solid curves are made for $\lambda_S=0$ while the dashed ones are for $\lambda_S=2$. For the lowest values of $M_S$ -- $i.e.$ where the BSM effects are the smallest -- the two-loop deviation $(\delta R)^{(2)}$ is only slightly larger than at one loop, however, as one increases $M_S$, the BSM contributions grow rapidly. This growth is stronger for larger values of $N$ and $\lambda_S$, and, \textit{crucially}, the universality of the BSM shift found at one loop is entirely lost at two loops. Looking at concrete numerical values of $(\delta R)^{(2)}$, for example for $M_S=500\gev$, we have on the one hand for $N=1$ and $\lambda_S=0$ a deviation of $\sim 94\%$, while on the other hand for $N=4$ and $\lambda_S=2$ we find $(\delta R)^{(2)}\sim 168\%$. Anticipating a little on the discussion in the next pages, we note that all parameter points in this figure~\ref{FIG:O(N)_deltaRnonCSI1} fulfill the criterion of tree-level perturbative unitarity~\cite{Lee:1977eg,Hashino:2016rvx}. 

\begin{figure}[h]
 \includegraphics[width=\textwidth]{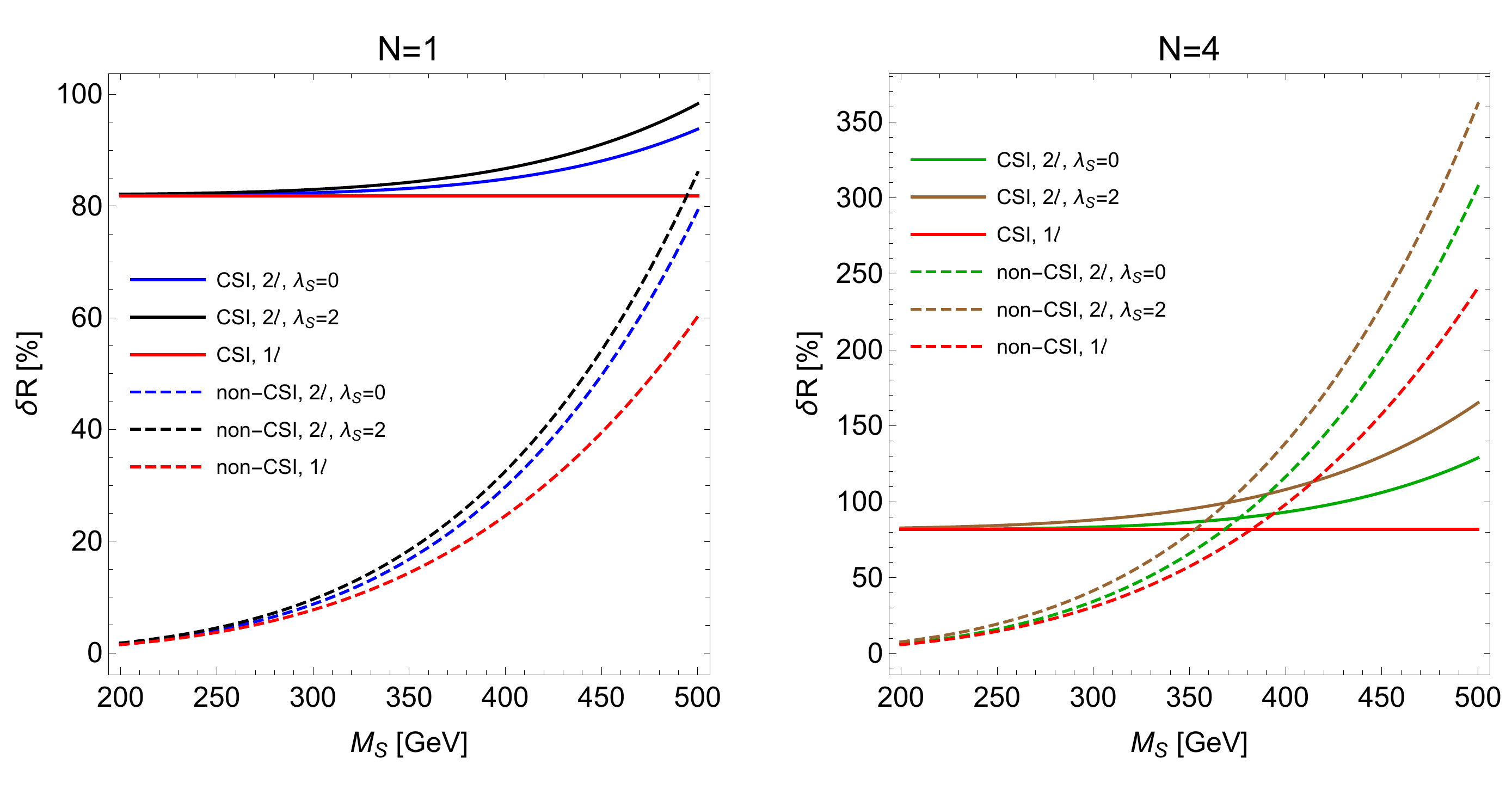}
 \caption{Comparison of the BSM deviations $\delta R$ computed in $O(N)$-symmetric models, with (solid curves) and without (dashed curves) CSI, as a function of the pole mass $M_S$ of the BSM scalars. Results for $N=1$ and $N=4$ are shown respectively in the left- and right-hand plots. Red curves are one-loop values, blue and green curves are two-loop results for $\lambda_S=0$ (respectively for $N=1$ and $N=4$), and black and brown curves are two-loop results for $\lambda_S=2$ (respectively for $N=1$ and $N=4$).}
 \label{FIG:O(N)_compareCSInonCSI}
\end{figure}

Another interesting comparison we can perform is to investigate the difference in the BSM deviation in $\hat\lambda_{hhh}$ between CSI and non-CSI versions of $O(N)$-symmetric models. For this purpose, we derive in appendix~\ref{APP:nonCSIO(N)} the dominant two-loop corrections to $\hat\lambda_{hhh}$ in a non-CSI variant of the $O(N)$-symmetric model, and for generic values of $N$ -- this calculation follows the procedure developed in Refs.~\cite{Braathen:2019pxr,Braathen:2019zoh} (note also that the case of $N=1$ was already treated in Ref.~\cite{Braathen:2019zoh}, under the name of ``Higgs singlet model''). We present in figure~\ref{FIG:O(N)_compareCSInonCSI} a comparison of our results for the BSM deviation $\delta R$ computed with respect to the SM result in both CSI and non-CSI $O(N)$-symmetric models, for $N=1$ (left side) and for $N=4$ (right side). Note that, as there is no mass term in the CSI $O(N)$-symmetric model, we must set $\tilde\mu_S=0$ in the result for the non-CSI version of the model in order to be able to compare both variants consistently.  
The solid curves in fig.~\ref{FIG:O(N)_compareCSInonCSI} are obtained for the versions of the models with CSI, while the results for the non-CSI variants are given by the dashed curves. Additionally, in both plots, the red curves show one-loop values, while the blue and black (green and brown) curves are the two-loop results for $N=1$ ($N=4$) for $\lambda_S=0$ and $\lambda_S=2$. 

One clearly notices that CSI and non-CSI variants of this BSM model behave quite differently. First, for low BSM masses, the BSM deviation in the non-CSI case is minute, while in the CSI case the deviation is already large ($80\%$ or so) -- this is because of the difference between the normal SM and CSI SM. Then, when $M_S$ increases, the non-CSI corrections grow much faster than their CSI counterparts, as in the non-CSI models the one-loop corrections are proportional to $M_S^4$ -- $c.f.$ equation~(\ref{EQ:nonCSIO(N)_lambdahhh_1l}). 

\begin{figure}[h]
 \includegraphics[width=\textwidth]{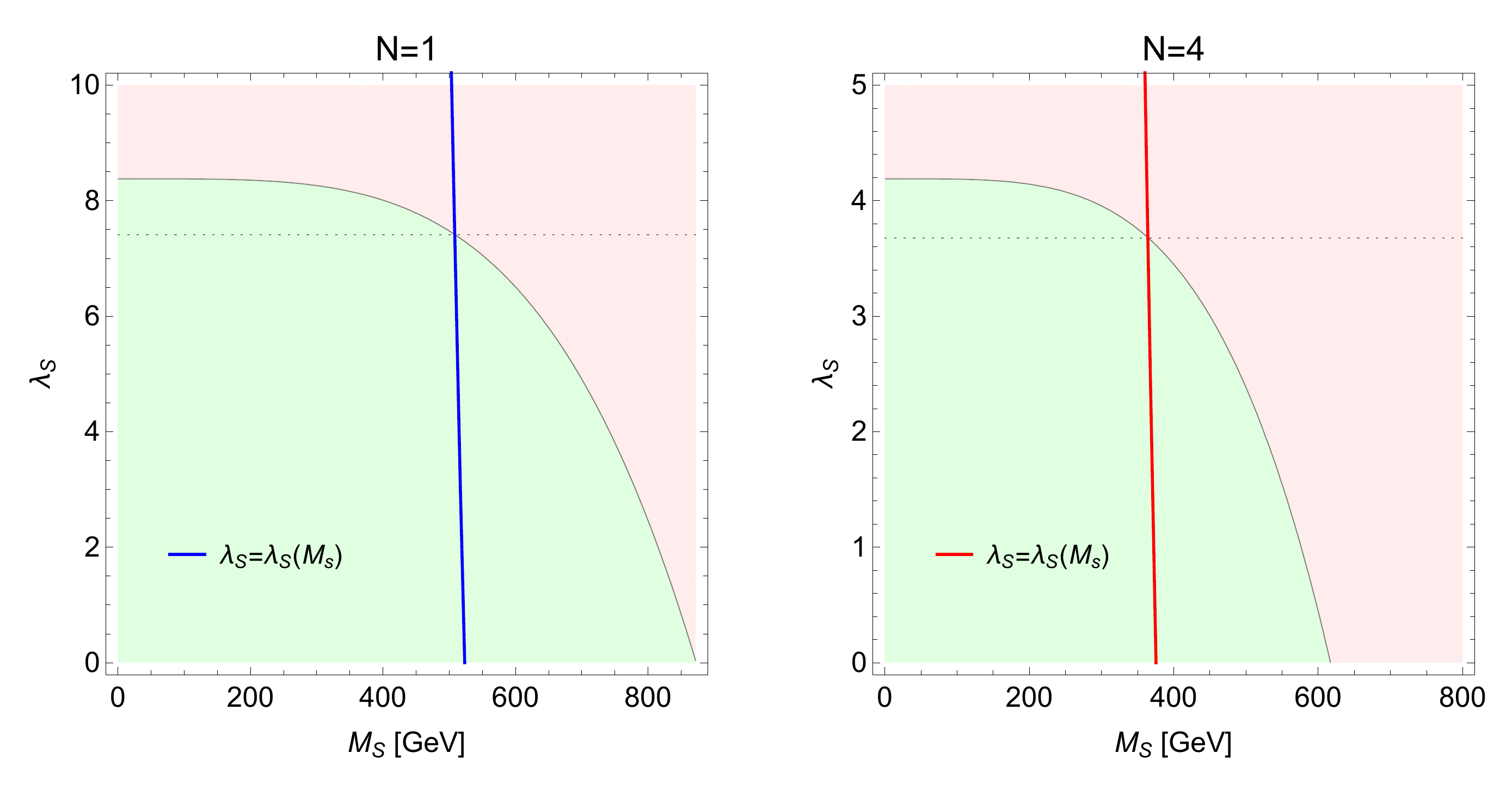}
 \caption{Regions of the $M_S$ and $\lambda_S$ parameter space of the $O(N)$-symmetric CSI  model allowed (light green) and excluded (light red) by the requirement of tree-level perturbative unitarity -- following Ref.~\cite{Hashino:2016rvx}. Additionally, the blue and red curves give the values of $\lambda_S$ computed at respectively one and two loops using equation~(\ref{EQ:O(N)_2l_massrelationOS}) as a function of $M_S$. The left pane is for $N=1$, while the right one is for $N=4$. }
 \label{FIG:O(N)_unitaritycontours}
\end{figure} 

Until this point, we have not taken into account the mass relation -- as derived in section~\ref{SEC:O(N)_massrel} -- nor the theoretical constraint from unitarity. For the latter, we choose to use as our criterion tree-level\footnote{We want to constrain the values of $\lambda_S$ and $M_S$, both of which appearing in $\hat\lambda_{hhh}$ only at two-loop order. Therefore, the difference from including higher-order (instead of tree-level) perturbative unitarity constraints should be formally a three-loop effect, and thus within the expected theoretical uncertainty of our result.} perturbative unitarity~\cite{Lee:1977eg}, and we can for the $O(N)$-symmetric model employ the results given in Ref.~\cite{Hashino:2016rvx}, in particular equation (22) therein. We show in figure~\ref{FIG:O(N)_unitaritycontours} the regions of parameter space --  in the $M_S$-$\lambda_S$ plane -- allowed (light green) and excluded (light red) under this criterion of tree-level perturbative unitarity, for both $N=1$ (left) and $N=4$ (right). Moreover, we superimpose on top of this plane the curves obtained when computing $\lambda_S$ as a function of $M_S$ (and $N$) using the mass relation we derived in eq.~(\ref{EQ:O(N)_2l_massrelationOS}) -- the blue line is for $N=1$ while the red one is for $N=4$. We recall also that negative values of $\lambda_S$ are forbidden, otherwise the potential would not be bounded from below. Although a wide range of BSM masses seems to be allowed by unitarity, once one takes into account the relation between masses in the CSI model -- or in other words once one imposes that $M_h=125\gev$ -- only a narrow interval remains. Concretely, we find for $N=1$ that $509\gev\lesssim M_S\lesssim 524\gev$, and for $N=4$ that $365\gev\lesssim M_S\lesssim 375\gev$. In fact, at one-loop order -- see the first line of eq.~(\ref{EQ:O(N)_2l_massrelationOS}) --  only a single value of $M_S$ is allowed for a given $N$, however once two-loop contributions are included the presence of a new parameter $\lambda_S$ allows moderate variations of $M_S$. Conversely, as it only appears from two loops, $\lambda_S$ is not constrained very severely: indeed values as large as $\lambda_S=7.4$ (3.7) are possible for $N=1$ ($N=4$). 

Another important criterion that we use to verify the validity of the parameter points we consider is the true-vacuum condition~\cite{Politzer:1978ic} -- $i.e.$ the requirement that the EW vacuum is the true minimum of the (two-loop) potential and corresponds to a lower value of $\veff$ than the origin of the potential (at $\varphi_h=v+h=0$). In CSI theories, as all states acquire their masses entirely from their coupling to the Higgs boson, all masses vanish at the origin $v+h=0$, and hence $\veff(\varphi_h=0)=0$ and we only need to make sure that $\veff(\varphi_h=v)$ is negative. The effective potential depends not only on the field $\varphi_h$, but also on the renormalisation scale $Q$ and therefore we examine the value of $\veff(\varphi_h=v)$ for $Q=v$ and $Q=M_S$ -- the two choices that we believe to be the most natural. We find for $N=1$ (where $M_S\sim 515\gev$) that the true-vacuum condition is fulfilled for $Q=M_S$ but not for $Q=v$, while for $N=4$ (where $M_S\sim 370\gev)$ the condition is verified for both values of $Q$. As it is unclear which value of $Q$ would be optimal (both values being around the EW scale anyway), the case of $N=1$ is not conclusively excluded by this criterion. 

\begin{figure}[h]
 \centering
 \includegraphics[width=\textwidth]{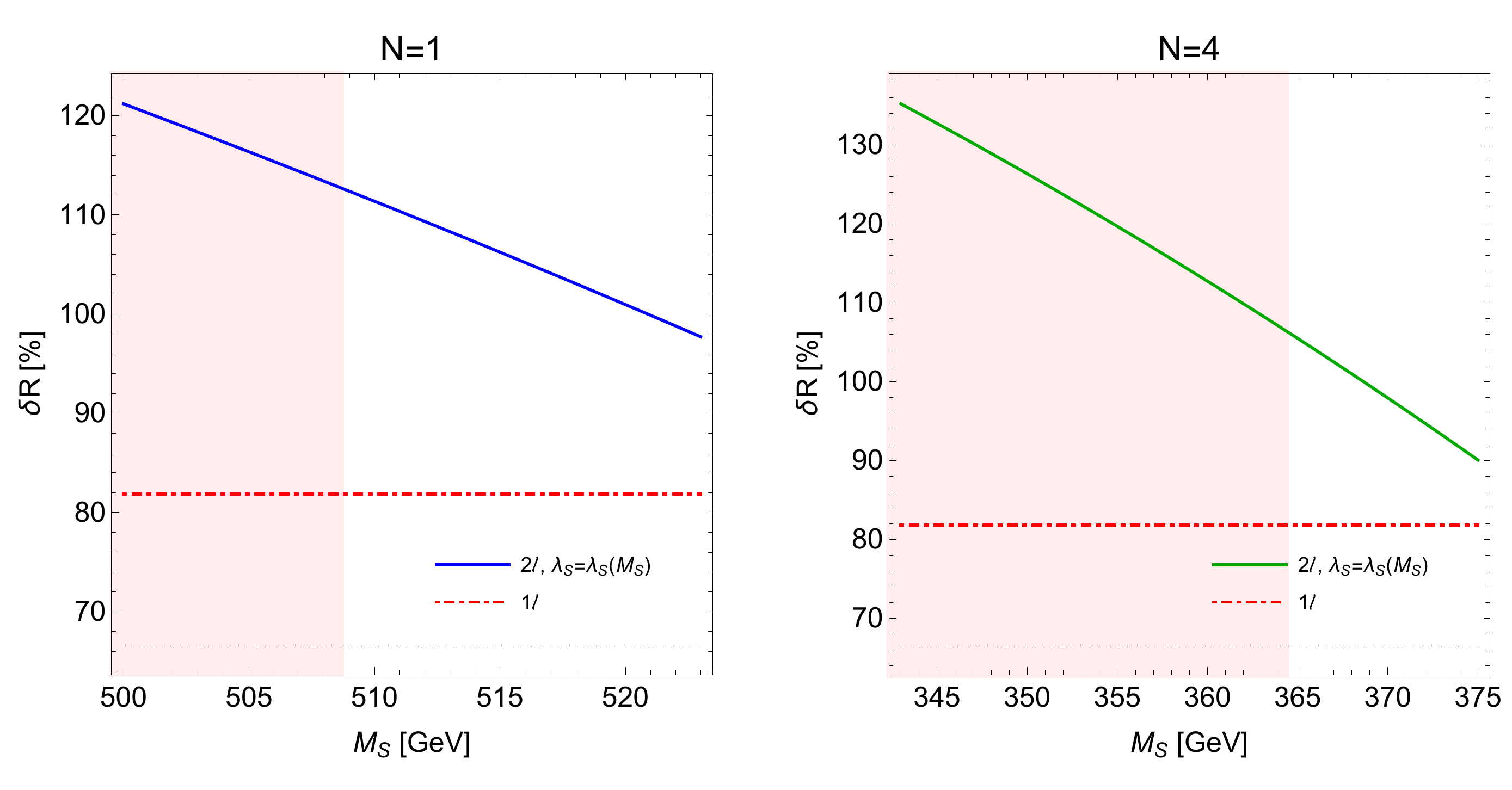}
 \caption{BSM deviation $\delta R$ -- defined in eq.~(\ref{EQ:deltaR_def}) -- of the Higgs trilinear coupling $\hat\lambda_{hhh}$ computed at two loops in the $O(N)$-symmetric CSI model with respect to its usual ($i.e.$ non-CSI) SM prediction. The blue curve (left side) is the two-loop result obtained for $N=1$, while the green curve (right side) is the two-loop result for $N=4$. Moreover, for this figure, $\lambda_S$ is calculated as a function of $M_S$ (for $N=1$ on the left side and $N=4$ on the right side) using equation~(\ref{EQ:O(N)_2l_massrelationOS}). The red dot-dashed line shows the one-loop result for $\delta R$, while the black dotted line is the comparison of the one-loop CSI result with the tree-level SM one. Finally, the light-red shaded regions are excluded by tree-level perturbative unitarity.}
 \label{FIG:O(N)_deltaRnonCSI_conslambdaS}
\end{figure}

Finally, we present in figure~\ref{FIG:O(N)_deltaRnonCSI_conslambdaS} the BSM deviations $\delta R$ we obtain once we take into account the mass relation and the constraint from perturbative unitarity. As for previous figures, the left and right panes are for $N=1$ and $N=4$ respectively, and the red dot-dashed lines show the one-loop result for $\delta R$ -- independent of $M_S$ and $N$. The blue and green curves give the two-loop results for $N=1$ and $N=4$. We emphasise that there is now only a single result for $\delta R$ at two loops because once $M_S$ and $N$ are fixed, the only remaining parameter -- $\lambda_S$ -- is fixed via equation~(\ref{EQ:O(N)_2l_massrelationOS}). The ranges of $M_S$ for these plots correspond to $\lambda_S$ varying between $0$ and $4\pi$, but the light-red shaded regions are excluded by the criterion of tree-level perturbative unitarity, as discussed above. 

Taking into consideration all these theoretical constraints, we find that the two-loop BSM deviation to $\hat\lambda_{hhh}$ must be within the intervals from $\sim 99\%$ to $\sim114\%$ and from $\sim 93\%$ to $\sim 109\%$ respectively for $N=1$ and $N=4$. This means that while the universality found at one loop for $(\delta R)^{(1)}$ is lost at two loops, only a limited range of values of $(\delta R)^{(2)}$ is possible. Moreover, our computation finds the Higgs trilinear coupling in CSI $O(N)$-symmetric model to be about twice as large as in the SM -- $(\delta R)^{(2)}\sim 100\%\pm 10\%$ -- and is larger than what is obtained at one loop, which makes the BSM deviation easier to access in experimental searches.

Before ending our discussion of the CSI $O(N)$-symmetric models, it is also important to comment on the experimental status of the parameter points we have considered. We have verified using \texttt{HiggsBounds}~\cite{Bechtle:2008jh,Bechtle:2011sb,Bechtle:2013wla,Bechtle:2015pma,Bechtle:2020pkv} that the points allowed by under the theoretical criteria discussed above are also allowed from the point of view of collider searches.  
The input values for \texttt{HiggsBounds} are produced by using the \textsc{Mathematica} package \texttt{SARAH}~\cite{Staub:2008uz,Staub:2009bi,Staub:2010jh,Staub:2012pb,Staub:2013tta} to create a \texttt{SPheno}-based~\cite{Porod:2003um,Porod:2011nf} spectrum generator for the CSI $O(N)$-symmetric models, with $N=1$ as well as $N=4$.

\section{CSI Two-Higgs-Doublet Model}
\label{SEC:CSI2HDM}

\subsection{Model definition}
We now turn to the case of a CSI version of the Two-Higgs-Doublet Model, as devised in Ref.~\cite{Lee:2012jn} (see also Ref.~\cite{Funakubo:1993jg}). We will be assuming here that CP is conserved in the Higgs sector, and thus we can write the tree-level potential of the model in terms of two $SU(2)_L$ doublets $\Phi_1$ and $\Phi_2$ of hypercharge $1/2$ as
\begin{align}
 \vtree=&\ \frac12\lambda_1|\Phi_1|^4+\frac12\lambda_2|\Phi_2|^4+\lambda_3|\Phi_1|^2|\Phi_2|^2+\lambda_4|\Phi_1^\dagger\Phi_2|^2+\frac12\lambda_5\left((\Phi_1^\dagger\Phi_2)^2+(\Phi_2^\dagger\Phi_1)^2\right)\,,
\end{align}
where all the quartic couplings $\lambda_i$ are real. 
To prevent the occurrence of tree-level flavour-changing neutral currents, we impose a discrete $\mathbb{Z}_2$ symmetry~\cite{Glashow:1976nt,Paschos:1976ay} under which the scalar doublets transform as  
\begin{equation}
 \Phi_1\xrightarrow{\mathbb{Z}_2} \Phi_1\,,\qquad\Phi_2\xrightarrow{\mathbb{Z}_2}-\Phi_2\,,
\end{equation}
while the fermions remain unaffected. 

We expand the two doublets as 
\begin{align}
 \Phi_i=\begin{pmatrix}
  w_i^+ \\ \frac{1}{\sqrt{2}}(v_i+h_i +iz_i)
 \end{pmatrix}\,,\qquad i=1,2\,,
\end{align}
where $v_i$ ($i=1,2$) are the vacuum expectation values of the neutral components of the doublets. These can be taken to be real -- owing to the assumption of CP conservation -- and furthermore verify the relation $v_1^2+v_2^2=v^2\simeq(246\gev)^2$. From the ratio of $v_1$ and $v_2$, we can also define the usual quantity $\tan\beta\equiv v_2/v_1$. 

Requiring both VEVs to be non-vanishing, we find for the tree-level tadpole equations
\begin{align}
 \lambda_1 v_1^2+\lambda_{345} v_2^2&=0\,,\nn\\
 \lambda_2 v_2^2+\lambda_{345} v_1^2&=0\,,
\end{align}
with the shorthand notation $\lambda_{345}\equiv\lambda_3+\lambda_4+\lambda_5$. 
These equations can be solved as
\begin{align}
 \lambda_1=-\lambda_{345}\tan^2\beta\,,\quad\text{and}\quad\lambda_2=-\lambda_{345}\cot^2\beta\,,
\end{align}
or equivalently as
\begin{align}
 \frac{\lambda_1}{\lambda_2}=\tan^4\beta\,,\quad\text{and}\quad \lambda_1\lambda_2=\lambda_{345}^2\,.
\end{align}

Next, mass eigenstates are obtained by rotating the gauge eigenstate fields with the angle $\beta$
\begin{align}
 \begin{pmatrix}
  w_1^\pm\\ w_2^\pm
 \end{pmatrix}=R(\beta)\begin{pmatrix}
  G^\pm\\H^\pm
 \end{pmatrix}\,,\quad
 \begin{pmatrix}
  z_1\\ z_2
 \end{pmatrix}=R(\beta)\begin{pmatrix}
  G\\A
 \end{pmatrix}\,,\quad
 \begin{pmatrix}
  h_1\\ h_2
 \end{pmatrix}=R(\beta)\begin{pmatrix}
  h\\H
 \end{pmatrix}\,,
\end{align} 
having defined
\begin{align}
 R(x)\equiv\begin{pmatrix}
  \cos x & -\sin x\\
  \sin x &  \cos x
 \end{pmatrix}\,.
\end{align}
Among these fields, $h,H$ are CP-even Higgs boson, $A$ is a CP-odd Higgs boson, and $H^\pm$ is the charged Higgs boson. As in the SM, $G$ and $G^\pm$ are respectively the neutral and charged would-be Goldstone bosons.

In this new basis, the tree-level scalar mass eigenvalues can be found to be 
\begin{align}
\label{EQ:CSI2HDM_treemasses}
 m_G^2&=m_{G^\pm}^2=m_h^2=0\,,\nn\\
 m_H^2=-\lambda_{345}v^2\,,\qquad m_A^2&=-\lambda_5v^2\,,\qquad m_{H^\pm}^2=-\frac12(\lambda_4+\lambda_5)v^2\,.
\end{align}
Interestingly, and to the difference of the usual (non-CSI) 2HDM, the CP-even mass matrix is already diagonal after the rotation of angle $\beta$. In other words, no additional rotation of angle $\alpha-\beta$ is necessary, and the CSI-2HDM is naturally \textit{aligned} (at tree level). Moreover, finding $m_h=0$ also tells us that $h$ is indeed the flat direction along which we want to work in what follows.

We also follow the common choice of trading the free Lagrangian quartic couplings $\lambda_3,\,\lambda_4,\,\lambda_5$ for the tree-level masses of the scalars. From eq.~(\ref{EQ:CSI2HDM_treemasses}), one has
\begin{align}
 \lambda_3=-\frac{m_H^2-2m_{H^\pm}^2}{v^2}\,,\qquad
 \lambda_4=\frac{m_A^2-2m_{H^\pm}^2}{v^2}\,,\qquad
 \lambda_5=-\frac{m_A^2}{v^2}\,.
\end{align}
We should note also that in the following, we neglect loop-induced deviations from the alignment found at tree level. Such effects have been studied for instance in Ref.~\cite{Lane:2019dbc} in the CSI 2HDM (see also Ref.~\cite{Braathen:2017izn} for the usual 2HDM) and were found to be relatively small for most of the parameter space, so that we consider them subleading before the two-loop corrections from the BSM scalar states. This enables us to write the effective potential under the form of equation~(\ref{EQ:form_Veff_1+2l}) and thus to follow the calculational procedure described in section~\ref{SEC:GW}. 

As in the usual 2HDM, the two scalar doublets in the CSI-2HDM can couple to quarks and leptons. Under the $\mathbb{Z}_2$ symmetry, four types of coupling assignments are possible, corresponding to the four types of 2HDMs~\cite{Barger:1989fj,Grossman:1994jb,Aoki:2009ha}. We note however that, for the major part of the discussion in this section, we do not need to specify a type as we only consider effects from the top quark -- only when verifying experimental constraints do we assume interactions of type I in order to have less severe limits from flavour physics than for instance for types II and Y~\cite{Misiak:2017bgg}.

\subsection{Dominant two-loop corrections to the Higgs trilinear coupling}
At one loop, the result for $\lambda_{hhh}$ is model-independent and was discussed in section~\ref{SEC:GW}. We can now consider the dominant two-loop contributions to the effective potential, and in turn to the Higgs trilinear coupling. We choose in this section to present results for the case of degenerate BSM scalar masses $M_H=M_A=M_{H^\pm}\equiv M_\Phi$, because the expressions for general values of the BSM scalar masses are quite long and cumbersome, and moreover because this will not affect the physics we discuss here. We provide the complete expressions for $\vtwo$, $\lambda_{hhh}$, and $\hat\lambda_{hhh}$ in appendix~\ref{APP:CSI2HDM_lambdahhh}. We should also note that as the BSM scalar masses will only appear in the Higgs trilinear coupling starting from the two-loop order, it does not make a difference whether we use pole or tree-level masses for the BSM scalars -- the difference in $\lambda_{hhh}$ from using one or the other is formally of three-loop order.

\begin{figure}[h]
 \centering
 \includegraphics[width=.8\textwidth]{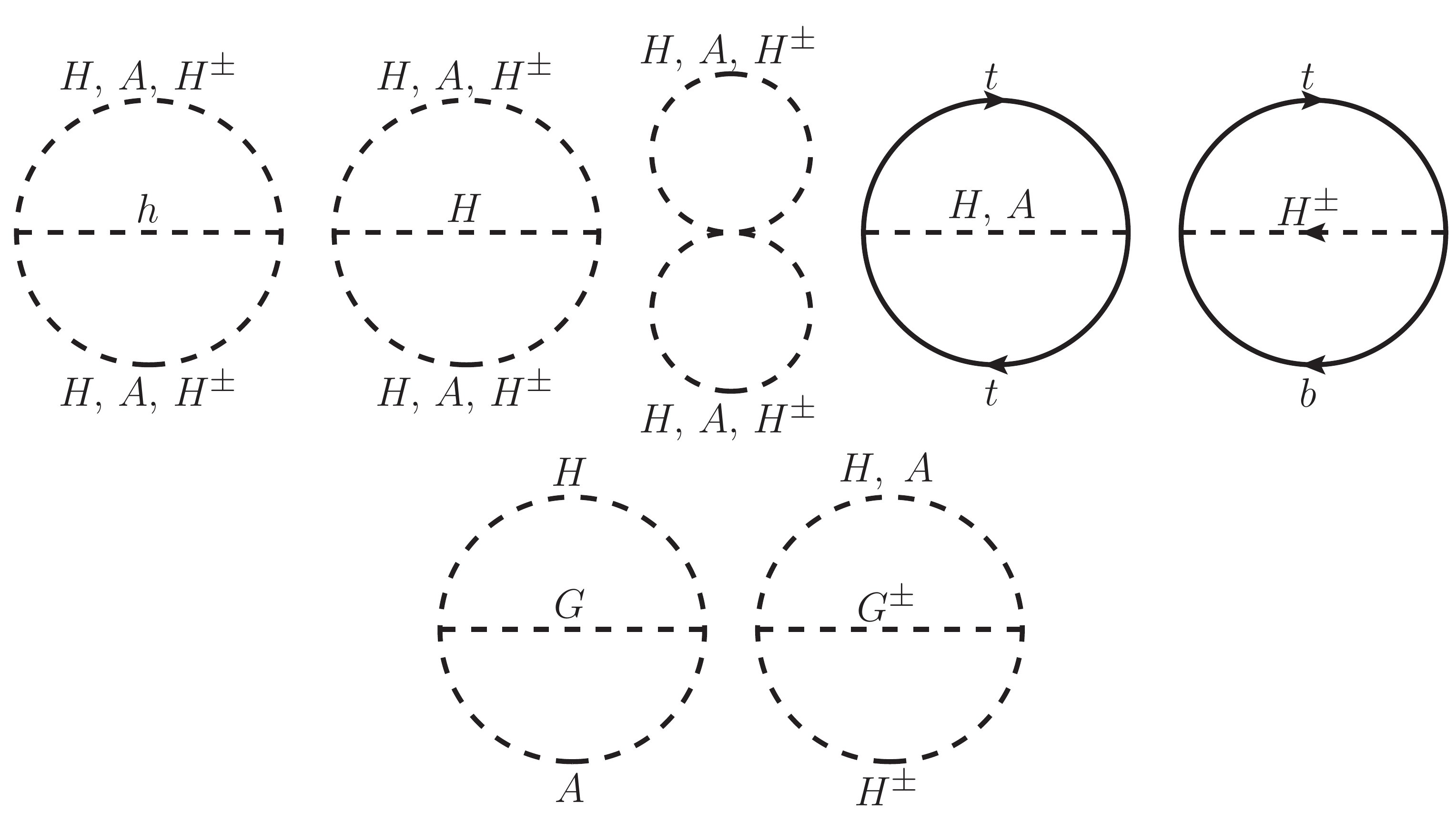}
 \caption{New dominant two-loop BSM diagrams contributing to the effective potential of the CSI-2HDM -- in addition to the leading SM diagrams of figure~\ref{FIG:SM2ldiags}. The diagrams in the second line all vanish in the limit of degenerate BSM scalar masses. }
 \label{FIG:CSI2HDM2ldiags}
\end{figure}

The BSM diagrams contributing to leading order to $\vtwo$ are shown in figure~\ref{FIG:CSI2HDM2ldiags}. In the limit of degenerate masses, their expressions read
\begin{align}
 \vtwo_{SSS}(h)=&-\frac{4m_\Phi^4(v+h)^2}{v^4}I(m_\Phi^2(h),m_\Phi^2(h),0)-\frac{6m_\Phi^4\cot^22\beta(v+h)^2}{v^4}I(m_\Phi^2(h),m_\Phi^2(h),m_\Phi^2(h))\,,\nn\\
 \vtwo_{SS}(h)=&\ \frac{12m_\Phi^2\cot^22\beta}{v^2}J(m_\Phi^2(h))^2\,,\nn\\
 \vtwo_{FFS}(h)=&\ \frac{6 m_t^2\cot^2\beta}{v^2}\Big[J(m_t^2(h))^2 - 3 J(m_\Phi^2(h)) J(m_t^2(h)) + (m_t^2(h) - m_\Phi^2(h)) I(0, m_\Phi^2(h),m_t^2(h))\nn\\
 &\hspace{2cm} + (2m_t^2(h) - m_\Phi^2(h)) I(m_\Phi^2(h), m_t^2(h), m_t^2(h)) \Big]\,.
\end{align}
From these equations we can extract the BSM logarithm-squared terms, and including also the SM-like contributions, we find in total for $\ctl$
\begin{align}
\label{EQ:CSI2HDM_C2}
 \ctl=&\ \frac{24g_3^2m_t^4}{v^4}-\frac{9m_t^6}{v^6}+\frac{m_\Phi^6}{v^6}(4+21\cot^22\beta)+ \frac{6 m_t^2\cot^2\beta}{v^6}\bigg(m_\Phi^4- 3m_\Phi^2m_t^2 - \frac32m_t^4 \bigg)\,.
\end{align}
Applying equation~(\ref{EQ:res_2l_lambdahhh}), we obtain the dominant two-loop corrections to $\lambda_{hhh}$, involving $g_3$, $m_t$, $m_\Phi$, and $\tan\beta$ as
\begin{align}
\label{EQ:CSI2HDM_2lres}
 \lambda_{hhh}=&\ \frac{5[M_h^2]_{\veff}}{v}+\frac{1}{(16\pi^2)^2}\bigg[\frac{768g_3^2m_t^4}{v^3}-\frac{288m_t^6}{v^5}+\frac{32m_\Phi^6}{v^5}(4+21\cot^22\beta)\nn\\
 &\hspace{3.5cm}+\frac{192m_t^2m_\Phi^4\cot^2\beta}{v^5}\left(1-\frac{3m_t^2}{m_\Phi^2}-\frac{3m_t^4}{2m_\Phi^4}\right)\bigg]\,.
\end{align}

Lastly, we can -- like in the previous sections -- express the one-loop result $5[M_h^2]_{\veff}/v$ in terms of the Higgs pole mass and the physical Higgs VEV, and take into account finite WFR effects, to obtain an OS result. The only 2HDM-specific corrections come from the momentum-dependent BSM scalar contributions to the Higgs self-energy. These can be found to be
\begin{align}
 \Pi_{hh}^{(1)}(p^2)\supset-\sum_{\mathclap{\Phi=H,A,H^\pm}}\frac{2n_\Phi m_\Phi^4}{v^2}B_0(p^2,m_\Phi^2,m_\Phi^2)\,,
\end{align} 
where $n_{H,A}=1$ and $n_{H^\pm}=2$. From this, we have
\begin{align}
 \frac{d\Pi_{hh}^{(1)}}{dp^2}\bigg|_{p^2=0}\supset-\frac13\sum_{\mathclap{\Phi=H,A,H^\pm}}\frac{n_\Phi m_\Phi^2}{v^2}\,. 
\end{align}
Finally, we obtain in terms of OS-renormalised parameters
\begin{align}
\label{EQ:CSI2HDM_lambdahhh_OS}
 \hat\lambda_{hhh}=&\ \frac{5M_h^2}{\vphys}+\frac{1}{16\pi^2}\frac{5M_h^2}{\vphys}\bigg[\frac{7}{2}\frac{M_t^2}{\vphys^2}-\frac{2}{3}\frac{M_\Phi^2}{\vphys^2}\bigg]\nn\\
 &+\frac{1}{(16\pi^2)^2}\bigg[\frac{768g_3^2M_t^4}{\vphys^3}-\frac{288M_t^6}{\vphys^5}+\frac{32M_\Phi^6}{\vphys^5}(4+21\cot^22\beta)\nn\\
 &\hspace{2cm}+\frac{192M_t^2M_\Phi^4\cot^2\beta}{\vphys^5}\left(1-\frac{3M_t^2}{M_\Phi^2}-\frac{3M_t^4}{2M_\Phi^4}\right)\bigg]\,.
\end{align}

\subsection{Mass relation}

Next, we also consider the mass relation for the case of the CSI-2HDM, which will serve to relate the BSM scalar masses $M_H,\,M_A,\,M_{H^\pm}$ and $\tan\beta$. As in the previous section, we present here only expressions in the limit of degenerate scalar masses, and leave the complete results for appendix~\ref{APP:CSI2HDM_massrelation}. To derive corrections to $\lambda_{hhh}$, we have already used the expression of $\ctl$ given in eq.~(\ref{EQ:CSI2HDM_C2}). We must now also derive results for $B$ at one- and two-loop orders. First, at one loop it is straightforward to calculate
\begin{align}
 \bol=&\ \frac{1}{4v^4}\big(4m_\Phi^4-12m_t^4+6m_W^4+3m_Z^4\big)\,.
\end{align}

At two loops, we have
\begin{align}
 \btl=&\ \frac{4g_3^2m_t^4}{v^4}\left[12\log\frac{m_t^2}{v^2}-16\right]+\frac{6m_t^6}{v^6}\left[-3\log\frac{m_t^2}{v^2}+8\right]+\btl_{SS}+\btl_{SSS}+\btl_{F\bar{F}S}\,,
\end{align}
where the first two terms are SM-like contributions and the latter three -- $\btl_{SS},\,\btl_{SSS},\,\btl_{F\bar{F}S}$ -- are the BSM ones. These can be found to be in the limit of degenerate BSM masses
\begin{align}
 \btl_{SS}=&\ \frac{24m_\Phi^6\cot^22\beta}{v^6}\left(\log\frac{m_\Phi^2}{v^2}-1\right)\,,\nn\\
 \btl_{SSS}=&\ \frac{2m_\Phi^6}{v^6}(4+9\cot^22\beta)\left(\log \frac{m_\Phi^2}{v^2}-2\right)\,,\nn\\
 \btl_{F\bar{F}S}=&\ \frac{6m_t^2\cot^2\beta}{v^6}\bigg\{ 2m_\Phi^4\left(\log\frac{m_\Phi^2}{v^2}-2\right)-6m_t^2m_\Phi^2\left(\log\frac{m_\Phi^2}{v^2}-1\right)-m_t^4\left(3\log\frac{m_t^2}{v^2}-8\right)\bigg\}\,.
\end{align}

Applying equation~(\ref{EQ:2l_massrelation}), the mass relation at two loops is, in terms of \msbar-renormalised quantities,
\begin{align}
\label{EQ:CSI2HDM_massrelation_msbar}
 8\pi^2v^2[M_h^2]_{\veff}=&\ 4m_\Phi^4-12m_t^4+6m_W^4+3m_Z^4\nn\\
 &+\frac{m_t^4}{4\pi^2}\bigg[g_3^2\left(48\log\frac{m_t^2}{Q^2}-40\right)+\frac{m_t^2}{v^2}\left(-18\log\frac{m_t^2}{Q^2}+39\right)\bigg]\nn\\
 &+\frac{1}{4\pi^2}\Bigg\{\frac{24m_\Phi^6\cot^22\beta}{v^2}\left(\log\frac{m_\Phi^2}{Q^2}-\frac12\right)+\frac{2m_\Phi^6}{v^2}[4+9\cot^22\beta]\left(\log\frac{m_\Phi^2}{Q^2}-\frac32\right)\nn\\
 &\hspace{1.5cm}+\frac{6m_t^2\cot^2\beta}{v^2}\bigg[2m_\Phi^4\left(\log\frac{m_\Phi^2}{Q^2}-\frac32\right)-6m_t^2m_\Phi^2\left(2\log\frac{m_\Phi^2}{Q^2}-\frac12\right)\nn\\
 &\hspace{4cm}-m_t^4\left(3\log\frac{m_t^2}{Q^2}-\frac{13}{2}\right)\bigg]\Bigg\}\,.
\end{align} 
The first line in this equation is the one-loop result, while the additional lines are the two-loop corrections. Among these, the first (square) brackets are the SM-like contributions, and the remains terms (in the curly brackets) are the BSM terms, proper to the CSI-2HDM.  

To convert equation~(\ref{EQ:CSI2HDM_massrelation_msbar}) from the \msbar to the OS scheme, we require as previously the correction to the Higgs VEV -- $c.f$ equation~(\ref{EQ:VEVconv}) and the top-quark and Higgs-boson one-loop self-energies, as well as the self-energies for the BSM scalars $H,\,A,\,H^\pm$ (also at one loop). All necessary expressions are provided in appendix~\ref{APP:CSI2HDM_massrelation}. With these, we are able to obtain the following result, in terms only of physical parameters
\begin{align}
\label{EQ:CSI2HDM_2l_massrelationOS}
 \frac{4\sqrt{2}\pi^2}{G_F}M_h^2=&\ 4M_\Phi^4-12M_t^4+6M_W^4+3M_Z^4+M_h^2\left[\frac72M_t^2-\frac23M_\Phi^2\right]+\frac{3M_t^4}{8\pi^2}\left[16g_3^2-\frac{6M_t^2}{\vphys^2}\right]\nn\\
 &+\frac{M_\Phi^6}{4\pi^2\vphys^2}\bigg[4+3\bigg(7-2\sqrt{3}\pi\bigg)\cot^22\beta\bigg]+\frac{3M_t^2\cot^2\beta}{2\pi^2\vphys^2} \Bigg\{- M_\Phi^4 - 3 M_t^2 M_\Phi^2 + \frac{13}{2} M_t^4\nn\\
 &\hspace{.5cm}- 
 2 M_t^2 (M_\Phi^2 - 2 M_t^2) \mathfrak{Re}\bigg[f_B\left(\frac{M_\Phi^2}{2M_t^2} - \sqrt{\frac{M_\Phi^2}{4M_t^4}-1} \right) + f_B\left(\frac{M_\Phi^2}{2M_t^2} + \sqrt{\frac{M_\Phi^2}{4M_t^4}-1} \right)\bigg] \nn\\
 &\hspace{.5cm}- M_\Phi^2 (M_\Phi^2 - 2 M_t^2) \mathfrak{Re}\bigg[f_B\left(\frac12 - \sqrt{\frac14 - \frac{M_t^2}{M_\Phi^2}}\right) + f_B\left(\frac12 + \sqrt{\frac14 - \frac{M_t^2}{M_\Phi^2}}\right)\bigg]\Bigg\}\,,
\end{align}
where the function $f_B$ is defined in equation~(\ref{EQ:def_fB}), and $\mathfrak{Re}[x]$ denotes the real part of a given quantity $x$.

\subsection{Numerical study}

\begin{figure}[h]
 \centering
 \includegraphics[width=.8\textwidth]{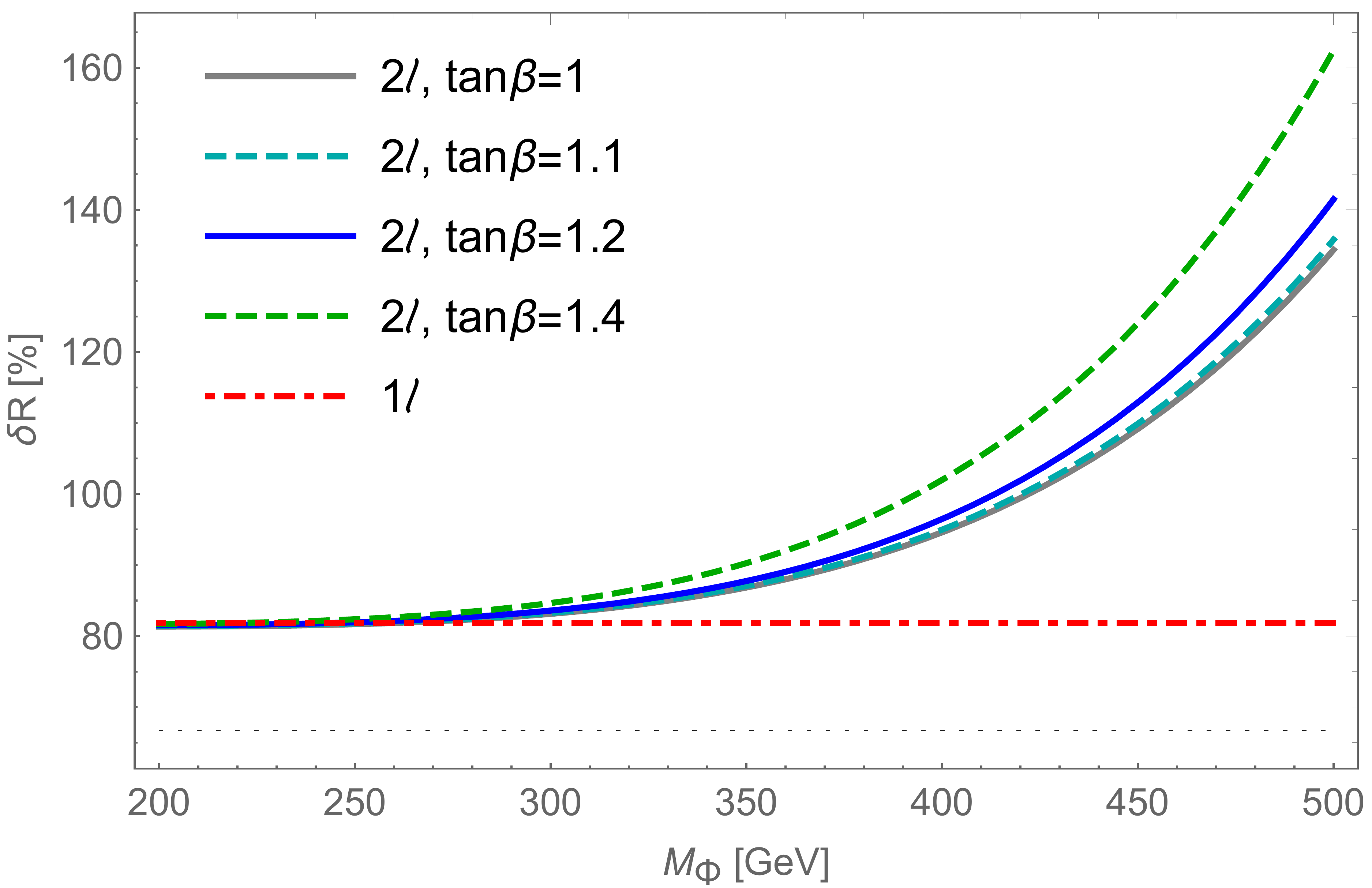}
 \caption{BSM deviation $\delta R$ -- defined in eq.~(\ref{EQ:deltaR_def}) -- of the Higgs trilinear coupling $\hat\lambda_{hhh}$ computed at one loop (dot-dashed line) and at two loops (solid and dashed curves) in the CSI-2HDM with respect to its SM prediction as a function of the degenerate mass of the BSM scalars $M_\Phi=M_H=M_A=M_{H^\pm}$. The different grey, teal, blue, and green two-loop curves correspond respectively to $\tan\beta$ equal to 1, 1.1, 1.2, and 1.4. As in the previous section, the black dotted line is obtained by comparing the one-loop CSI result for $\lambda_{hhh}$ to the SM tree-level result.}
 \label{FIG:CSI2HDM_deltaRnonCSI1}
\end{figure}

We have now derived all the results we require to study numerically the leading two-loop corrections to the Higgs trilinear coupling in the CSI-2HDM. We follow here the same outline as we did for the $O(N)$-symmetric model, and we provide results for the same type of BSM deviation $\delta R$ -- as defined in eq.~(\ref{EQ:deltaR_def}) -- comparing the prediction for the Higgs trilinear coupling in the CSI-2HDM to that in the usual (non-CSI) SM. Moreover, we will consider the three BSM scalars $H,\,A,\,H^\pm$ to be degenerate in mass -- this ensures that the $\rho$ parameter remains close to one, and furthermore this conveniently reduces the number of BSM mass scales to one. 

\begin{figure}[h]
 \includegraphics[width=\textwidth]{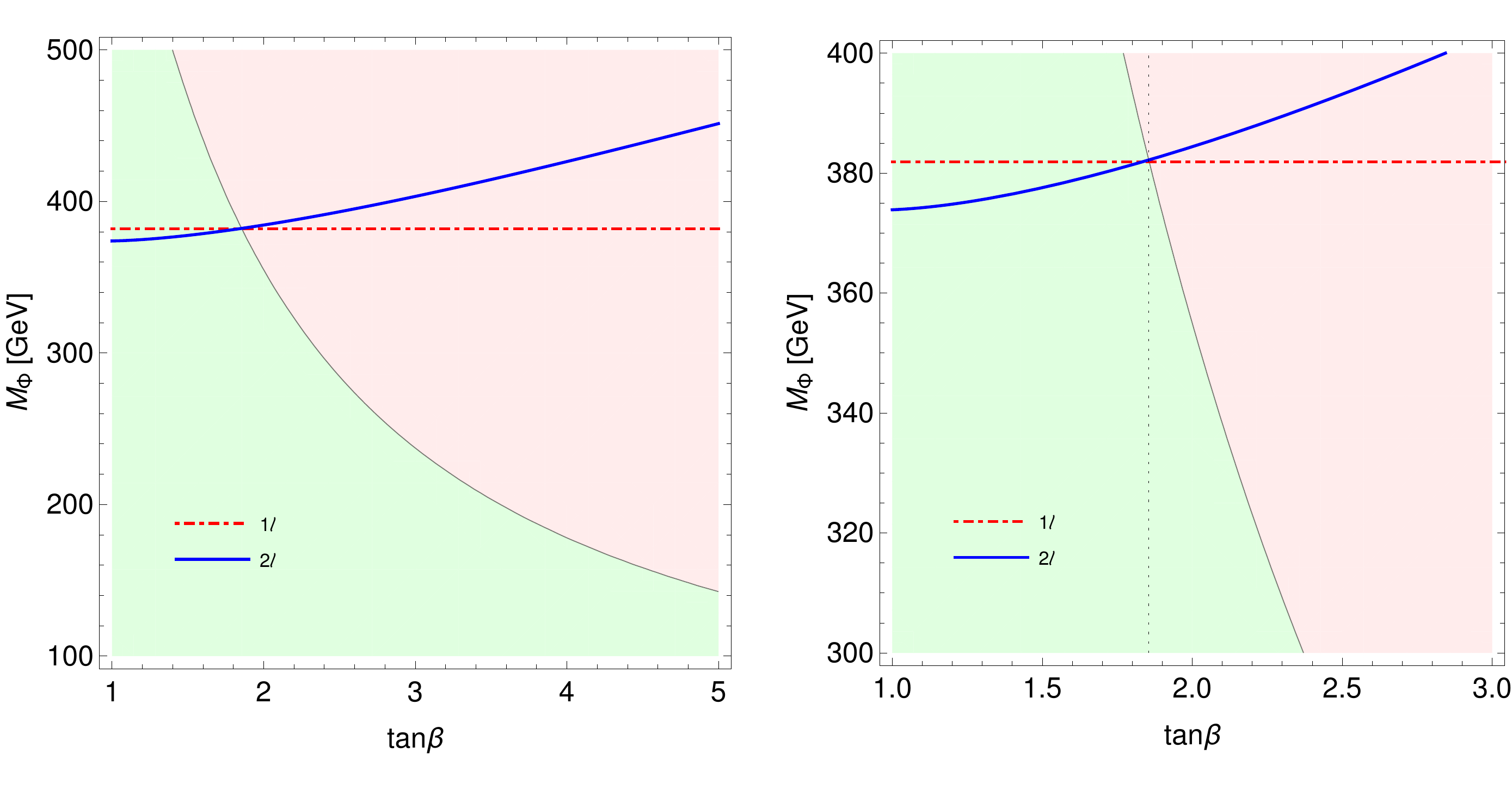}\
 \caption{Regions of the $\tan\beta$ and $M_\Phi$ parameter space of the CSI-2HDM allowed (light green) and excluded (light red) by the requirement of tree-level perturbative unitarity~\cite{Kanemura:1993hm}. Additionally, the red and blue curves give the values of $M_\Phi$ computed at respectively one and two loops using equation~(\ref{EQ:CSI2HDM_2l_massrelationOS}) as a function of $\tan\beta$. The right-hand side plot is an enlargement of the left-hand one. }
 \label{FIG:CSI2HDM_unitaritycontours}
\end{figure}

We present first in figure~\ref{FIG:CSI2HDM_deltaRnonCSI1} the deviation $\delta R$ at one loop (red line) and two loops (grey, teal, blue, and green curves) for different values of $\tan\beta$. We employ here the expression in equation~(\ref{EQ:CSI2HDM_lambdahhh_OS}), without for now taking into account the relation between masses. As we had seen for the $N$-scalar model, while at one loop the BSM deviation in $\hat\lambda_{hhh}$ has a universal value of $\delta R\sim 82\%$, the two-loop corrections introduce new dependences on both $M_\Phi$ and $\tan\beta$, thereby spoiling the universality of $\delta R$. We consider in this plot different values of $\tan\beta$, ranging from $\tan\beta=1$ to $\tan\beta=1.4$, which we find to be the largest value of $\tan\beta$ for which tree-level perturbative unitarity~\cite{Lee:1977eg,Kanemura:1993hm} is maintained all the way to $M_\Phi=500\gev$ -- see figure~\ref{FIG:CSI2HDM_unitaritycontours} and the discussion in the following. Considering for instance $M_\Phi=500\gev$, we find $(\delta R)^{(2)}=134\%$ for $\tan\beta=1$, and $(\delta R)^{(2)}=162\%$ for $\tan\beta=1.4$, in stark contrast to the one-loop result. 

At this point, we should comment briefly on our interpretation of the quantity $\tan\beta$. While $\tan\beta$ is originally defined as the ratio of the VEVs of the two scalar doublets in the model, it appears in two-loop corrections to the Higgs trilinear coupling via the dependence of four-point scalar interactions at tree level and of the coupling between BSM scalars and the top quark. It should be emphasised however that because $\tan\beta$ appears only from the two-loop order, any scheme difference in how its value could be extracted from some experimental data (in the event a 2HDM Higgs sector is discovered) only has an impact from three-loop order and beyond -- $i.e.$ below the level of accuracy to which we are working.

\begin{figure}[h]
 \includegraphics[width=.8\textwidth]{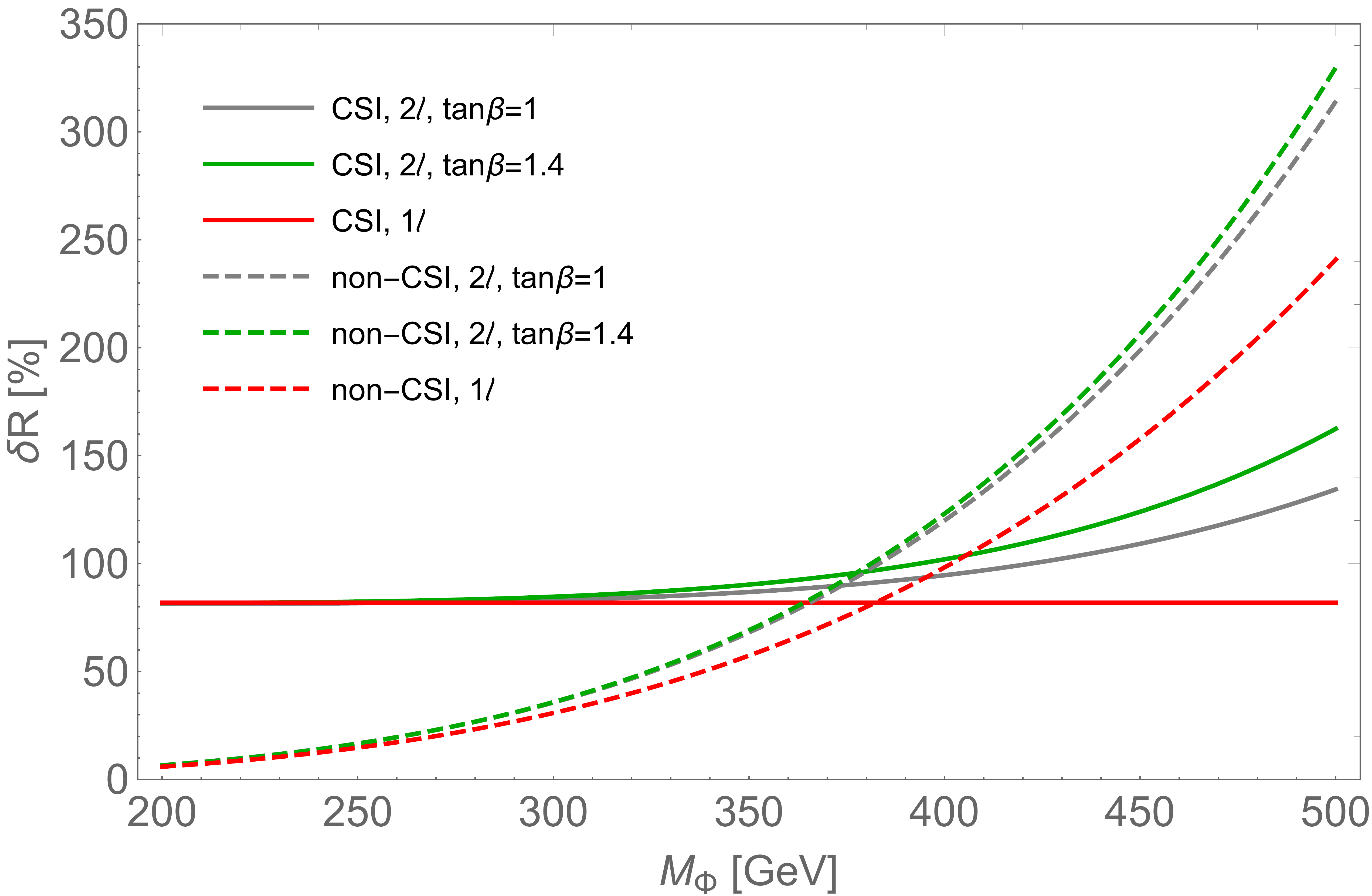}
 \caption{Comparison of the BSM deviations $\delta R$ computed in versions of the 2HDM, with (solid curves) and without (dashed curves) CSI, as a function of the degenerate pole mass $M_\Phi$ of the BSM scalars. Red curves present one-loop values, while grey and green curves are two-loop results for $\tan\beta=1$ and $\tan\beta=1.4$ respectively.}
 \label{FIG:2HDM_compareCSInonCSI}
\end{figure}

Next, it is also interesting to compare predictions for the Higgs trilinear coupling in variants of the 2HDM with or without CSI. For this purpose, we present in figure~\ref{FIG:2HDM_compareCSInonCSI} the BSM deviations in $\hat\lambda_{hhh}$ computed with respect to the usual-SM prediction in both variants of the 2HDM. For the usual 2HDM, we use the analytic expression for $\hat\lambda_{hhh}$ derived in Ref.~\cite{Braathen:2019pxr,Braathen:2019zoh} -- see specifically equation~(5.15) in Ref.~\cite{Braathen:2019zoh} -- in which we set the mass term $\tilde M$ (representing the scale of the soft breaking of the $\mathbb{Z}_2$ symmetry acting on the Higgs doublets) to zero to allow the comparison with the CSI scenario. 

We can observe that the Higgs trilinear coupling behaves very differently in CSI and non-CSI versions of the 2HDM. The main discrepancy occurs at one-loop level: indeed, while in the CSI-2HDM $(\delta R)^{(1)}$ is constant, in the usual (non-CSI) 2HDM, one-loop corrections involving the BSM scalars are minute in the low-mass range, but grow rapidly (as $M_\Phi^4$) and reach $\simeq 240\%$ for $M_\Phi=500\gev$. At two loops, the radiative corrections to $\hat\lambda_{hhh}$ behave quite similarly in the CSI and non-CSI scenarios -- the main difference arising from effects due to the scheme conversion of the one-loop BSM corrections in the non-CSI 2HDM case ($c.f.$ Ref.~\cite{Braathen:2019zoh}).  

At this point, we should discuss the theoretical constraints on the BSM parameters -- $M_\Phi$ and $\tan\beta$ -- coming from unitarity and from the mass relation in eq.~(\ref{EQ:CSI2HDM_2l_massrelationOS}). For the former, we choose to take tree-level perturbative unitarity~\cite{Lee:1977eg} as our criterion, and we employ results from Ref.~\cite{Kanemura:1993hm} (see also Ref.~\cite{Akeroyd:2000wc}). Figure~\ref{FIG:CSI2HDM_unitaritycontours} shows the allowed (light green) and excluded (light red) regions of the CSI-2HDM parameter space, in the $M_\Phi-\tan\beta$ plane. Additionally, the values of $M_\Phi$ that are extracted from the equation~(\ref{EQ:CSI2HDM_2l_massrelationOS}) are given by the dashed red (at one loop) and solid blue (at two loops) curves. On the one hand, at one loop $\tan\beta$ does not appear in the mass relation and hence at a constant value is found for $M_\Phi$. On the other hand, at two loops, the result for $M_\Phi$ obtained from eq.~(\ref{EQ:CSI2HDM_2l_massrelationOS}) becomes a function of $\tan\beta$. Along the two-loop (blue) curve, we find that the maximal possible value of $\tan\beta$ is 1.855, which corresponds to a maximal value of the degenerate mass of the BSM scalars $M_\Phi\leq 382.2 \gev$. As for the $O(N)$-symmetric models, we have also verified that the true-vacuum condition is respected in the CSI-2HDM: we have found that the value of the two-loop potential is, at the EW minimum, indeed lower than the value of the potential at the origin, and this is true both for $Q=v$ and $Q=M_\Phi\sim 378\gev$. 

\begin{figure}[h]
 \centering
 \includegraphics[width=.8\textwidth]{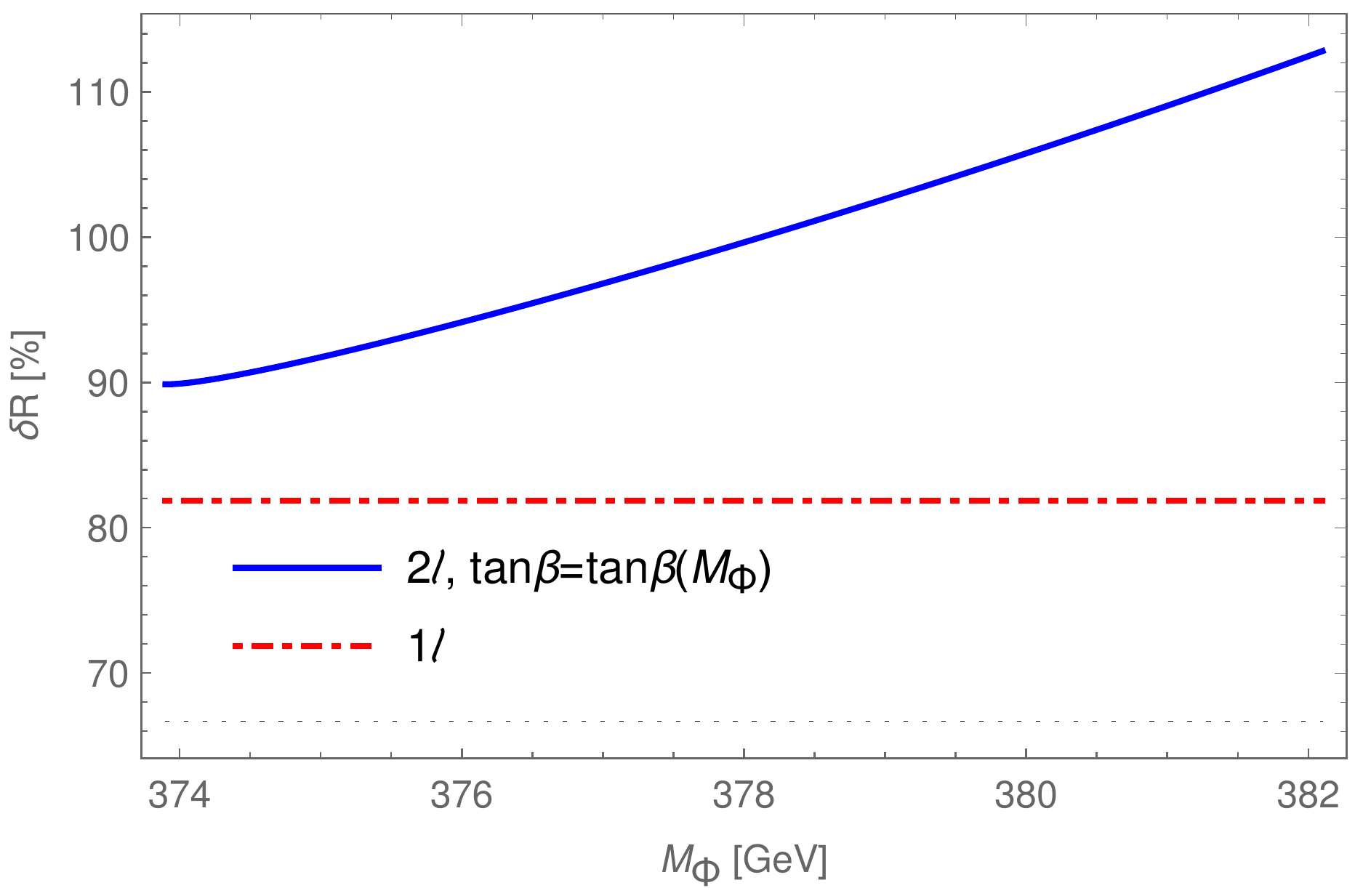}
 \caption{BSM deviation $\delta R$ -- defined in eq.~(\ref{EQ:deltaR_def}) -- of the Higgs trilinear coupling $\hat\lambda_{hhh}$ computed at two loops in the CSI-2HDM with respect to its SM prediction as a function of the degenerate mass of the BSM scalars $M_\Phi=M_H=M_A=M_{H^\pm}$. The value of $\tan\beta$ in the point in this figure is computed as a function of $M_\Phi$ using equation~(\ref{EQ:CSI2HDM_2l_massrelationOS}). Finally, the maximal value of $M_\Phi$ (equivalently the maximal value of $\tan\beta$) in this plot is constrained by the requirement of tree-level perturbative unitarity -- $c.f.$ figure~\ref{FIG:CSI2HDM_unitaritycontours}. The red dot-dashed and black dotted lines show the comparison of the one-loop CSI value of $\lambda_{hhh}$ with respectively the one-loop effective-potential and tree-level results in the SM.} 
 \label{FIG:CSI2HDM_deltaRnonCSIconstanbeta}
\end{figure}

Having considered the constraints on $M_\Phi$ and $\tan\beta$, we can finally investigate the theoretically-allowed values of the Higgs trilinear coupling. Figure~\ref{FIG:CSI2HDM_deltaRnonCSIconstanbeta} therefore presents our results for the deviation $\delta R$ at one loop (red dot-dashed line) and at two loops (blue solid curve), once $\tan\beta$ is extracted as a function of $M_\Phi$ from the mass relation in eq.~(\ref{EQ:CSI2HDM_2l_massrelationOS}). As can be seen in figure~\ref{FIG:CSI2HDM_unitaritycontours}, the possible values of $M_\Phi$ are all within the narrow interval from $373.9\gev$ to $382.1\gev$. We find that $\delta R$ varies between $89.9\%$ at the lowest, and $112.5\%$ at most. In other words, as for the $O(N)$-symmetric model earlier, we obtain a larger deviation of $\hat\lambda_{hhh}$ from its deviation at two loops, and the total BSM deviation is (approximately) in the range $100\pm10\%$. 

We have verified that the corresponding parameter points are not excluded by experimental searches for BSM scalars. Assuming Yukawa interactions of type I (to benefit from less severe constraints from flavour physics than for types II or Y, see $e.g.$ Ref.~\cite{Misiak:2017bgg}), we create a \texttt{SPheno}-based spectrum generator for the CSI-2HDM using \texttt{SARAH}, which we in turn used to provide inputs for \texttt{HiggsBounds} to verify the experimental status of the parameter choices we took. We ensured that the Lagrangian inputs provided to \texttt{SPheno} yielded the corrected spectrum of scalar masses -- a 125-GeV SM-like Higgs boson and BSM scalars at 374 GeV, with an allowed error of 1 GeV -- using the \textsc{Mathematica} package \texttt{SSP}~\cite{Staub:2011dp}. 

\section{Discussion}
\label{SEC:Discussion}
The computations presented in this paper allow studying the Higgs trilinear coupling in models with classical scale invariance to the same level of accuracy as for non-CSI extensions of the SM -- as done for instance in Refs.~\cite{Braathen:2019pxr,Braathen:2019zoh,Senaha:2018xek} (and appendix~\ref{APP:nonCSIO(N)} of this work). Classical scale invariance is an attractive concept for model building, and may relate to the solution to the hierarchy problem. It is therefore of paramount importance to find new ways to investigate CSI models, in complement of collider and dark matter searches -- see $e.g.$ Refs~\cite{Endo:2015ifa,Endo:2015nba,Endo:2016koi,Helmboldt:2016mpi,Fujitani:2017gma,Lane:2019dbc,Brooijmans:2020yij}.  

Due to the simple form of the potential at one loop, containing only two free quantities -- recall equation~(\ref{EQ:form_Veff_1l}) -- $\hat\lambda_{hhh}$ is found to be predicted universally at the one-loop level, in \textit{all} CSI models. This universality of $(\hat\lambda_{hhh})^{(1)}$ is lifted by the inclusion of two-loop effects, which thus allows distinguishing different scenarios of CSI models. We found that the new corrections at two loops are quite significant, leading to a further 5-30\% deviation of $\hat\lambda_{hhh}$ from the (non-CSI) SM prediction, compared to the one-loop CSI result. We should emphasise that this is by no means a problem from the point of view of the validity of the perturbative loop expansion: indeed the two-loop corrections to $\hat\lambda_{hhh}$ involve new parameters that are not present at one loop -- namely $M_S,\,\lambda_S,\,N$ for the $O(N)$-symmetric models and $M_H,\,M_A,\,M_{H^\pm},\,\tan\beta$ in the CSI-2HDM. Furthermore, as there are numerous arguments in favour of extended scalar sectors, it is interesting to note that the two-loop contributions to the Higgs trilinear coupling receives from scalars are always positive, which makes it deviate further from the SM value so that it is more easily accessible in collider searches than what is expected from the one-loop result. 

We have also paid particular attention to the relation among masses arising in theories with CSI, which is often over-looked in the literature. More than being simply bounded from above, the masses of BSM states in CSI models are actually strongly constrained by the requirement of reproducing the correct 125-GeV mass of the Higgs boson -- which is entirely generated at loop level. In the scenarios considered in the main text of this paper, we imposed additional constraints -- a global $O(N)$ symmetry for the $N$-scalar model, and mass degeneracy of the BSM scalars in the CSI-2HDM -- which resulted in severe bounds on the allowed BSM masses.\footnote{Admittedly, if these restrictions were to be loosened (as $e.g.$ in appendix~\ref{APP:CSI2HDM}), there would be more freedom to vary the BSM scalar masses. Nevertheless, the masses would remain related so as to ensure the correctness of the Higgs mass. } In turn, this allows us to find predictions for $\hat\lambda_{hhh}$ within a narrow range: in both types of BSM models, we obtain $90\%\lesssim (\delta R)^{(2)}\lesssim 115\%$. Looking back at figures~\ref{FIG:O(N)_compareCSInonCSI} and~\ref{FIG:2HDM_compareCSInonCSI}, it is worth noting that the values obtained for $(\delta R)^{(2)}$ for the allowed BSM mass ranges are very close to those found for the same masses in the non-CSI counterparts of the scenarios. Taking the example of the 2HDM, while in the CSI version of the model we have $90\%\lesssim (\delta R)^{(2)}\lesssim 110\%$, in the non-CSI variant and taking the non-decoupling limit ($i.e.$ $\tilde M=0$) the value of $(\delta R)^{(2)}$ computed at $M_\Phi\simeq 374\gev$ varies from $\sim 90\%$ (for $\tan\beta=1$) to $105\%$ (for $\tan\beta=1.9$, close to the upper limit from perturbative unitarity). This implies that the magnitude of a potential BSM deviation in the Higgs trilinear coupling may not provide a clear indication of whether the associated new physics exhibits classical scale invariance -- instead, a better criterion might come from whether masses and couplings fulfill the mass relations imposed by CSI.
 
Rather than considering only a single type of observable, a powerful strategy to investigate CSI theories, and to discriminate theories with or without CSI, would be to study the correlation between several observables. In particular, synergies between measurements of the Higgs trilinear coupling and direct searches of new particles at colliders may prove crucial to ascertain whether Nature exhibits classical scale invariance or not. Indeed, we should recall that due to the relation among masses in CSI models, the new BSM states cannot be arbitrarily heavy (even in non-degenerate scenarios) and will be within reach of experiments in a foreseeable future (see for instance the discussions for the CSI-2HDM in Refs.~\cite{Lane:2019dbc,Brooijmans:2020yij}). In the eventuality that a new scalar state is found in direct searches, determining the value of the Higgs trilinear coupling and verifying whether the relation among masses is fulfilled or not can potentially provide clear evidence of the CSI nature of the underlying theory. Furthermore, if one considers a model with charged scalars ($e.g.$ $H^\pm$ in the CSI-2HDM), sizeable one-loop corrections to the Higgs-to-two-photon ($h\gamma\gamma$) coupling can also be expected, as studied in Ref.~\cite{Hashino:2015nxa}. Such effects could also help confirm the existence (or absence) of CSI in BSM models. In this context, depending on the achieved accuracy of the determination of the $h\gamma\gamma$ coupling at future colliders -- possibly down to percent level, see $e.g.$ Refs.~\cite{Fujii:2015jha,Fujii:2017vwa} for prospects at the ILC -- it may become important (if not unavoidable) to compute theory predictions for this coupling at two-loop level in order to match the precision of experimental results and allow consistent comparisons between theory and experiment. We leave this endeavour for future work.

A caveat that should be mentioned is that the scenarios considered in this work require relatively large scalar quartic couplings -- once again, this comes from the need to generate the correct mass of the Higgs boson, in the absence of any BSM mass term in the Lagrangian. Consequently, under renormalisation-group running, we can expect these couplings to grow rapidly and encounter Landau poles well before the Planck scale -- in a similar way as what occurs in the non-decoupling limit of (non-CSI) extensions of the SM. This is consistent with our expectation that the assumption of CSI at the electroweak scale can a priori not be extended up to the Planck scale, and therefore does not solve the hierarchy problem by itself.

Finally, we should also point out the importance of computing the Higgs trilinear coupling to high precision when studying the behaviour of a number of BSM phenomena occuring in models with extended scalar sectors -- in particular the possibility of a strong-first order electroweak phase transition and the spectrum of gravitational waves that may be produced during the former. In complementarity with collider searches, the future measurement of primordial gravitational waves (GW) at the LISA and DECIGO space-based interferometers will provide a new way to probe the value of the Higgs trilinear coupling and the shape of the Higgs potential. The synergy between the two -- collider and GW measurements -- was studied at one loop for $O(N)$-symmetric CSI models in Ref.~\cite{Hashino:2016rvx} (see also Refs.~\cite{Kakizaki:2015wua,Hashino:2016xoj,Hashino:2018wee} for similar studies in non-CSI theories).
Given the large deviation from the SM in the one-loop predictions for $\lambda_{hhh}$, one expects in CSI models the EWPT to be of strong first order ($c.f.$ Ref~\cite{Kanemura:2004ch}). However, two-loop corrections to the Higgs trilinear coupling certainly affect the strength of the EWPT to some extent, although for the time being this type of effects have only been studied for a non-CSI IDM in Ref.~\cite{Senaha:2018xek} -- the EWPT was then found to be slightly weaker once two-loop corrections to $\hat\lambda_{hhh}$ where included. In turn, this also modifies the spectrum of gravitational waves produced during the EWPT if it is of strong first-order nature. There is therefore strong motivation to consider two-loop corrections to the Higgs trilinear coupling at finite temperatures in CSI theories.

\section{Summary}
\label{SEC:Conclusion}
In this paper, we have performed the first explicit calculation of leading two-loop corrections to the Higgs trilinear coupling in models with classical scale invariance, using the effective-potential approximation. For the wide range of CSI models in which the scalon/Higgs direction does not mix with other states -- so that we can compute $\veff$ as in eq.~(\ref{EQ:form_Veff_1+2l}) -- we have derived general two-loop results relating the coefficients of the effective potential to the corrections to the Higgs trilinear coupling. Importantly, we find that while at one loop the prediction for $\hat\lambda_{hhh}$ is the same in all CSI theories, this is not any more the case at two loops. 

We focused our investigations on two particular types of CSI theories: first an $N$-scalar model with a global $O(N)$ symmetry (where we took either $N=1$ or $N=4$ for numerical applications), and next a CSI-2HDM. For both models, we computed the leading corrections at two loops involving the BSM scalars -- and in the CSI-2HDM, the top quark as well. Because these expressions are derived from $\vtwo$, which is computed in the \msbar scheme, results are obtained at first in the \msbar scheme. Therefore, we have included the necessary scheme conversion to rewrite the corrections to the Higgs trilinear coupling in terms of physical parameters (pole masses and physical Higgs VEV) and to take into account the effects from finite WFR. 
Once we turn to the numerical study of these models, we observe that the two-loop corrections clearly lift the degeneracy in $\hat\lambda_{hhh}$ among different BSM masses and parameters. Furthermore, the results for $\hat\lambda_{hhh}$ in CSI and non-CSI variants of the same model strongly differ in their theoretical behaviours, mainly because of the difference at one loop -- constant (mass-independent) value in the CSI case, and corrections proportional to the fourth power of the BSM mass(es) in the non-CSI case. 

However, when we also consider the constraints on the allowed ranges of the BSM parameters coming from the criterion of tree-level perturbative unitarity and from the mass relation in CSI theories, the possible values of $\hat\lambda_{hhh}$ become quite restricted, namely $90\%\lesssim (\delta R)^{(2)}\lesssim 110-115\%$. In other words, while the Higgs trilinear coupling at two loops is not universally predicted in all CSI models (without mixing), its allowed values are still severely limited by the classical scale invariance. Nevertheless, the additional \textit{positive} deviation of order $5-30\%$ of the Higgs trilinear coupling from its SM prediction at two loops can make the BSM discrepancy easier to find at future experiments, and is of the same order of magnitude as the expected accuracy of the determination of $\lambda_{hhh}$ at future colliders (as discussed in the introduction). Therefore, two-loop calculations will in the future prove necessary for CSI theories in order to consistently compare theoretical predictions with experimental results. 

While it may be difficult to distinguish between CSI scenarios and non-CSI scenarios (in the non-decoupling limit) simply from the size of the Higgs trilinear coupling, our work demonstrates the important role of the mass relation that strongly constrains masses and parameters in CSI theories, and may prove to be one of the most powerful tools to probe the CSI nature of a potential BSM discovery. Additionally, as we have mentioned in the previous section, the synergy of the measurement of the spectrum of primordial gravitational waves produced during the EWPT together with that of the Higgs trilinear coupling can also help distinguish scenarios with or without CSI~\cite{Hashino:2016rvx}.

Finally, in appendix~\ref{APP:generic} we have provided generic \msbar results for the coefficients $\btl$ and $\ctl$ (see eq.~(\ref{EQ:form_Veff_1+2l})) applicable to all CSI models without mixing, and which served to cross-check some of our calculations. We hope these can be of use for the community to study further CSI scenarios. 

\section*{Acknowledgements} 
This work is supported by JSPS, Grant-in-Aid for Scientific Research, No. 16H06492, 18F18022, 18F18321 and 20H00160. This work is also partly supported by the Deutsche Forschungsgemeinschaft (DFG, German Research Foundation) under Germany’s Excellence Strategy – EXC 2121 “Quantum Universe” – 390833306.

\vfill 
\pagebreak
\appendix
\allowdisplaybreaks

\section{Loop functions}
\label{APP:loopfn}
This appendix summarises notations and definitions of loop functions used throughout this paper.  

First of all, the loop factor is defined as 
\begin{equation}
\label{EQ:loopfactor}
 \kappa=\frac{1}{16\pi^2}\,.
\end{equation}
Next, we denote the regularisation and renormalisation scales respectively $\mu$ and $Q$. These two scales are related as
\begin{align}
 Q^2=4\pi e^{-\gamma_E}\mu^2\,,
\end{align}
where $\gamma_E\simeq0.577$ is the Euler constant. For convenience, we also make extensive use of the notation
\begin{equation}
\label{EQ:logbar}
 \llog x\equiv\log\frac{x}{Q^2}\,.
\end{equation} 

\subsection{One-loop functions}
At one-loop order, we employ the following two integrals
\begin{align}
\label{EQ:loopfnAB}
 \mathbf{J}(x)&\equiv-16\pi^2 \frac{\mu^{2\epsilon}}{i(2\pi)^d}\int_k \frac{d^dk}{k^2-x}\,,\nn\\
 \mathbf{B}(p^2,x,y)&\equiv16\pi^2 \frac{\mu^{2\epsilon}}{i(2\pi)^d} \int_k\frac{d^dk}{(k^2-x)((p-k)^2-y)}\,,
\end{align}
defined in dimensional regularisation, with $d=4-2\epsilon$. 
Their finite (and $\epsilon$-independent) part yield the usual Passarino-Veltmann functions \cite{Passarino:1978jh}
\begin{align}
 J(x)&\equiv\lim_{\eps\to0}\bigg[\mathbf{J}(x)+\frac{x}{\eps}\bigg]=x(\llog x-1)\,,\nn\\
 B_0(p^2,x,y)&\equiv\lim_{\eps\to0}\bigg[\mathbf{B}(p^2,x,y)-\frac{1}{\eps}\bigg]=-\llog p^2-f_B(x_+)-f_B(x_-),
\end{align}
where
\begin{align}
\label{EQ:def_fB}
 f_B(x)&=\log(1-x)-x\log\left(1-\frac{1}{x}\right)-1\,,\nn\\
 x_\pm&=\frac{p^2+x-y\pm\sqrt{(p^2+x-y)^2-4p^2x}}{2p^2}\,.
\end{align}
Some simple limits of particular interest include
\begin{align}
 B_0(x,0,0)&=B_0(x,0,x)=2-\llog x\,,\nn\\
 B_0(x,x,x)&=2-\frac{\pi}{\sqrt{3}}-\llog x\,,\nn\\
 B_0(x,0,y)&=2-\llog y+\left(\frac{y}{x}-1\right)\log\left(1-\frac{x}{y}\right)\,.
\end{align}
Finally, when computing contributions from finite WFR, it is useful to expand the $B_0$ function for low momentum $p^2$ with the relation
\begin{align}
 B_0(p^2,x,x)\underset{p^2\ll x}{=}-\llog x+\frac{p^2}{6x}+\mathcal{O}\left(\frac{p^4}{x^2}\right)\,.
\end{align}

\subsection{Two-loop functions}
When considering two-loop corrections to the effective potential, we also come to encounter the sunrise integral $\mathbf{I}$ (see $e.g.$ Ref.~\cite{Ford:1992pn}) defined as
\begin{align}
 \mathbf{I}(x,y,z)\equiv (16\pi^2)^2\frac{\mu^{4\epsilon}}{(2\pi)^{2d}}\int_{k_1}\int_{k_2}\frac{d^dk_1d^dk_2}{(k_1^2-x)(k_2^2-y)((k_1+k_2)^2-z)}\,.
\end{align}
Its finite part is
\begin{align}
 I(x,y,z)\equiv\lim_{\epsilon\to0}\left[\mathbf{I}(x,y,z)-\frac{\mathbf{J}(x)+\mathbf{J}(y)+\mathbf{J}(z)}{\epsilon}-\frac12\left(\frac{1}{\epsilon^2}-\frac{1}{\epsilon}\right)(x+y+z)\right]\,,
\end{align} 
with $\mathbf{J}$ the one-loop integral from eq.~(\ref{EQ:loopfnAB}). Expressions for $I$, equivalent up to relations among dilogarithms can be found in many references (see for instance Refs.~\cite{Ford:1992pn,Martin:2001vx,Martin:2003qz,Degrassi:2009yq}). Ref.~\cite{Degrassi:2009yq} provides the result in a convenient way, which we reproduce here
\begin{align}
\label{EQ:expr_I}
 I(x,y,z)=&\ \frac12\big[(x-y-z)\llog y \llog z + (y-z-x) \llog x\llog z + (z-x-y) \llog x \llog y\big]\nn\\
          &+2\big[x\llog x+y \llog y+ z \llog z\big]-\frac52(x+y+z)-\frac{\Delta(x,y,z)}{2z}\Phi(x,y,z)\,,
\end{align}
with
\begin{align}
\label{EQ:def_Delta}
 \Delta(x,y,z)=&\ x^2+y^2+z^2-2(xy+xz+yz)\,,\\
\label{EQ:def_Phi} \Phi(x,y,z)=&\ \frac{1}{\lambda(x,y,z)}\bigg[2\log X_+\log X_--\log\frac{x}{z}\log\frac{y}{z}-2\big(\mathrm{Li}_2(X_+)+\mathrm{Li}_2(X_-)\big)+\frac{\pi^2}{3}\bigg]\,.
\end{align}
$\mathrm{Li}_2(z)\equiv-\int_0^z \log(1-t)/t dt$ is the dilogarithm function, and 
\begin{align}
 \lambda(x,y,z)=&\ \frac{1}{z}\big[\big(z-x-y\big)^2-4xy\big]^{1/2}\,,\nn\\
 X_\pm=&\ \frac12\bigg[1\pm\frac{x-y}{z}-\lambda(x,y,z)\bigg]\,.
\end{align}
Note that the definition of the function $\Phi$ in eq.~(\ref{EQ:def_Phi}) is only valid for $x,y<z$, and the other branches of the function can be found with the relations
\begin{align}
 \Phi(x,y,z)=\Phi(y,x,z)\,,\qquad\text{and}\qquad  x\ \Phi(x,y,z)=z\ \Phi(z,y,x)\,.
\end{align}
Furthermore, a number of useful limits are included in Refs.~\cite{Martin:2003qz,Braathen:2016cqe}, among which
\begin{align}
 I(x,x,x)=&\ \frac32x\left[-\llog^2x+4\llog x-5-\frac{i}{\sqrt{3}}\bigg(\frac{\pi^2}{9}-4\ \mathrm{Li}_2\left(\frac12-\frac{i\sqrt{3}}{2}\right)\bigg)\right]\,,\nn\\
 I(x,x,0)=&\ x\big[-\llog^2 x+4\llog x-5\big]\,,\nn\\
 I(x,0,0)=&\ \frac12x\left[-\llog^2x+4\llog x-5-\frac{\pi^2}{3}\right]\,.
\end{align}

\section{Detailed results for the $N$-scalar model}
\label{APP:O(N)}
\subsection{With classical scale invariance}
\label{APP:CSIO(N)}
We present in this appendix intermediate results that we employed for the \msbar to OS scheme conversion of our expressions for the CSI $O(N)$-symmetric model. The different parameters we need conversions for are $v$, $m_t$, $[M_h^2]_{\veff}$, and $m_S$. First of all, for the Higgs VEV, we can use the same result as in the SM, $c.f.$ equation~(\ref{EQ:VEVconv}).

Next, the one-loop self-energy of the top quark is the same as in the SM (the BSM singlet scalars do not couple to it) and it reads  
\begin{align}
\Pi_{tt}^{(1)}(p^2=m_t^2)=&\ \frac43g_3^2m_t^2(8-6\llog m_t^2)+\frac{m_t^4}{v^2}(-8+3\llog m_t^2)\,.
\end{align}

For the conversion from the Higgs curvature to pole mass and the Higgs WFR, we only require the momentum-dependent contributions to the one-loop Higgs self-energy -- momentum-independent terms cancel out when taking the difference $\Pi_{hh}(p^2=m_h^2)-\Pi_{hh}(p^2=0)$ and moreover their derivative with respect to $p^2$ is also zero of course. The momentum-dependent corrections are
\begin{align}
 \Pi^{(1)}_{hh}(p^2)\bigg|_{p^2\text{-dep.}}=\frac{6m_t^2}{v^2}\bigg[(4m_t^2-p^2)B_0(p^2,m_t^2,m_t^2)\bigg]-\frac{2Nm_S^4}{v^2}B_0(p^2,m_S^2,m_S^2)\,,
\end{align}
where the first term is the SM-like correction from the top quark and the second term is the BSM contribution from the scalars $S_i$. 

Finally, we find for the one-loop self-energy of the BSM scalars 
\begin{align}
\label{EQ:O(N)sym_S_selfenergy}
 \Pi^{(1)}_{SS}(p^2)=(N+2)\lambda_SJ(m_S^2)-4\lambda_{\Phi S}^2v^2B_0(p^2,0,m_S^2)\,.
\end{align}

\subsection{Without classical scale invariance}
\label{APP:nonCSIO(N)}

In order to compare predictions for the Higgs trilinear coupling in $O(N)$-symmetric models with and without CSI, we derive in this appendix the leading two-loop contributions to $\hat\lambda_{hhh}$ in the non-CSI version of these models. Note that the case for $N=1$ was already considered in Ref.~\cite{Braathen:2019zoh} -- where it was referred to as the "Higgs-Singlet Model" (HSM). 

If we do not impose classical scale invariance, the tree-level scalar potential in a theory invariant under a global $O(N)$ symmetry reads
\begin{align}
 \vtree=&\ \mu^2|\Phi|^2+\frac12\mu_S^2\vec{S}^2+\lambda |\Phi|^4+\lambda_{\Phi S}\vec{S}^2|\Phi|^2+\frac{1}{4}\lambda_S(\vec{S}^2)^2\,.
\end{align} 
Due to the unbroken $O(N)$ symmetry, the singlets $S_i$ do not acquire VEVs, and the only tadpole equation gives at tree level
\begin{equation}
 \mu^2=-\lambda v^2\,.
\end{equation}
The field-dependent tree-level masses read
\begin{align}
 m_h^2(h)=\mu^2+3\lambda(v+h)^2\,,\quad m_G^2(h)=m_{G^\pm}^2(h)=\mu^2+\lambda(v+h)^2\,,\quad m_{S_i}^2(h)=\mu_S^2+\lambda_{\Phi S}(v+h)^2\,.
\end{align}
The dominant two-loop contributions to the effective potential in this model are the same as shown in figure~\ref{FIG:O(N)2ldiags}, and the expressions are (like for the CSI version of the model):
\begin{align}
 \vtwo=&\ \vtwo_{hSS}+\vtwo_{SS}\nn\\
 \vtwo_{hSS}=&-\sum_{i=1}^N\lambda_{\Phi S}^2 (v+h)^2 I(0,m_{S_i}^2(h),m_{S_i}^2(h))\nn\\
 =&-N\lambda_{\Phi S}^2 (v+h)^2 I(m_S^2(h),m_S^2(h),0)\nn\\
 \vtwo_{SS}=&\ \sum_{i=1}^N \frac{1}{8}\lambda_{S_iS_iS_iS_i}J(m_{S_i}^2(h))^2+\sum_{i=1}^N\sum_{j=1,j\neq i}^N\frac{1}{8}\lambda_{S_iS_iS_jS_j}J(m_{S_i}^2(h))J(m_{S_j}^2(h))\nn\\
 =&\ \frac14N(N+2)\lambda_SJ(m_S^2(h))^2\nn\\
\end{align}
Following the calculation method explained in detail in Ref.~\cite{Braathen:2019zoh} and using the differential operator $\mathcal{D}_3$ defined therein, we find for the dominant two-loop BSM corrections to the Higgs trilinear coupling $\lambda_{hhh}$ 
\begin{align}
 \delta^{(2)}\lambda_{hhh}=&\ \frac{1}{(16\pi^2)^2} \mathcal{D}_3\vtwo\big|_\text{min.}\nn\\
 =&\ \frac{1}{(16\pi^2)^2}\bigg\{\frac{16 N m_S^4}{v^5} \left(1 - \frac{\mu_S^2}{m_S^2}\right)^4 \left[-m_S^2 -2 \mu_S^2 + (2 m_S^2 + \mu_S^2) \llog m_S^2\right]\nn\\
 &\hspace{2cm}+\frac{4N(N+2)\lambda_Sm_S^4}{v^3} \left(1- \frac{\mu_S^2}{m_S^2}\right)^3 [1 + 2 \llog m_S^2]\bigg\}\,.
\end{align}
For the leading one-loop BSM corrections, we have simply
\begin{align}
\label{EQ:nonCSIO(N)_lambdahhh_1l}
 \delta^{(1)}\lambda_{hhh}=&\ \frac{1}{16\pi^2}\mathcal{D}_3\vone\big|_\text{min.}=\frac{1}{16\pi^2}\frac{4Nm_S^4}{v^3}\left(1-\frac{\mu_S^2}{m_S^2}\right)^3\,.
\end{align}
To convert this result to the on-shell scheme, we need the one-loop self-energy of the scalars $S_i$, given in eq.~(\ref{EQ:O(N)sym_S_selfenergy}), as well as the finite counter-term for $\mu_S^2$ (to ensure proper decoupling of the OS result when taking $\tilde\mu_S^2\to\infty$ as discussed at length in Ref.~\cite{Braathen:2019zoh}). The latter can be obtained straightforwardly by adapting the result in eq.~(5.29) of Ref.~\cite{Braathen:2019zoh} (originally for the HSM) to our case. We find
\begin{align}
 \delta^\text{OS}\mu_S^2=\frac{1}{16\pi^2}(N+2)\lambda_S\tilde\mu_S^2(\llog M_S^2-1)\,.
\end{align}
Combining all these results, we obtain for the dominant one-loop and two-loop BSM contributions to the Higgs trilinear coupling, expressed in the OS scheme
\begin{align}
 (16\pi^2)\,  \delta^{(1)}\hat\lambda_{hhh}=&\ \frac{4NM_S^4}{\vphys^3}\left(1-\frac{\tilde\mu_S^2}{M_S^2}\right)^3-\frac{NM_S^2M_h^2}{2\vphys^3}\left(1-\frac{\tilde\mu_S^2}{M_S^2}\right)^2\,,\nn\\
 (16\pi^2)^2\delta^{(2)}\hat\lambda_{hhh}=&\ \frac{48 NM_S^6}{\vphys^5} \left(1- \frac{\tilde\mu_S^2}{M_S^2}\right)^4+\frac{12N (N+2) \lambda_SM_S^4}{\vphys^3}\left(1 - \frac{\tilde\mu_S^2}{M_S^2}\right)^3\nn\\
 &+\frac{42NM_S^4M_t^2}{\vphys^5}\left(1-\frac{\tilde\mu_S^2}{M_S^2}\right)^3+\frac{24NM_S^2M_t^4}{\vphys^5}\left(1-\frac{\tilde\mu_S^2}{M_S^2}\right)^2-\frac{2N^2M_S^6}{\vphys^5}\left(1-\frac{\tilde\mu_S^2}{M_S^2}\right)^5\,.
\end{align}
The last three terms ($i.e.$ the second line) come from WF and VEV renormalisations. Setting $N=1$ and taking into account the slightly different convention for $\lambda_S$, we do recover the same result as in equation~(5.31) of Ref.~\cite{Braathen:2019zoh}.

\section{Detailed results for the CSI-2HDM}
\label{APP:CSI2HDM}
In this appendix, we generalise the results of the section~\ref{SEC:CSI2HDM} for the case where the masses of the three BSM scalars are not degenerate. 

\subsection{Results for $\vtwo$ and $\hat\lambda_{hhh}$}
\label{APP:CSI2HDM_lambdahhh}
First of all, the leading two-loop BSM corrections to $\veff$ of the CSI-2HDM read
\begin{align}
 \vtwo_{SSS}(h)=&-\sum_{\Phi=H,A,H^\pm}\frac{n_\Phi m_\Phi^4(v+h)^2}{v^4}I(0,m_\Phi^2(h),m_\Phi^2(h))\nn\\
                &-\sum_{\Phi=A,H^\pm}\frac{n_\Phi m_H^4\cot^22\beta(v+h)^2}{v^4}I(m_H^2(h),m_\Phi^2(h),m_\Phi^2(h))\nn\\
                &-\frac{3m_H^4\cot^22\beta(v+h)^2}{v^4}I(m_H^2(h),m_H^2(h),m_H^2(h))\nn\\
                &-\frac{(v+h)^2}{2v^4}\bigg[(m_H^2-m_A^2)^2I(m_H^2(h),m_A^2(h),0)+2(m_H^2-m_{H^\pm}^2)^2I(m_H^2(h),m_{H^\pm}^2(h),0)\nn\\
                &\hspace{2cm}+2(m_A^2-m_{H^\pm}^2)^2I(m_A^2(h),m_{H^\pm}^2(h),0)\bigg]\\
  \vtwo_{SS}(h)=&\ \frac{m_H^2 \cot^22\beta}{v^2} \bigg[\frac32 J(m_H^2(h))^2 + \frac32 J(m_A^2(h))^2 + 4 J(m_{H^\pm}^2(h))^2 +  J(m_A^2(h)) J(m_H^2(h)) \nn\\
                &\hspace{4cm} + 2 J(m_H^2(h)) J(m_{H^\pm}^2(h)) + 2 J(m_A^2(h)) J(m_{H^\pm}^2(h))\bigg]\\
  \vtwo_{FFS}(h)=&-\frac32y_t^2c_\beta^2\bigg[2J(m_t^2(h))J(m_H^2(h))-J(m_t^2(h))^2-(4m_t^2(h)-m_H^2(h))I(m_H^2(h),m_t^2(h),m_t^2(h))\bigg]\nn\\
                 &-\frac32y_t^2c_\beta^2\bigg[2J(m_t^2(h))J(m_A^2(h))-J(m_t^2(h))^2+m_A^2(h)I(m_A^2(h),m_t^2(h),m_t^2(h))\bigg]\nn\\
                 &-3y_t^2c_\beta^2\bigg[J(m_t^2(h))J(m_{H^\pm}^2(h))-(m_t^2(h)-m_{H^\pm}^2(h))I(m_{H^\pm}^2(h),m_t^2(h),0)\bigg]
\end{align}

Turning next to the $\log^2$ terms $\ctl$, we have first for the scalar sunrise diagrams
\begin{align}
\label{EQ:CSI2HDM_C2SSS}
 \ctl_{SSS}=&\ \frac{m_H^6+m_A^6+2m_{H^\pm}^6}{v^6}+\frac{m_H^4\cot^22\beta}{v^6}\big[6m_H^2+m_A^2+2m_{H^\pm}^2\big]\nn\\
 &+\frac14\frac{(m_H^2-m_A^2)^2(m_H^2+m_A^2)}{v^6}+\frac12\frac{(m_H^2-m_{H^\pm}^2)^2(m_H^2+m_{H^\pm}^2)}{v^6}\nn\\
 &+\frac12\frac{(m_A^2-m_{H^\pm}^2)^2(m_A^2+m_{H^\pm}^2)}{v^6}
\end{align} 
Second, for the eight-shaped diagrams we have
\begin{align}
\label{EQ:CSI2HDM_C2SS}
 \ctl_{SS}&=\ \frac{m_H^2\cot^22\beta}{2v^6}\left[3m_H^4+3m_A^4+8m_{H^\pm}^4+2m_H^2m_A^2+4m_{H^\pm}^2(m_H^2+m_A^2)\right]\,.
\end{align}
Finally, for the diagrams with top quarks, we find
\begin{align}
\label{EQ:CSI2HDM_C2FFSgen}
 \ctl_{F\bar{F}S}&=\frac{3m_t^2\cot^2\beta}{2v^6}\big[m_H^4+m_A^4+2m_{H^\pm}^4-2m_t^2\big(3m_H^2+m_A^2+2m_{H^\pm}^2\big)-6m_t^4\big]\,.
\end{align}

Grouping all these results together, and using eq.~(\ref{EQ:res_2l_lambdahhh}), we find
\begin{align}
 \lambda_{hhh}=\frac{5[M_h^2]_{\veff}}{v}+\frac{1}{(16\pi^2)^2}\bigg[&\frac{768g_3^2m_t^4}{v^3}-\frac{288m_t^6}{v^5}+\frac{56m_H^6+56m_A^6+96m_{H^\pm}^6}{v^5}\nn\\
 &-\frac{8m_H^4m_A^2+8m_H^2m_A^4+16m_H^4m_{H^\pm}^2+16m_H^2m_{H^\pm}^4+16m_A^4m_{H^\pm}^2+16m_A^2m_{H^\pm}^4}{v^5}\nn\\
 &+\frac{32m_H^4\cot^22\beta}{v^5}\big(6m_H^2+m_A^2+2m_{H^\pm}^2\big)\nn\\
 &+\frac{16m_H^2\cot^22\beta}{v^5}\left(3m_H^4+3m_A^4+8m_{H^\pm}^4+2m_H^2m_A^2+4m_{H^\pm}^2(m_H^2+m_A^2)\right)\nn\\
 &+\frac{48m_t^2\cot^2\beta}{v^5}\big(m_H^4+m_A^4+2m_{H^\pm}^4-2m_t^2\big(3m_H^2+m_A^2+2m_{H^\pm}^2\big)-6m_t^4\big)\bigg]\,.
\end{align}
By taking $m_H=m_A=m_{H^\pm}=m_\Phi$, we can straightforwardly recover eq.~(\ref{EQ:CSI2HDM_2lres}). 

In order to convert this expression from the \msbar to the OS scheme, we need to take into account the finite renormalisation of the Higgs VEV -- as given in equation~(\ref{EQ:VEVconv}), as well as the correction to the Higgs mass and the effect of finite WFR. The latter two are calculated from the momentum-dependent part of the Higgs self-energy, which in the CSI-2HDM receives new contributions from each of the BSM scalars. In total, it reads
\begin{align}
\label{EQ:CSI2HDM_Higgsself}
 \Pi_{hh}^{(1)}(p^2)\big|_{p^2\text{-dep.}}=&\ \frac{6m_t^2}{v^2}\bigg[(4m_t^2-p^2)B_0(p^2,m_t^2,m_t^2)\bigg]\nn\\
 &-\frac{2m_H^4}{v^2}B_0(p^2,m_H^2,m_H^2)-\frac{2m_A^4}{v^2}B_0(p^2,m_A^2,m_A^2)-\frac{4m_{H^\pm}^4}{v^2}B_0(p^2,m_{H^\pm}^2,m_{H^\pm}^2)\,.
\end{align} 
This enables us to obtain finally in the OS scheme
\begin{align}
 \hat\lambda_{hhh}=&\ \frac{5M_h^2}{\vphys}+\frac{1}{16\pi^2}\frac{5M_h^2}{\vphys}\bigg[\frac{7}{2}\frac{M_t^2}{\vphys^2}-\frac{1}{6}\frac{M_H^2+M_A^2+2M_{H^\pm}^2}{\vphys^2}\bigg]\nn\\
 &+\frac{1}{(16\pi^2)^2}\bigg[\frac{768g_3^2M_t^4}{\vphys^3}-\frac{288M_t^6}{\vphys^5}+\frac{56M_H^6+56M_A^6+96M_{H^\pm}^6}{\vphys^5}\nn\\
 &\hspace{1.75cm}-\frac{8(M_H^4M_A^2+M_H^2M_A^4+2M_H^4M_{H^\pm}^2+2M_H^2M_{H^\pm}^4+2M_A^4M_{H^\pm}^2+2M_A^2M_{H^\pm}^4)}{\vphys^5}\nn\\
 &\hspace{1.75cm}+\frac{32M_H^4\cot^22\beta}{\vphys^5}\big(6M_H^2+M_A^2+2M_{H^\pm}^2\big)\nn\\
 &\hspace{1.75cm}+\frac{16M_H^2\cot^22\beta}{\vphys^5}\left(3M_H^4+3M_A^4+8M_{H^\pm}^4+2M_H^2M_A^2+4M_{H^\pm}^2(M_H^2+M_A^2)\right)\nn\\
 &\hspace{1.75cm}+\frac{48M_t^2\cot^2\beta}{\vphys^5}\big(M_H^4+M_A^4+2M_{H^\pm}^4-2M_t^2\big(3M_H^2+M_A^2+2M_{H^\pm}^2\big)-6M_t^4\big)\bigg]\,.
\end{align}

\subsection{Results for the mass relation}
\label{APP:CSI2HDM_massrelation}
\subsubsection{Expression in the \msbar scheme}
Finally, we also consider the mass relation for the CSI-2HDM with general BSM scalar masses, and therefore we start by deriving expressions for the quantities $\ctl,\,\bol,\,\btl$. First of all, we have that
\begin{align}
 \ctl=\frac{24g_3^2m_t^4}{v^4}-\frac{9m_t^6}{v^6}+\ctl_{SS}+\ctl_{SSS}+\ctl_{F\bar{F}S}\,,
\end{align}
with $\ctl_{SS}$, $\ctl_{SSS}$, and $\ctl_{F\bar{F}S}$ given respectively in equations~(\ref{EQ:CSI2HDM_C2SSS}), (\ref{EQ:CSI2HDM_C2SS}), and (\ref{EQ:CSI2HDM_C2FFSgen}). Next, $\bol$ is simply
\begin{align}
 \bol=&\ \frac{1}{4v^4}\big(m_H^4+m_A^4+2m_{H^\pm}^4-12m_t^4+6m_W^4+3m_Z^4\big)\,.
\end{align} 
The expression for $\btl$ is significantly longer, and can be written as
\begin{align}
 \btl=&\ \frac{4g_3^2m_t^4}{v^4}\left[12\log\frac{m_t^2}{v^2}-16\right]+\frac{6m_t^6}{v^6}\left[-3\log\frac{m_t^2}{v^2}+8\right]+\btl_{SS}+\btl_{SSS}+\btl_{F\bar{F}S}\,,
\end{align}
with 
\begin{align}
 \btl_{SS}=&\ \frac{3m_H^6\cot^22\beta}{v^6}\left(\log\frac{m_H^2}{v^2}-1\right)+\frac{3m_H^2m_A^4\cot^22\beta}{v^6}\left(\log\frac{m_A^2}{v^2}-1\right)\nn\\
 &+\frac{8m_H^2m_{H^\pm}^4\cot^22\beta}{v^6}\left(\log\frac{m_{H^\pm}^2}{v^2}-1\right)+\frac{m_H^4m_A^2\cot^22\beta}{v^6}\left(\log\frac{m_H^2}{v^2}+\log\frac{m_A^2}{v^2}-2\right)\nn\\
 &+\frac{2m_H^4m_{H^\pm}^2\cot^22\beta}{v^6}\left(\log\frac{m_H^2}{v^2}+\log\frac{m_{H^\pm}^2}{v^2}-2\right)\nn\\
 &+\frac{2m_H^2m_A^2m_{H^\pm}^2\cot^22\beta}{v^6}\left(\log\frac{m_A^2}{v^2}+\log\frac{m_{H^\pm}^2}{v^2}-2\right)\,,\nn\\
 \btl_{SSS}=&\ \frac{2m_H^6}{v^6}\left(\log \frac{m_H^2}{v^2}-2\right)+\frac{2m_A^6}{v^6}\left(\log \frac{m_A^2}{v^2}-2\right)+\frac{4m_{H^\pm}^6}{v^6}\left(\log \frac{m_{H^\pm}^2}{v^2}-2\right)\nn\\
 &+\frac{12m_H^6\cot^22\beta}{v^6}\left(\log \frac{m_H^2}{v^2}-2\right)+\frac{2m_H^4m_A^2\cot^22\beta}{v^6}\left(\log \frac{m_A^2}{v^2}-2\right)\nn\\
 &+\frac{4m_H^4m_{H^\pm}^2\cot^22\beta}{v^6}\left(\log \frac{m_{H^\pm}^2}{v^2}-2\right)\nn\\
 &+\frac1{2v^6}\bigg\{(m_H^2-m_A^2)^2\bigg[m_H^2\left(\log\frac{m_H^2}{v^2}-2\right)+m_A^2\left(\log\frac{m_A^2}{v^2}-2\right)\bigg]\nn\\
 &\hspace{1.3cm}+2(m_H^2-m_{H^\pm}^2)^2\bigg[m_H^2\left(\log\frac{m_H^2}{v^2}-2\right)+m_{H^\pm}^2\left(\log\frac{m_{H^\pm}^2}{v^2}-2\right)\bigg]\nn\\
 &\hspace{1.3cm}+2(m_A^2-m_{H^\pm}^2)^2\bigg[m_A^2\left(\log\frac{m_A^2}{v^2}-2\right)+m_{H^\pm}^2\left(\log\frac{m_{H^\pm}^2}{v^2}-2\right)\bigg]\bigg\}\,,
\end{align}
and
\begin{align}
 \btl_{F\bar{F}S}=&\ \btl_{ttH}+\btl_{ttA}+\btl_{tbH^\pm}\,,\text{ with}\nn\\
 \btl_{ttH}=&\ \frac{3m_t^2\cot^2\beta}{v^6}\bigg\{(4m_t^2-m_H^2)\bigg[2m_t^2\left(2-\log\frac{m_t^2}{v^2}\right)+m_H^2\left(2-\log\frac{m_H^2}{v^2}\right)\bigg]\nn\\
 &\hspace{2.5cm}+2m_t^4\left(\log\frac{m_t^2}{v^2}-1\right)-2m_t^2m_H^2\left(\log\frac{m_t^2}{v^2}+\log\frac{m_H^2}{v^2}-2\right)\bigg\}\nn\\
 =&\ \frac{3m_t^2\cot^2\beta}{v^6}\bigg\{ m_H^4\left(\log\frac{m_H^2}{v^2}-2\right)-2m_t^2m_H^2\left(3\log\frac{m_H^2}{v^2}-4\right)-2m_t^4\left(3\log\frac{m_t^2}{v^2}-7\right)\bigg\}\,,\nn\\
 \btl_{ttA}=&\ \frac{3m_t^2\cot^2\beta}{v^6}\bigg\{-m_A^2\bigg[2m_t^2\left(2-\log\frac{m_t^2}{v^2}\right)+m_A^2\left(2-\log\frac{m_A^2}{v^2}\right)\bigg]\nn\\
 &\hspace{2.5cm}+2m_t^4\left(\log\frac{m_t^2}{v^2}-1\right)-2m_t^2m_A^2\left(\log\frac{m_t^2}{v^2}+\log\frac{m_A^2}{v^2}-2\right)\bigg\}\nn\\
 =&\ \frac{3m_t^2\cot^2\beta}{v^6}\bigg[m_A^4\left(\log\frac{m_A^2}{v^2}-2\right)-2m_t^2m_A^2\log\frac{m_A^2}{v^2}+2m_t^4\left(\log\frac{m_t^2}{v^2}-1\right)\bigg]\,,\nn\\
 \btl_{tbH^\pm}=&\ \frac{6m_t^2\cot^2\beta}{v^6}\bigg\{m_{H^\pm}^4\left(\log\frac{m_{H^\pm}^2}{v^2}-2\right)-2m_t^2m_{H^\pm}^2\left(\log\frac{m_{H^\pm}^2}{v^2}-1\right)-m_t^4\left(\log\frac{m_t^2}{v^2}-2\right)\bigg\}\,.
\end{align}
As an intermediate step, we next calculate the sum $\btl+\ctl(1+2\log {v^2}/{Q^2})$ for the different contributions -- $SS$, $SSS$, $ttH$, $ttA$, and $tbH^\pm$ -- and we denote this quantity as $D^{(2)}$. We obtain 
\begin{align}
 D^{(2)}_{SS}\equiv&\ \btl_{SS}+\ctl_{SS}\left(1+2\log\frac{v^2}{Q^2}\right)\nn\\
 =&\ \frac{3m_H^6\cot^22\beta}{v^6}\left(\log\frac{m_H^2}{Q^2}-\frac12\right)+\frac{3m_H^2m_A^4\cot^22\beta}{v^6}\left(\log\frac{m_A^2}{Q^2}-\frac12\right)\nn\\
 &+\frac{8m_H^2m_{H^\pm}^4\cot^22\beta}{v^6}\left(\log\frac{m_{H^\pm}^2}{Q^2}-\frac12\right)+\frac{m_H^4m_A^2\cot^22\beta}{v^6}\left(\log\frac{m_H^2}{Q^2}+\log\frac{m_A^2}{Q^2}-1\right)\nn\\
 &+\frac{2m_H^4m_{H^\pm}^2\cot^22\beta}{v^6}\left(\log\frac{m_H^2}{Q^2}+\log\frac{m_{H^\pm}^2}{Q^2}-1\right)\nn\\
 &+\frac{2m_H^2m_A^2m_{H^\pm}^2\cot^22\beta}{v^6}\left(\log\frac{m_A^2}{Q^2}+\log\frac{m_{H^\pm}^2}{Q^2}-1\right)\,,
\end{align}

\begin{align}
 D^{(2)}_{SSS}\equiv&\ \btl_{SSS}+\ctl_{SSS}\left(1+2\log\frac{v^2}{Q^2}\right)\nn\\
 =&\ \frac{2m_H^6}{v^6}\left(\log \frac{m_H^2}{Q^2}-\frac32\right)+\frac{2m_A^6}{v^6}\left(\log \frac{m_A^2}{Q^2}-\frac32\right)+\frac{4m_{H^\pm}^6}{v^6}\left(\log \frac{m_{H^\pm}^2}{Q^2}-\frac32\right)\nn\\
 &+ \frac{12m_H^6\cot^22\beta}{v^6}\left(\log \frac{m_H^2}{Q^2}-\frac32\right)+\frac{2m_H^4m_A^2\cot^22\beta}{v^6}\left(\log \frac{m_A^2}{Q^2}-\frac32\right)\nn\\
 &+\frac{4m_H^4m_{H^\pm}^2\cot^22\beta}{v^6}\left(\log \frac{m_{H^\pm}^2}{Q^2}-\frac32\right)\nn\\
 &+\frac1{2v^6}\bigg\{(m_H^2-m_A^2)^2\bigg[m_H^2\left(\log\frac{m_H^2}{Q^2}-\frac32\right)+m_A^2\left(\log\frac{m_A^2}{Q^2}-\frac32\right)\bigg]\nn\\
 &\hspace{1.5cm}+2(m_H^2-m_{H^\pm}^2)^2\bigg[m_H^2\left(\log\frac{m_H^2}{Q^2}-\frac32\right)+m_{H^\pm}^2\left(\log\frac{m_{H^\pm}^2}{Q^2}-\frac32\right)\bigg]\nn\\
 &\hspace{1.5cm}+2(m_A^2-m_{H^\pm}^2)^2\bigg[m_A^2\left(\log\frac{m_A^2}{Q^2}-\frac32\right)+m_{H^\pm}^2\left(\log\frac{m_{H^\pm}^2}{Q^2}-\frac32\right)\bigg]\bigg\}\,,
\end{align}

\begin{align}
 D^{(2)}_{ttH}\equiv&\ \btl_{ttH}+\ctl_{ttH}\left(1+2\log\frac{v^2}{Q^2}\right)\nn\\
 =&\ \frac{3m_t^2\cot^2\beta}{v^6}\bigg[ m_H^4\left(\log\frac{m_H^2}{Q^2}-\frac32\right)-6m_t^2m_H^2\left(\log\frac{m_H^2}{Q^2}-\frac56\right)-6m_t^4\left(\log\frac{m_t^2}{Q^2}-\frac{11}{6}\right)\bigg]\,,
\end{align}

\begin{align}
 D^{(2)}_{ttA}\equiv&\ \btl_{ttA}+\ctl_{ttA}\left(1+2\log\frac{v^2}{Q^2}\right)\nn\\
 =&\ \frac{3m_t^2\cot^2\beta}{v^6}\bigg[m_A^4\left(\log\frac{m_A^2}{Q^2}-\frac32\right)-2m_t^2m_A^2\left(\log\frac{m_A^2}{Q^2}+\frac12\right)+2m_t^4\left(\log\frac{m_t^2}{Q^2}-\frac12\right)\bigg]\,,
\end{align}

\begin{align}
 D^{(2)}_{tbH^\pm}\equiv&\ \btl_{tbH^\pm}+\ctl_{tbH^\pm}\left(1+2\log\frac{v^2}{Q^2}\right)\nn\\
 =&\ \frac{6m_t^2\cot^2\beta}{v^6}\bigg[m_{H^\pm}^4\left(\log\frac{m_{H^\pm}^2}{Q^2}-\frac32\right)-2m_t^2m_{H^\pm}^2\left(\log\frac{m_{H^\pm}^2}{Q^2}-\frac12\right)-m_t^4\left(\log\frac{m_t^2}{Q^2}-\frac32\right)\bigg]\,.
\end{align}
 
The mass relation at two loops, for general masses and in terms of \msbar parameters, is then
\begin{align}
\label{EQ:CSI2HDM_massrel_nondegmsbar}
 8\pi^2v^2[M_h^2]_{\veff}=&\ m_H^4+m_A^4+2m_{H^\pm}^4-12m_t^4+6m_W^4+3m_Z^4\nn\\
 &+\frac{4m_t^4}{16\pi^2}\bigg[g_3^2\left(48\log\frac{m_t^2}{Q^2}-40\right)+\frac{m_t^2}{v^2}\left(-18\log\frac{m_t^2}{Q^2}+39\right)\bigg]\nn\\
 &+\frac{4v^4}{16\pi^2}\bigg[D^{(2)}_{SS}+D^{(2)}_{SSS}+D^{(2)}_{ttH}+D^{(2)}_{ttA}+D^{(2)}_{tbH^\pm}\bigg]\,,
\end{align}
with the quantities $D^{(2)}$ given in the previous equations. 

\subsubsection{Intermediate results for the \msbar$\to$ OS conversion}
At this point, we wish to express the result in equation~(\ref{EQ:CSI2HDM_massrel_nondegmsbar}) in terms of pole masses and of the physical Higgs VEV. 

First, for the conversion from the curvature to the pole mass of the Higgs boson, we employ again the expression given in eq.~(\ref{EQ:CSI2HDM_Higgsself}). 

Next, for the one-loop self-energy of the top quark, we have 
\begin{align}
 \Pi_{tt}^{(1)}(p^2=m_t^2)=&\ \frac43g_3^2m_t^2(8-6\llog m_t^2)+\frac{m_t^4}{v^2}(-8+3\llog m_t^2)\nn\\
 &-\frac{m_t^2\cot^2\beta}{v^2}\bigg[2J(m_t^2)-J(m_H^2)-J(m_A^2)-J(m_{H^\pm}^2)+(4m_t^2-m_H^2)B_0(m_t^2,m_t^2,m_H^2)\nn\\
  &\hspace{2.25cm}-m_A^2B_0(m_t^2,m_t^2,m_A^2)+(m_t^2-m_{H^\pm}^2)B_0(m_t^2,0,m_{H^\pm}^2)\bigg] 
\end{align}
where the terms in the square brackets are BSM contributions involving $H$, $A$, and $H^\pm$.

Finally, the self-energies of the three BSM scalars can be found to be 
\begin{align}
 \Pi^{(1)}_{HH}(p^2) =&\ \frac{2m_H^2}{v^2}\cot^22\beta\Big[3J(m_H^2)+J(m_A^2)+2J(m_{H^\pm}^2)\Big]\nn\\
                &-\frac{4m_H^4}{v^2}B_0(p^2,0,m_H^2)-\frac{18m_H^4}{v^2}\cot^22\beta B_0(p^2,m_H^2,m_H^2)\nn\\
                &-\frac{(m_H^2-m_A^2)^2}{v^2}B_0(p^2,0,m_A^2)-\frac{2(m_H^2-m_{H^\pm}^2)^2}{v^2}B_0(p^2,0,m_{H^\pm}^2)\nn\\
                &-\frac{2m_H^4}{v^2}\cot^22\beta \big[B_0(p^2,m_A^2,m_A^2)+2B_0(p^2,m_{H^\pm}^2,m_{H^\pm}^2)\big]\nn\\
                &-\frac{12m_t^2\cot^2\beta}{v^2}\bigg[J(m_t^2)-\bigg(2m_t^2- \frac{p^2}{2}\bigg)B_0(p^2,m_t^2,m_t^2)\bigg]\,,\nn\\
 \Pi^{(1)}_{AA}(p^2) =&\ \frac{2m_H^2}{v^2}\cot^22\beta\Big[J(m_H^2)+3J(m_A^2)+2J(m_{H^\pm}^2)\Big]\nn\\
                &-\frac{4m_A^4}{v^2}B_0(p^2,0,m_A^2)-\frac{4m_H^4}{v^2}\cot^22\beta B_0(p^2,m_A^2,m_H^2)\nn\\
                &-\frac{(m_A^2-m_H^2)^2}{v^2}B_0(p^2,0,m_H^2)-\frac{2(m_A^2-m_{H^\pm}^2)^2}{v^2}B_0(p^2,0,m_{H^\pm}^2)\nn\\
                &-\frac{12m_t^2\cot^2\beta}{v^2}\big[J(m_t^2)+ \frac{p^2}{2}B_0(p^2,m_t^2,m_t^2)\big]\,,\nn\\
 \Pi^{(1)}_{H^+H^-}(p^2)=&\ \frac{2m_H^2}{v^2}\cot^22\beta\Big[J(m_H^2)+J(m_A^2)+4J(m_{H^\pm}^2)\Big]\nn\\
                   &-\frac{4m_{H^\pm}^4}{v^2}B_0(p^2,0,m_{H^\pm}^2)-\frac{4m_H^4}{v^2}\cot^22\beta B_0(p^2,m_{H^\pm}^2,m_H^2)\nn\\
                   &-\frac{(m_{H^\pm}^2-m_H^2)^2}{v^2}B_0(p^2,0,m_H^2)-\frac{(m_{H^\pm}^2-m_A^2)^2}{v^2}B_0(p^2,0,m_A^2)\nn\\ 
                   &-\frac{6m_t^2\cot^2\beta}{v^2}\big[J(m_t^2)+(p^2-m_t^2)B_0(p^2,0,m_t^2)\big]\,.
\end{align}

\subsubsection{Expression in the OS scheme}
Using the previous results, and after some steps of tedious but straightforward algebra, we obtain for the mass relation in terms of physical parameters
\begin{align}
\label{EQ:CSI2HDM_2l_massrelationOSnondeg}
 \frac{4\sqrt{2}\pi^2}{G_F}M_h^2=&\ M_H^4+M_A^4+2M_{H^\pm}^4-12M_t^4+6M_W^4+3M_Z^4\nn\\
 &+M_h^2\left[\frac72M_t^2-\frac16(M_H^2+M_A^2+2M_{H^\pm}^2)\right]+\frac{3M_t^4}{8\pi^2}\left[16g_3^2-\frac{6M_t^2}{\vphys^2}\right]\nn\\
 &+\frac{1}{4\pi^2\vphys^2}\big[M_H^6+M_A^6+2M_{H^\pm}^6\big]+\frac{1}{16\pi^2\vphys^2}(M_H^2-M_A^2)^2(M_H^2+M_A^2)\nn\\
 &+\frac{1}{8\pi^2\vphys^2}\big[(M_H^2-M_{H^\pm}^2)^2(M_H^2+M_{H^\pm}^2)+(M_A^2-M_{H^\pm}^2)^2(M_A^2+M_{H^\pm}^2)\big]\nn\\
 &+\mathfrak{Re}(\aleph_S)+\mathfrak{Re}(\aleph_t)\,,
\end{align}
with the quantities $\aleph_S$ and $\aleph_t$ defined as
\begin{align}
 \aleph_S\equiv&\ \frac{M_H^6\cot^22\beta}{4\pi^2\vphys^2}\bigg[\frac32-3\sqrt{3}\pi-f_B\left(\frac12-\sqrt{\frac{1}{4}-\frac{M_A^2}{M_H^2}}\right)-f_B\left(\frac12+\sqrt{\frac14-\frac{M_A^2}{M_H^2}}\right)\nn\\
 &\hspace{3cm}-2f_B\left(\frac12-\sqrt{\frac14-\frac{M_{H^\pm}^2}{M_H^2}}\right)-2f_B\left(\frac12+\sqrt{\frac14-\frac{M_{H^\pm}^2}{M_H^2}}\right)\bigg]\nn\\
 &+\frac{M_H^4M_A^2\cot^22\beta}{2\pi^2\vphys^2}\bigg[-1-f_B\left(\frac{M_H^2}{2M_A^2}-\sqrt{\frac{M_H^4}{4M_A^4}-1}\right)-f_B\left(\frac{M_H^2}{2M_A^2}+\sqrt{\frac{M_H^4}{4M_A^4}-1}\right)\bigg]\nn\\
 &+\frac{M_H^4M_{H^\pm}^2\cot^22\beta}{\pi^2\vphys^2}\bigg[-1-f_B\left(\frac{M_H^2}{2M_{H^\pm}^2}-\sqrt{\frac{M_H^4}{4M_{H^\pm}^4}-1}\right)-f_B\left(\frac{M_H^2}{2M_{H^\pm}^2}+\sqrt{\frac{M_H^4}{4M_{H^\pm}^4}-1}\right)\bigg]\nn\\
 &+\frac{3M_H^2M_A^4\cot^22\beta}{8\pi^2\vphys^2}+\frac{M_H^2M_{H^\pm}^4\cot^22\beta}{\pi^2\vphys^2}+\frac{M_H^2M_A^2M_{H^\pm}^2\cot^22\beta}{2\pi^2\vphys^2}\,\nn\\
 \aleph_t\equiv&\ \frac{3M_t^6\cot^2\beta}{2\pi^2\vphys^2}\bigg[\log\frac{M_{H^\pm}^2}{M_t^2}+\frac{13}{2}-\left(\frac{M_{H^\pm}^2}{M_t^2}-1\right)\log\left(1-\frac{M_t^2}{M_{H^\pm}^2}\right)\nn\\
 &\hspace{2cm}+4\left(f_B\left(\frac{M_{H^\pm}^2}{2M_t^2}-\sqrt{\frac{M_{H^\pm}^4}{4M_t^4}-1}\right)+f_B\left(\frac{M_{H^\pm}^2}{2M_t^2}+\sqrt{\frac{M_{H^\pm}^4}{4M_t^4}-1}\right)\right)\bigg]\nn\\
 &+\frac{3M_H^2M_t^2\cot^2\beta}{4\pi^2\vphys^2}\bigg[M_t^2-\frac32M_H^2+\left(4M_t^2-M_H^2\right)\left(f_B\left(\frac12-\sqrt{\frac14-\frac{M_t^2}{M_H^2}}\right)+f_B\left(\frac12+\sqrt{\frac14-\frac{M_t^2}{M_H^2}}\right)\right)\nn\\
 &\hspace{2.5cm}-2M_t^2\left(f_B\left(\frac{M_H^2}{2M_t^2}-\sqrt{\frac{M_H^4}{4M_t^4}-1}\right)+f_B\left(\frac{M_H^2}{2M_t^2}+\sqrt{\frac{M_H^4}{4M_t^4}-1}\right)\right)\bigg]\nn\\
 &+\frac{3M_A^2M_t^2\cot^2\beta}{4\pi^2\vphys^2}\bigg[-5M_t^2-\frac32M_A^2-M_A^2\left(f_B\left(\frac12-\sqrt{\frac14-\frac{M_t^2}{M_A^2}}\right)+f_B\left(\frac12+\sqrt{\frac14-\frac{M_t^2}{M_A^2}}\right)\right)\nn\\
 &\hspace{3cm}-2M_t^2\left(f_B\left(\frac{M_A^2}{2M_t^2}-\sqrt{\frac{M_A^4}{4M_t^4}-1}\right)+f_B\left(\frac{M_A^2}{2M_t^2}+\sqrt{\frac{M_A^4}{4M_t^4}-1}\right)\right)\bigg]\nn\\
 &+\frac{3M_{H^\pm}^2M_t^2\cot^2\beta}{2\pi^2\vphys^2}\bigg[(2M_t^2-M_{H^\pm}^2)\log\frac{M_t^2}{M_{H^\pm}^2}-M_t^2+\frac12M_{H^\pm}^2\nn\\
 &\hspace{3cm}+(M_{H^\pm}^2-M_t^2)\left(\frac{M_t^2}{M_{H^\pm}^2}-1\right)\log\left(1-\frac{M_{H^\pm}^2}{M_t^2}\right)+(M_{H^\pm}^2-M_t^2)\log\left(1-\frac{M_t^2}{M_{H^\pm}^2}\right)\bigg]\,.
\end{align}

\section{Results for generic theories}
\label{APP:generic}
In this last appendix, we provide generic expressions for the coefficients $\btl$ and $\ctl$ -- $c.f.$ eq.~(\ref{EQ:form_Veff_1+2l}) -- which we extract from the formulae given in Ref.~\cite{Martin:2001vx} for the two-loop effective potential in general renormalisable theories (note that we employ the \msbar expressions). The expressions presented here can in principle be applied in order to study corrections to the Higgs trilinear coupling as well as the relation between masses in any CSI theory where the ``scalon'' direction does not mix with other states. 

The couplings are defined among real scalars $\phi_i$, Dirac fermions $\psi_I$, and gauge bosons $A^a_\mu$ (and also ghosts $\omega^a$) as follows
\begin{align}
\label{EQ:genericcoup_def}
 \lagr_S=&-\frac16\lambda_{ijk}\phi_i\phi_j\phi_k-\frac1{24}\lambda_{ijkl}\phi_i\phi_j\phi_k\phi_l\,,\nn\\
 \lagr_{FS}=&-y_{IJk}\bar\psi_I\mathcal{P}_L\psi_J\phi_k-(y_{IJk})^*\bar\psi_J\pr\psi_I\phi_k\,,\nn\\
 \lagr_{VS}=&\ \frac12g^{abi}A_\mu^aA^{\mu\ b}\phi_i+\frac14g^{abij}A_\mu^aA^{\mu\ b}\phi_i\phi_j+g^{aij}A^a_\mu\phi_i\partial^\mu\phi_j\,,\nn\\
 \lagr_{FV}=&\ g^{IJa}\bar\psi_I\slashed{A}^a\psi_J\,,\nn\\
 \lagr_\text{gauge}=&\ g^{abc}A^a_\mu A^b_\nu\partial^\mu A^{c\nu}-\frac14g^{abe}g^{cde}A^{a\mu}A^{b\nu}A^c_\mu A^d_\nu+g^{abc}A^a_\mu\omega^b\partial^\mu \overline{\omega}^c\,.
\end{align}
Note that we work in the Landau gauge, and hence we do not need to care about couplings between scalars and ghosts. 

Our notations for the different contributions to the effective potential correspond to those in Ref.~\cite{Martin:2001vx}, and we refer the reader to equations (4.2) to (4.21) in this reference for all definitions. We simply list in the following the results we obtain (in the \msbar scheme) for the coefficients $\btl$ and $\ctl$ for each type of contribution. Note that we do not give results for $\atl$ as we do not require at any point in our analysis, and also because they can be obtained straightforwardly by the replacement $Q\to v$ in the expressions of Ref.~\cite{Martin:2001vx}.

\subsubsection{Scalar eight-shaped diagrams $\vtwo_{SS}$}
\begin{align}
 \btl_{SS}=&\ \frac18\lambda_{iijj} \frac{m_i^2m_j^2}{v^4}\left[\log\frac{m_i^2}{v^2}+\log\frac{m_j^2}{v^2}-2\right]\,,\nn\\
 \ctl_{SS}=&\ \frac18\lambda_{iijj} \frac{m_i^2m_j^2}{v^4}\,.
\end{align}

\subsubsection{Scalar sunrise diagrams $\vtwo_{SSS}$}
\begin{align}
 \btl_{SSS}=& -\frac1{12}\left(\frac{\lambda_{ijk}}{v^2}\right)^2\bigg\{m_i^2\bigg[2-\log\frac{m_i^2}{v^2}\bigg]+m_j^2\bigg[2-\log\frac{m_j^2}{v^2}\bigg]+m_k^2\bigg[2-\log\frac{m_k^2}{v^2}\bigg]\bigg\}\,,\nn\\
 \ctl_{SSS}=&\ \frac1{24}\left(\frac{\lambda_{ijk}}{v^2}\right)^2\big[m_i^2+m_j^2+m_k^2\big]\,.
\end{align}

\subsubsection{Scalar-fermion sunrise diagrams $\vtwo_{F\bar FS}\equiv\vtwo_{FFS}+\vtwo_{\bar{F}\bar{F}S}$}
\begin{align}
 \btl_{F\bar FS}=&\ \frac{1}{2v^4}\bigg(\big[y_{IJk}y_{JIk}+\hc\big]m_Im_J+|y_{IJk}|^2(m_I^2+m_J^2-m_k^2)\bigg)\nn\\
 &\hspace{1cm}\times\bigg\{m_I^2\left(2-\log\frac{m_I^2}{v^2}\right)+m_J^2\left(2-\log\frac{m_J^2}{v^2}\right)+m_k^2\left(2-\log\frac{m_k^2}{v^2}\right)\bigg\}\nn\\
 &+\frac{|y_{IJk}|^2}{2v^4}\bigg\{m_I^2m_J^2\left(\log\frac{m_I^2}{v^2}+\log\frac{m_J^2}{v^2}-2\right)-m_I^2m_k^2\left(\log\frac{m_I^2}{v^2}+\log\frac{m_k^2}{v^2}-2\right)\nn\\
 &\hspace{1.75cm}-m_J^2m_k^2\left(\log\frac{m_J^2}{v^2}+\log\frac{m_k^2}{v^2}-2\right)\bigg\}\,,\nn\\
 \ctl_{F\bar FS}=&-\frac{1}{4v^4}\bigg(\big[y_{IJk}y_{JIk}+\hc\big]m_Im_J+|y_{IJk}|^2(m_I^2+m_J^2-m_k^2)\bigg)\big[m_I^2+m_J^2+m_k^2\big]\nn\\
 &+\frac{|y_{IJk}|^2}{2v^4}\bigg\{m_I^2m_J^2-m_I^2m_k^2-m_J^2m_k^2\bigg\}\,.
\end{align}

\subsubsection{Scalar-gauge-boson eight-shaped diagrams $\vtwo_{VS}$}
\begin{align}
 \btl_{VS}=&\ \frac14g^{aaii}\frac{m_a^2m_i^2}{v^4}\left[3\log\frac{m_a^2}{v^2}+3\log\frac{m_i^2}{v^2}-4\right]\,,\nn\\
 \ctl_{VS}=&\ \frac34g^{aaii}\frac{m_a^2m_i^2}{v^4}\,.
\end{align}

\subsubsection{Sunrise diagrams involving two scalars and a gauge boson $\vtwo_{SSV}$}
\begin{align}
 \btl_{SSV}=\ \frac{1}{4}\left(\frac{g^{aij}}{v^2}\right)^2\Bigg\{&-\frac{\Delta(m_i^2,m_j^2,m_a^2)}{m_a^2}\bigg[m_i^2\bigg(2-\log\frac{m_i^2}{v^2}\bigg)+m_j^2\bigg(2-\log\frac{m_j^2}{v^2}\bigg)+m_a^2\bigg(2-\log\frac{m_a^2}{v^2}\bigg)\bigg]\nn\\
 &+\frac{(m_i^2-m_j^2)^2}{m_a^2}\bigg[m_i^2\bigg(2-\log\frac{m_i^2}{v^2}\bigg)+m_j^2\bigg(2-\log\frac{m_j^2}{v^2}\bigg)\bigg]\nn\\
 &+m_i^2(m_j^2-m_i^2-m_a^2)\bigg[\log\frac{m_i^2}{v^2}+\log\frac{m_a^2}{v^2}-2\bigg]\nn\\
 &+m_j^2(m_i^2-m_j^2-m_a^2)\bigg[\log\frac{m_j^2}{v^2}+\log\frac{m_a^2}{v^2}-2\bigg]\nn\\
 &+m_i^2m_j^2\bigg(\log\frac{m_i^2}{v^2}+\log\frac{m_j^2}{v^2}-2\bigg)+2m_a^2\bigg(m_i^2+m_j^2-\frac13m_a^2\bigg)\Bigg\}\,,\nn\\
 \ctl_{SSV}=\ \frac{1}{4}\left(\frac{g^{aij}}{v^2}\right)^2\bigg[&\frac{\Delta(m_i^2,m_j^2,m_a^2)}{2m_a^2}(m_i^2+m_j^2+m_a^2)-\frac{(m_i^2-m_j^2)^2}{2m_a^2}(m_i^2+m_j^2)\nn\\
 &+m_i^2(m_j^2-m_i^2-m_a^2)+m_j^2(m_i^2-m_j^2-m_a^2)+m_i^2m_j^2\bigg]\,.
\end{align}

\subsubsection{Sunrise diagrams involving one scalar and two gauge bosons $\vtwo_{VVS}$}
\begin{align}
 \btl_{VVS}=&\left(\frac{g^{abi}}{4v^2m_am_b}\right)^2\Bigg\{\big[-m_a^4-m_b^4-m_i^4-10m_a^2m_b^2+2(m_a^2+m_b^2)m_i^2\big]\nn\\
 &\quad\times\bigg( \frac12\big[(m_a^2-m_b^2-m_i^2)\left(\log \frac{m_b^2}{v^2} +\log \frac{m_i^2}{v^2}\right) + (m_b^2-m_i^2-m_a^2) \left(\log \frac{m_a^2}{v^2}+\log \frac{m_i^2}{v^2}\right)\nn\\
 &\qquad + (m_i^2-m_a^2-m_b^2)\left( \log \frac{m_a^2}{v^2} +\log \frac{m_b^2}{v^2}\right)\big]+2\big[m_a^2+m_b^2 + m_i^2\big]\bigg)\nn\\
          &\hspace{3cm}+(m_a^2-m_i^2)^2\bigg[m_a^2\left(2-\log\frac{m_a^2}{v^2}\right)+m_i^2\left(2-\log \frac{m_i^2}{v^2}\right)\bigg]\nn\\
          &\hspace{3cm}+(m_b^2-m_i^2)^2\bigg[m_b^2\left(2-\log\frac{m_b^2}{v^2}\right)+m_i^2\left(2-\log \frac{m_i^2}{v^2}\right)\bigg]\nn\\
          &\hspace{3cm}-m_i^6\bigg(2-\log \frac{m_i^2}{v^2}\bigg)\nn\\
          &\hspace{3cm}+(m_i^2-m_a^2-m_b^2)m_a^2m_b^2\left(\log\frac{m_a^2}{v^2}+\log\frac{m_b^2}{v^2}-2\right)\nn\\
          &\hspace{3cm}+m_a^2m_b^2m_i^2\left(\log\frac{m_a^2}{v^2}+\log\frac{m_i^2}{v^2}-2\right)+m_a^2m_b^2m_i^2\left(\log\frac{m_b^2}{v^2}+\log\frac{m_i^2}{v^2}-2\right)\Bigg\}\nn\\
          &+\left(\frac{g^{abi}}{2v^2}\right)^2\Bigg\{\frac12m_a^2+\frac12m_b^2+2m_i^2\Bigg\}\,,\nn\\
 \ctl_{VVS}=&\left(\frac{g^{abi}}{4v^2m_am_b}\right)^2\Bigg\{\frac12\big[m_a^4+m_b^4+m_i^4+10m_a^2m_b^2-2(m_a^2+m_b^2)m_i^2\big] \big[m_a^2+m_b^2+m_i^2 \big]\nn\\
          &\hspace{3cm}-\frac12(m_a^2-m_i^2)^2(m_a^2+m_i^2)-\frac12(m_b^2-m_i^2)^2(m_b^2+m_i^2)-\frac12m_i^6\nn\\
          &\hspace{3cm}+(3m_i^2-m_a^2-m_b^2)m_a^2m_b^2\Bigg\}\,.
\end{align}

\subsubsection{Fermion-gauge boson diagrams $\vtwo_{FFV}$ and $\vtwo_{\bar{F}\bar{F}V}$}
\begin{align}
 \btl_{FFV}=&\ \frac12\frac{|g^{IJa}|^2}{m_a^2v^4}\Bigg\{\Big(m_I^4+m_J^4-2m_a^4-2m_I^2m_J^2+m_I^2m_a^2+m_J^2m_a^2\Big)\nn\\
 &\hspace{0.75cm}\times \bigg[m_I^2\left(2-\log \frac{m_I^2}{v^2}\right)+m_J^2\left(2-\log \frac{m_J^2}{v^2}\right)+m_a^2\left(2-\log \frac{m_a^2}{v^2}\right)\bigg]\nn\\
          &\hspace{1.5cm}-(m_I^2-m_J^2)^2\bigg[m_I^2\left(2-\log \frac{m_I^2}{v^2}\right)+m_J^2\left(2-\log \frac{m_J^2}{v^2}\right)\bigg]\nn\\
          &\hspace{1.5cm}+(m_I^2-m_J^2-2m_a^2)m_I^2m_a^2\left(\log\frac{m_I^2}{v^2}+\log\frac{m_a^2}{v^2}-2\right)\nn\\
          &\hspace{1.5cm}+(m_J^2-m_I^2-2m_a^2)m_J^2m_a^2\left(\log\frac{m_J^2}{v^2}+\log\frac{m_a^2}{v^2}-2\right)\nn\\
          &\hspace{1.5cm}+2m_a^2m_I^2m_J^2\left(\log\frac{m_I^2}{v^2}+\log\frac{m_J^2}{v^2}-2\right)\Bigg\}\nn\\
          &-2\left(m_I^2+m_J^2-\frac13m_a^2\right)m_a^2-2m_I^4-2m_J^4\,,\nn\\
 \ctl_{FFV}=&\ \frac12\frac{|g^{IJa}|^2}{m_a^2v^4}\Bigg\{\frac12\Big(m_I^4+m_J^4-2m_a^4-2m_I^2m_J^2+m_I^2m_a^2+m_J^2m_a^2\Big)(m_I^2+m_J^2+m_a^2)\nn\\
          &\hspace{1.5cm}+\frac{1}{2}(m_I^2-m_J^2)^2(m_I^2+m_J^2)+(m_I^2-m_J^2-2m_a^2)m_I^2m_a^2\nn\\
          &\hspace{1.5cm}+(m_J^2-m_I^2-2m_a^2)m_J^2m_a^2+2m_a^2m_I^2m_J^2\Bigg\}\,.
\end{align}

\begin{align}
 \btl_{\bar{F}\bar{F}V}=&\ \frac12\frac{(g^{IJa})^2m_Im_J}{v^4}\Bigg\{3(m_I^2-m_J^2-m_a^2)\left(\log \frac{m_J^2}{v^2}+ \log \frac{m_a^2}{v^2}\right) \nn\\
 &\hspace{2.75cm}+ 3(m_J^2-m_a^2-m_I^2) \left(\log \frac{m_I^2}{v^2}+\log \frac{m_a^2}{v^2}\right) \nn\\
 &\hspace{2.75cm}+ 3(m_a^2-m_I^2-m_J^2) \left(\log \frac{m_I^2}{v^2} +\log \frac{m_J^2}{v^2}\right)+8m_I^2+8m_J^2 +12 m_a^2\Bigg\}\,,\nn\\
 \ctl_{\bar{F}\bar{F}V}=&\ -\frac32\frac{(g^{IJa})^2}{v^4}m_Im_J(m_I^2+m_J^2+m_a^2)\,.
\end{align}

\subsubsection{Pure gauge diagrams $\vtwo_\text{gauge}$}
\begin{align}
 \btl_\text{gauge}=\frac{(g^{abc})^2}{48m_a^2m_b^2m_c^2v^4}\Bigg\{&\Big(-m_a^8-8m_a^6m_b^2-8m_a^6m_c^2+32m_a^4m_b^2m_c^2+18m_b^4m_c^4\Big)\nn\\
 &\times\bigg[m_a^2\bigg(2-\log\frac{m_a^2}{v^2}\bigg)+m_b^2\bigg(2-\log\frac{m_b^2}{v^2}\bigg)+m_c^2\bigg(2-\log \frac{m_c^2}{v^2}\bigg)\bigg]\nn\\
 &+(m_b^2-m_c^2)^2(m_b^4+10m_b^2m_c^2+m_c^4)\nn\\
 &\hspace{3cm}\times\bigg[m_b^2\bigg(2-\log\frac{m_b^2}{v^2}\bigg)+m_c^2\bigg(2-\log \frac{m_c^2}{v^2}\bigg)\bigg]\nn\\
 &+m_a^6(2m_b^2m_c^2-m_a^4)\bigg(2-\log \frac{m_a^2}{v^2}\bigg) \nn\\
 &+\Big(m_a^4-9m_b^4-9m_c^4+9m_a^2m_b^2+9m_a^2m_c^2+14m_b^2m_c^2\Big)\nn\\
 &\hspace{3cm}\times m_a^2m_b^2m_c^2 \left(\log\frac{m_b^2}{v^2}+\log\frac{m_c^2}{v^2}-2\right)\nn\\
 &+\left(22m_b^2+22m_c^2-\frac{16}{3}m_a^2\right)m_a^2m_b^2m_c^2 \Bigg\}\nn\\
 &\hspace{-1cm}+(a\leftrightarrow b)+(a\leftrightarrow c)\,,\nn\\
 \ctl_\text{gauge}=-\frac{(g^{abc})^2}{96m_a^2m_b^2m_c^2v^4}\Bigg\{&\Big(-m_a^8-8m_a^6m_b^2-8m_a^6m_c^2+32m_a^4m_b^2m_c^2+18m_b^4m_c^4\Big)(m_a^2+m_b^2+m_c^2)\nn\\
 &+(m_b^2-m_c^2)^2(m_b^4+10m_b^2m_c^2+m_c^4)(m_b^2+m_c^2)+m_a^6(2m_b^2m_c^2-m_a^4) \nn\\
 &-2\Big(m_a^4-9m_b^4-9m_c^4+9m_a^2m_b^2+9m_a^2m_c^2+14m_b^2m_c^2\Big)m_a^2m_b^2m_c^2 \Bigg\}\nn\\
 &\hspace{-1cm}+(a\leftrightarrow b)+(a\leftrightarrow c)\,.
\end{align}
Here $(a\leftrightarrow b)$ and $(a\leftrightarrow c)$ denote the exchange of indices $a$ and $b$, or $a$ and $c$.

\bibliographystyle{utphys}
\bibliography{CSI}

\providecommand{\href}[2]{#2}\begingroup\raggedright\begin{thebibliography}{100}

\bibitem{Chatrchyan:2012xdj}
{\bf CMS} Collaboration, S.~Chatrchyan {\em et al.}, {\em {Observation of a new
  boson at a mass of 125 GeV with the CMS experiment at the LHC}}.
  \href{http://dx.doi.org/10.1016/j.physletb.2012.08.021}{Phys. Lett. {\bf
  B716} (2012)  30--61},
\href{http://arxiv.org/abs/1207.7235}{{\tt arXiv:1207.7235 [hep-ex]}}.

\bibitem{Aad:2012tfa}
{\bf ATLAS} Collaboration, G.~Aad {\em et al.}, {\em {Observation of a new
  particle in the search for the Standard Model Higgs boson with the ATLAS
  detector at the LHC}}.
  \href{http://dx.doi.org/10.1016/j.physletb.2012.08.020}{Phys. Lett. {\bf
  B716} (2012)  1--29},
\href{http://arxiv.org/abs/1207.7214}{{\tt arXiv:1207.7214 [hep-ex]}}.

\bibitem{Aad:2019mbh}
{\bf ATLAS} Collaboration, G.~Aad {\em et al.}, {\em {Combined measurements of
  Higgs boson production and decay using up to $80$ fb$^{-1}$ of proton-proton
  collision data at $\sqrt{s}=$ 13 TeV collected with the ATLAS experiment}}.
  \href{http://dx.doi.org/10.1103/PhysRevD.101.012002}{Phys. Rev. D {\bf 101}
  (2020) no.~1, 012002}, \href{http://arxiv.org/abs/1909.02845}{{\tt
  arXiv:1909.02845 [hep-ex]}}.

\bibitem{Sirunyan:2018koj}
{\bf CMS} Collaboration, A.~M. Sirunyan {\em et al.}, {\em {Combined
  measurements of Higgs boson couplings in proton\textendash{}proton collisions
  at $\sqrt{s}=13\,\text {Te}\text {V} $}}.
  \href{http://dx.doi.org/10.1140/epjc/s10052-019-6909-y}{Eur. Phys. J. C {\bf
  79} (2019) no.~5, 421}, \href{http://arxiv.org/abs/1809.10733}{{\tt
  arXiv:1809.10733 [hep-ex]}}.

\bibitem{Appelquist:1974tg}
T.~Appelquist and J.~Carazzone, {\em {Infrared Singularities and Massive
  Fields}}.
\href{http://dx.doi.org/10.1103/PhysRevD.11.2856}{Phys. Rev. {\bf D11} (1975)
  2856}.

\bibitem{Gunion:2002zf}
J.~F. Gunion and H.~E. Haber, {\em {The CP conserving two Higgs doublet model:
  The Approach to the decoupling limit}}.
  \href{http://dx.doi.org/10.1103/PhysRevD.67.075019}{Phys. Rev. {\bf D67}
  (2003)  075019},
\href{http://arxiv.org/abs/hep-ph/0207010}{{\tt arXiv:hep-ph/0207010
  [hep-ph]}}.

\bibitem{Dev:2014yca}
P.~S. Bhupal~Dev and A.~Pilaftsis, {\em {Maximally Symmetric Two Higgs Doublet
  Model with Natural Standard Model Alignment}}.
  \href{http://dx.doi.org/10.1007/JHEP11(2015)147,
  10.1007/JHEP12(2014)024}{JHEP {\bf 12} (2014)  024},
  \href{http://arxiv.org/abs/1408.3405}{{\tt arXiv:1408.3405 [hep-ph]}}.
[Erratum: JHEP11,147(2015)].

\bibitem{Deshpande:1977rw}
N.~G. Deshpande and E.~Ma, {\em {Pattern of Symmetry Breaking with Two Higgs
  Doublets}}. \href{http://dx.doi.org/10.1103/PhysRevD.18.2574}{Phys. Rev. D
  {\bf 18} (1978)  2574}.

\bibitem{Silveira:1985rk}
V.~Silveira and A.~Zee, {\em {SCALAR PHANTOMS}}.
\href{http://dx.doi.org/10.1016/0370-2693(85)90624-0}{Phys. Lett. {\bf 161B}
  (1985)  136--140}.

\bibitem{Barbieri:2006dq}
R.~Barbieri, L.~J. Hall, and V.~S. Rychkov, {\em {Improved naturalness with a
  heavy Higgs: An Alternative road to LHC physics}}.
  \href{http://dx.doi.org/10.1103/PhysRevD.74.015007}{Phys. Rev. {\bf D74}
  (2006)  015007},
\href{http://arxiv.org/abs/hep-ph/0603188}{{\tt arXiv:hep-ph/0603188
  [hep-ph]}}.

\bibitem{Benakli:2018vqz}
K.~Benakli, M.~D. Goodsell, and S.~L. Williamson, {\em {Higgs alignment from
  extended supersymmetry}}.
  \href{http://dx.doi.org/10.1140/epjc/s10052-018-6125-1}{Eur. Phys. J. {\bf
  C78} (2018) no.~8, 658},
\href{http://arxiv.org/abs/1801.08849}{{\tt arXiv:1801.08849 [hep-ph]}}.

\bibitem{Benakli:2018vjk}
K.~Benakli, Y.~Chen, and G.~Lafforgue-Marmet, {\em {R-symmetry for Higgs
  alignment without decoupling}}.
  \href{http://dx.doi.org/10.1140/epjc/s10052-019-6676-9}{Eur. Phys. J. {\bf
  C79} (2019) no.~2, 172},
\href{http://arxiv.org/abs/1811.08435}{{\tt arXiv:1811.08435 [hep-ph]}}.

\bibitem{Coyle:2019exn}
N.~M. Coyle and C.~E. Wagner, {\em {Dynamical Higgs field alignment in the
  NMSSM}}. \href{http://dx.doi.org/10.1103/PhysRevD.101.055037}{Phys. Rev. D
  {\bf 101} (2020) no.~5, 055037}, \href{http://arxiv.org/abs/1912.01036}{{\tt
  arXiv:1912.01036 [hep-ph]}}.

\bibitem{Aiko:2020atr}
M.~Aiko and S.~Kanemura, {\em {New scenario for aligned Higgs couplings
  originated from the twisted custodial symmetry at high energies}}.
  \href{http://arxiv.org/abs/2009.04330}{{\tt arXiv:2009.04330 [hep-ph]}}.


\bibitem{Aiko:2020ksl}
M.~Aiko, S.~Kanemura, M.~Kikuchi, K.~Mawatari, K.~Sakurai and K.~Yagyu, {\em {Probing extended Higgs sectors by the synergy between direct searches at the LHC and precision tests at future lepton colliders}}.
 \href{http://arxiv.org/abs/2010.15057}{{\tt arXiv:2010.15057 [hep-ph]}}.

\bibitem{Veltman:1976rt}
M.~Veltman, {\em {Second Threshold in Weak Interactions}}. Acta Phys. Polon. B
  {\bf 8} (1977)  475.

\bibitem{Fayet:1977yc}
P.~Fayet, {\em {Spontaneously Broken Supersymmetric Theories of Weak,
  Electromagnetic and Strong Interactions}}.
  \href{http://dx.doi.org/10.1016/0370-2693(77)90852-8}{Phys. Lett. B {\bf 69}
  (1977)  489}.

\bibitem{Kaplan:1983fs}
D.~B. Kaplan and H.~Georgi, {\em {SU(2) x U(1) Breaking by Vacuum
  Misalignment}}. \href{http://dx.doi.org/10.1016/0370-2693(84)91177-8}{Phys.
  Lett. B {\bf 136} (1984)  183--186}.

\bibitem{Kaplan:1983sm}
D.~B. Kaplan, H.~Georgi, and S.~Dimopoulos, {\em {Composite Higgs Scalars}}.
  \href{http://dx.doi.org/10.1016/0370-2693(84)91178-X}{Phys. Lett. B {\bf 136}
  (1984)  187--190}.

\bibitem{ATLAS:2020uqv}
{\bf ATLAS} Collaboration, {\em {Summary Plots for Heavy Particle Searches and
  Long-lived Particle Searches}}.

\bibitem{ATLAS:2020ujn}
{\bf ATLAS} Collaboration, {\em {SUSY July 2020 Summary Plot Update}}.

\bibitem{Bardeen:1995kv}
W.~A. Bardeen, ``{On naturalness in the standard model},'' in {\em {Ontake
  Summer Institute on Particle Physics}}.
\newblock 8, 1995.

\bibitem{Coleman:1973jx}
S.~R. Coleman and E.~J. Weinberg, {\em {Radiative Corrections as the Origin of
  Spontaneous Symmetry Breaking}}.
  \href{http://dx.doi.org/10.1103/PhysRevD.7.1888}{Phys. Rev. D {\bf 7} (1973)
  1888--1910}.

\bibitem{Gildener:1976ih}
E.~Gildener and S.~Weinberg, {\em {Symmetry Breaking and Scalar Bosons}}.
  \href{http://dx.doi.org/10.1103/PhysRevD.13.3333}{Phys. Rev. D {\bf 13}
  (1976)  3333}.

\bibitem{Foot:2007as}
R.~Foot, A.~Kobakhidze, and R.~R. Volkas, {\em {Electroweak Higgs as a
  pseudo-Goldstone boson of broken scale invariance}}.
  \href{http://dx.doi.org/10.1016/j.physletb.2007.06.084}{Phys. Lett. B {\bf
  655} (2007)  156--161}, \href{http://arxiv.org/abs/0704.1165}{{\tt
  arXiv:0704.1165 [hep-ph]}}.

\bibitem{Foot:2007iy}
R.~Foot, A.~Kobakhidze, K.~L. McDonald, and R.~R. Volkas, {\em {A Solution to
  the hierarchy problem from an almost decoupled hidden sector within a
  classically scale invariant theory}}.
  \href{http://dx.doi.org/10.1103/PhysRevD.77.035006}{Phys. Rev. D {\bf 77}
  (2008)  035006}, \href{http://arxiv.org/abs/0709.2750}{{\tt arXiv:0709.2750
  [hep-ph]}}.

\bibitem{Meissner:2007xv}
K.~A. Meissner and H.~Nicolai, {\em {Effective action, conformal anomaly and
  the issue of quadratic divergences}}.
  \href{http://dx.doi.org/10.1016/j.physletb.2007.12.035}{Phys. Lett. B {\bf
  660} (2008)  260--266}, \href{http://arxiv.org/abs/0710.2840}{{\tt
  arXiv:0710.2840 [hep-th]}}.

\bibitem{Iso:2009ss}
S.~Iso, N.~Okada, and Y.~Orikasa, {\em {Classically conformal $B^-$ L extended
  Standard Model}}.
  \href{http://dx.doi.org/10.1016/j.physletb.2009.04.046}{Phys. Lett. B {\bf
  676} (2009)  81--87}, \href{http://arxiv.org/abs/0902.4050}{{\tt
  arXiv:0902.4050 [hep-ph]}}.

\bibitem{Iso:2012jn}
S.~Iso and Y.~Orikasa, {\em {TeV Scale B-L model with a flat Higgs potential at
  the Planck scale: In view of the hierarchy problem}}.
  \href{http://dx.doi.org/10.1093/ptep/pts099}{PTEP {\bf 2013} (2013)  023B08},
  \href{http://arxiv.org/abs/1210.2848}{{\tt arXiv:1210.2848 [hep-ph]}}.

\bibitem{Lane:2019dbc}
K.~Lane and E.~Pilon, {\em {Phenomenology of the new light Higgs bosons in
  Gildener-Weinberg models}}.
  \href{http://dx.doi.org/10.1103/PhysRevD.101.055032}{Phys. Rev. D {\bf 101}
  (2020) no.~5, 055032}, \href{http://arxiv.org/abs/1909.02111}{{\tt
  arXiv:1909.02111 [hep-ph]}}.

\bibitem{Sakharov:1967dj}
A.~D. Sakharov, {\em {Violation of CP Invariance, C asymmetry, and baryon
  asymmetry of the universe}}.
  \href{http://dx.doi.org/10.1070/PU1991v034n05ABEH002497}{Pisma Zh. Eksp.
  Teor. Fiz. {\bf 5} (1967)  32--35}.
[Usp. Fiz. Nauk161,no.5,61(1991)].

\bibitem{Kuzmin:1985mm}
V.~A. Kuzmin, V.~A. Rubakov, and M.~E. Shaposhnikov, {\em {On the Anomalous
  Electroweak Baryon Number Nonconservation in the Early Universe}}.
\href{http://dx.doi.org/10.1016/0370-2693(85)91028-7}{Phys. Lett. {\bf 155B}
  (1985)  36}.

\bibitem{Cohen:1993nk}
A.~G. Cohen, D.~B. Kaplan, and A.~E. Nelson, {\em {Progress in electroweak
  baryogenesis}}.
  \href{http://dx.doi.org/10.1146/annurev.ns.43.120193.000331}{Ann. Rev. Nucl.
  Part. Sci. {\bf 43} (1993)  27--70},
\href{http://arxiv.org/abs/hep-ph/9302210}{{\tt arXiv:hep-ph/9302210
  [hep-ph]}}.

\bibitem{Grojean:2004xa}
C.~Grojean, G.~Servant, and J.~D. Wells, {\em {First-order electroweak phase
  transition in the standard model with a low cutoff}}.
  \href{http://dx.doi.org/10.1103/PhysRevD.71.036001}{Phys. Rev. {\bf D71}
  (2005)  036001},
\href{http://arxiv.org/abs/hep-ph/0407019}{{\tt arXiv:hep-ph/0407019
  [hep-ph]}}.

\bibitem{Kanemura:2004ch}
S.~Kanemura, Y.~Okada, and E.~Senaha, {\em {Electroweak baryogenesis and
  quantum corrections to the triple Higgs boson coupling}}.
  \href{http://dx.doi.org/10.1016/j.physletb.2004.12.004}{Phys. Lett. {\bf
  B606} (2005)  361--366},
\href{http://arxiv.org/abs/hep-ph/0411354}{{\tt arXiv:hep-ph/0411354
  [hep-ph]}}.

\bibitem{Grojean:2006bp}
C.~Grojean and G.~Servant, {\em {Gravitational Waves from Phase Transitions at
  the Electroweak Scale and Beyond}}.
  \href{http://dx.doi.org/10.1103/PhysRevD.75.043507}{Phys. Rev. D {\bf 75}
  (2007)  043507}, \href{http://arxiv.org/abs/hep-ph/0607107}{{\tt
  arXiv:hep-ph/0607107}}.

\bibitem{Kanemura:2002vm}
S.~Kanemura, S.~Kiyoura, Y.~Okada, E.~Senaha, and C.~P. Yuan, {\em {New physics
  effect on the Higgs selfcoupling}}.
  \href{http://dx.doi.org/10.1016/S0370-2693(03)00268-5}{Phys. Lett. {\bf B558}
  (2003)  157--164},
\href{http://arxiv.org/abs/hep-ph/0211308}{{\tt arXiv:hep-ph/0211308
  [hep-ph]}}.

\bibitem{Kanemura:2004mg}
S.~Kanemura, Y.~Okada, E.~Senaha, and C.-P. Yuan, {\em {Higgs coupling
  constants as a probe of new physics}}.
  \href{http://dx.doi.org/10.1103/PhysRevD.70.115002}{Phys. Rev. D {\bf 70}
  (2004)  115002}, \href{http://arxiv.org/abs/hep-ph/0408364}{{\tt
  arXiv:hep-ph/0408364}}.

\bibitem{Kanemura:2015fra}
S.~Kanemura, M.~Kikuchi, and K.~Yagyu, {\em {Radiative corrections to the Higgs
  boson couplings in the model with an additional real singlet scalar field}}.
  \href{http://dx.doi.org/10.1016/j.nuclphysb.2016.04.005}{Nucl. Phys. {\bf
  B907} (2016)  286--322},
\href{http://arxiv.org/abs/1511.06211}{{\tt arXiv:1511.06211 [hep-ph]}}.

\bibitem{Kanemura:2016lkz}
S.~Kanemura, M.~Kikuchi, and K.~Yagyu, {\em {One-loop corrections to the Higgs
  self-couplings in the singlet extension}}.
  \href{http://dx.doi.org/10.1016/j.nuclphysb.2017.02.004}{Nucl. Phys. {\bf
  B917} (2017)  154--177},
\href{http://arxiv.org/abs/1608.01582}{{\tt arXiv:1608.01582 [hep-ph]}}.

\bibitem{He:2016sqr}
S.-P. He and S.-h. Zhu, {\em {One-Loop Radiative Correction to the Triple Higgs
  Coupling in the Higgs Singlet Model}}.
  \href{http://dx.doi.org/10.1016/j.physletb.2016.11.007}{Phys. Lett. {\bf
  B764} (2017)  31--37},
\href{http://arxiv.org/abs/1607.04497}{{\tt arXiv:1607.04497 [hep-ph]}}.

\bibitem{Kanemura:2017wtm}
S.~Kanemura, M.~Kikuchi, K.~Sakurai, and K.~Yagyu, {\em {Gauge invariant
  one-loop corrections to Higgs boson couplings in non-minimal Higgs models}}.
  \href{http://dx.doi.org/10.1103/PhysRevD.96.035014}{Phys. Rev. {\bf D96}
  (2017) no.~3, 035014},
\href{http://arxiv.org/abs/1705.05399}{{\tt arXiv:1705.05399 [hep-ph]}}.

\bibitem{Kanemura:2015mxa}
S.~Kanemura, M.~Kikuchi, and K.~Yagyu, {\em {Fingerprinting the extended Higgs
  sector using one-loop corrected Higgs boson couplings and future precision
  measurements}}.
  \href{http://dx.doi.org/10.1016/j.nuclphysb.2015.04.015}{Nucl. Phys. {\bf
  B896} (2015)  80--137},
\href{http://arxiv.org/abs/1502.07716}{{\tt arXiv:1502.07716 [hep-ph]}}.

\bibitem{Arhrib:2015hoa}
A.~Arhrib, R.~Benbrik, J.~El~Falaki, and A.~Jueid, {\em {Radiative corrections
  to the Triple Higgs Coupling in the Inert Higgs Doublet Model}}.
  \href{http://dx.doi.org/10.1007/JHEP12(2015)007}{JHEP {\bf 12} (2015)  007},
\href{http://arxiv.org/abs/1507.03630}{{\tt arXiv:1507.03630 [hep-ph]}}.

\bibitem{Kanemura:2016sos}
S.~Kanemura, M.~Kikuchi, and K.~Sakurai, {\em {Testing the dark matter scenario
  in the inert doublet model by future precision measurements of the Higgs
  boson couplings}}. \href{http://dx.doi.org/10.1103/PhysRevD.94.115011}{Phys.
  Rev. D {\bf 94} (2016) no.~11, 115011},
  \href{http://arxiv.org/abs/1605.08520}{{\tt arXiv:1605.08520 [hep-ph]}}.

\bibitem{Aoki:2012jj}
M.~Aoki, S.~Kanemura, M.~Kikuchi, and K.~Yagyu, {\em {Radiative corrections to
  the Higgs boson couplings in the triplet model}}.
  \href{http://dx.doi.org/10.1103/PhysRevD.87.015012}{Phys. Rev. {\bf D87}
  (2013) no.~1, 015012},
\href{http://arxiv.org/abs/1211.6029}{{\tt arXiv:1211.6029 [hep-ph]}}.

\bibitem{Hill:2014mqa}
C.~T. Hill, {\em {Is the Higgs Boson Associated with Coleman-Weinberg Dynamical
  Symmetry Breaking?}}
  \href{http://dx.doi.org/10.1103/PhysRevD.89.073003}{Phys. Rev. D {\bf 89}
  (2014) no.~7, 073003}, \href{http://arxiv.org/abs/1401.4185}{{\tt
  arXiv:1401.4185 [hep-ph]}}.

\bibitem{Hashino:2015nxa}
K.~Hashino, S.~Kanemura, and Y.~Orikasa, {\em {Discriminative phenomenological
  features of scale invariant models for electroweak symmetry breaking}}.
  \href{http://dx.doi.org/10.1016/j.physletb.2015.11.044}{Phys. Lett. B {\bf
  752} (2016)  217--220}, \href{http://arxiv.org/abs/1508.03245}{{\tt
  arXiv:1508.03245 [hep-ph]}}.

\bibitem{Agrawal:2019bpm}
P.~Agrawal, D.~Saha, L.-X. Xu, J.-H. Yu, and C.~Yuan, {\em {Determining the
  shape of the Higgs potential at future colliders}}.
  \href{http://dx.doi.org/10.1103/PhysRevD.101.075023}{Phys. Rev. D {\bf 101}
  (2020) no.~7, 075023}, \href{http://arxiv.org/abs/1907.02078}{{\tt
  arXiv:1907.02078 [hep-ph]}}.

\bibitem{Hashino:2016rvx}
K.~Hashino, M.~Kakizaki, S.~Kanemura, and T.~Matsui, {\em {Synergy between
  measurements of gravitational waves and the triple-Higgs coupling in probing
  the first-order electroweak phase transition}}.
  \href{http://dx.doi.org/10.1103/PhysRevD.94.015005}{Phys. Rev. D {\bf 94}
  (2016) no.~1, 015005}, \href{http://arxiv.org/abs/1604.02069}{{\tt
  arXiv:1604.02069 [hep-ph]}}.

\bibitem{ATL-PHYS-PUB-2019-009}
{\bf ATLAS Collaboration} Collaboration, {\em {Constraint of the Higgs boson
  self-coupling from Higgs boson differential production and decay
  measurements}} Tech. Rep. ATL-PHYS-PUB-2019-009, CERN, Geneva, Mar, 2019.
\newblock \url{http://cds.cern.ch/record/2667570}.

\bibitem{Ferrari:2018akh}
{\bf ATLAS} Collaboration, A.~Ferrari, ``{Searches for Higgs boson pair
  production with ATLAS},'' in {\em {13th Conference on the Intersections of
  Particle and Nuclear Physics (CIPANP 2018) Palm Springs, California, USA, May
  29-June 3, 2018}}.
\newblock 2018.
\newblock
\href{http://arxiv.org/abs/1809.08870}{{\tt arXiv:1809.08870 [hep-ex]}}.
\newblock

\bibitem{Aaboud:2018ftw}
{\bf ATLAS} Collaboration, M.~Aaboud {\em et al.}, {\em {Search for Higgs boson
  pair production in the $\gamma\gamma b\bar{b}$ final state with 13 TeV $pp$
  collision data collected by the ATLAS experiment}}.
  \href{http://dx.doi.org/10.1007/JHEP11(2018)040}{JHEP {\bf 11} (2018)  040},
\href{http://arxiv.org/abs/1807.04873}{{\tt arXiv:1807.04873 [hep-ex]}}.

\bibitem{Sirunyan:2018iwt}
{\bf CMS} Collaboration, A.~M. Sirunyan {\em et al.}, {\em {Search for Higgs
  boson pair production in the $\gamma\gamma\mathrm{b\overline{b}}$ final state
  in pp collisions at $\sqrt{s}=$ 13 TeV}}.
  \href{http://dx.doi.org/10.1016/j.physletb.2018.10.056}{Phys. Lett. {\bf
  B788} (2019)  7--36},
\href{http://arxiv.org/abs/1806.00408}{{\tt arXiv:1806.00408 [hep-ex]}}.

\bibitem{Sirunyan:2018two}
{\bf CMS} Collaboration, A.~M. Sirunyan {\em et al.}, {\em {Combination of
  searches for Higgs boson pair production in proton-proton collisions at
  $\sqrt{s} = $ 13 TeV}}. Submitted to: Phys. Rev. Lett. (2018)  ,
\href{http://arxiv.org/abs/1811.09689}{{\tt arXiv:1811.09689 [hep-ex]}}.

\bibitem{deBlas:2019rxi}
J.~de~Blas {\em et al.}, {\em {Higgs Boson Studies at Future Particle
  Colliders}}.
\href{http://arxiv.org/abs/1905.03764}{{\tt arXiv:1905.03764 [hep-ph]}}.

\bibitem{Cepeda:2019klc}
{\bf Physics of the HL-LHC Working Group} Collaboration, M.~Cepeda {\em et
  al.}, {\em {Higgs Physics at the HL-LHC and HE-LHC}}.
\href{http://arxiv.org/abs/1902.00134}{{\tt arXiv:1902.00134 [hep-ph]}}.

\bibitem{Chang:2019ncg}
J.~Chang, K.~Cheung, J.~S. Lee, and J.~Park, {\em {Probing the trilinear Higgs
  boson self-coupling at the high-luminosity LHC via multivariate analysis}}.
  \href{http://dx.doi.org/10.1103/PhysRevD.101.016004}{Phys. Rev. D {\bf 101}
  (2020) no.~1, 016004}, \href{http://arxiv.org/abs/1908.00753}{{\tt
  arXiv:1908.00753 [hep-ph]}}.

\bibitem{Homiller:2018dgu}
S.~Homiller and P.~Meade, {\em {Measurement of the Triple Higgs Coupling at a
  HE-LHC}}.
\href{http://arxiv.org/abs/1811.02572}{{\tt arXiv:1811.02572 [hep-ph]}}.

\bibitem{Goncalves:2018yva}
D.~Gonçalves, T.~Han, F.~Kling, T.~Plehn, and M.~Takeuchi, {\em {Higgs boson
  pair production at future hadron colliders: From kinematics to dynamics}}.
  \href{http://dx.doi.org/10.1103/PhysRevD.97.113004}{Phys. Rev. {\bf D97}
  (2018) no.~11, 113004},
\href{http://arxiv.org/abs/1802.04319}{{\tt arXiv:1802.04319 [hep-ph]}}.

\bibitem{Fujii:2015jha}
K.~Fujii {\em et al.}, {\em {Physics Case for the International Linear
  Collider}}.
\href{http://arxiv.org/abs/1506.05992}{{\tt arXiv:1506.05992 [hep-ex]}}.

\bibitem{Abramowicz:2016zbo}
H.~Abramowicz {\em et al.}, {\em {Higgs physics at the CLIC electron–positron
  linear collider}}.
  \href{http://dx.doi.org/10.1140/epjc/s10052-017-4968-5}{Eur. Phys. J. {\bf
  C77} (2017) no.~7, 475},
\href{http://arxiv.org/abs/1608.07538}{{\tt arXiv:1608.07538 [hep-ex]}}.

\bibitem{Charles:2018vfv}
{\bf CLICdp, CLIC} Collaboration, T.~K. Charles {\em et al.}, {\em {The Compact
  Linear Collider (CLIC) - 2018 Summary Report}}.
  \href{http://dx.doi.org/10.23731/CYRM-2018-002}{CERN Yellow Rep. Monogr. {\bf
  1802} (2018)  1--98},
\href{http://arxiv.org/abs/1812.06018}{{\tt arXiv:1812.06018
  [physics.acc-ph]}}.

\bibitem{Roloff:2019crr}
{\bf CLICdp} Collaboration, P.~Roloff, U.~Schnoor, R.~Simoniello, and B.~Xu,
  {\em {Double Higgs boson production and Higgs self-coupling extraction at
  CLIC}}.
\href{http://arxiv.org/abs/1901.05897}{{\tt arXiv:1901.05897 [hep-ex]}}.

\bibitem{Chang:2018uwu}
J.~Chang, K.~Cheung, J.~S. Lee, C.-T. Lu, and J.~Park, {\em {Higgs-boson-pair
  production $H(\rightarrow b\overline{b})H(\rightarrow\gamma\gamma)$ from
  gluon fusion at the HL-LHC and HL-100 TeV hadron collider}}.
\href{http://arxiv.org/abs/1804.07130}{{\tt arXiv:1804.07130 [hep-ph]}}.

\bibitem{vanderBij:1983bw}
J.~van~der Bij and M.~Veltman, {\em {Two Loop Large Higgs Mass Correction to
  the rho Parameter}}.
  \href{http://dx.doi.org/10.1016/0550-3213(84)90284-0}{Nucl. Phys. B {\bf 231}
  (1984)  205--234}.

\bibitem{Ford:1992pn}
C.~Ford, I.~Jack, and D.~R.~T. Jones, {\em {The Standard model effective
  potential at two loops}}.
  \href{http://dx.doi.org/10.1016/0550-3213(92)90165-8,
  10.1016/S0550-3213(97)00532-4}{Nucl. Phys. {\bf B387} (1992)  373--390},
  \href{http://arxiv.org/abs/hep-ph/0111190}{{\tt arXiv:hep-ph/0111190
  [hep-ph]}}.
[Erratum: Nucl. Phys.B504,551(1997)].

\bibitem{Martin:2001vx}
S.~P. Martin, {\em {Two Loop Effective Potential for a General Renormalizable
  Theory and Softly Broken Supersymmetry}}.
  \href{http://dx.doi.org/10.1103/PhysRevD.65.116003}{Phys. Rev. D {\bf 65}
  (2002)  116003}, \href{http://arxiv.org/abs/hep-ph/0111209}{{\tt
  arXiv:hep-ph/0111209}}.

\bibitem{Brucherseifer:2013qva}
M.~Brucherseifer, R.~Gavin, and M.~Spira, {\em {Minimal supersymmetric Higgs
  boson self-couplings: Two-loop $O(\alpha_{t}\alpha_{s})$ corrections}}.
  \href{http://dx.doi.org/10.1103/PhysRevD.90.117701}{Phys. Rev. {\bf D90}
  (2014) no.~11, 117701},
\href{http://arxiv.org/abs/1309.3140}{{\tt arXiv:1309.3140 [hep-ph]}}.

\bibitem{Muhlleitner:2015dua}
M.~Mühlleitner, D.~T. Nhung, and H.~Ziesche, {\em {The order $
  \mathcal{O}\left({\alpha}_t{\alpha}_s\right) $ corrections to the trilinear
  Higgs self-couplings in the complex NMSSM}}.
  \href{http://dx.doi.org/10.1007/JHEP12(2015)034}{JHEP {\bf 12} (2015)  034},
\href{http://arxiv.org/abs/1506.03321}{{\tt arXiv:1506.03321 [hep-ph]}}.

\bibitem{Senaha:2018xek}
E.~Senaha, {\em {Radiative Corrections to Triple Higgs Coupling and Electroweak
  Phase Transition: Beyond One-loop Analysis}}.
  \href{http://dx.doi.org/10.1103/PhysRevD.100.055034}{Phys. Rev. D {\bf 100}
  (2019) no.~5, 055034}, \href{http://arxiv.org/abs/1811.00336}{{\tt
  arXiv:1811.00336 [hep-ph]}}.

\bibitem{Braathen:2019pxr}
J.~Braathen and S.~Kanemura, {\em {On two-loop corrections to the Higgs
  trilinear coupling in models with extended scalar sectors}}.
  \href{http://dx.doi.org/10.1016/j.physletb.2019.07.021}{Phys. Lett. B {\bf
  796} (2019)  38--46}, \href{http://arxiv.org/abs/1903.05417}{{\tt
  arXiv:1903.05417 [hep-ph]}}.

\bibitem{Braathen:2019zoh}
J.~Braathen and S.~Kanemura, {\em {Leading two-loop corrections to the Higgs
  boson self-couplings in models with extended scalar sectors}}.
  \href{http://dx.doi.org/10.1140/epjc/s10052-020-7723-2}{Eur. Phys. J. C {\bf
  80} (2020) no.~3, 227}, \href{http://arxiv.org/abs/1911.11507}{{\tt
  arXiv:1911.11507 [hep-ph]}}.

\bibitem{Lee:2012jn}
J.~S. Lee and A.~Pilaftsis, {\em {Radiative Corrections to Scalar Masses and
  Mixing in a Scale Invariant Two Higgs Doublet Model}}.
  \href{http://dx.doi.org/10.1103/PhysRevD.86.035004}{Phys. Rev. D {\bf 86}
  (2012)  035004}, \href{http://arxiv.org/abs/1201.4891}{{\tt arXiv:1201.4891
  [hep-ph]}}.

\bibitem{Jackiw:1974cv}
R.~Jackiw, {\em {Functional evaluation of the effective potential}}.
  \href{http://dx.doi.org/10.1103/PhysRevD.9.1686}{Phys. Rev. D {\bf 9} (1974)
  1686}.

\bibitem{Lane:2018ycs}
K.~Lane and W.~Shepherd, {\em {Natural stabilization of the Higgs
  boson\textquoteright{}s mass and alignment}}.
  \href{http://dx.doi.org/10.1103/PhysRevD.99.055015}{Phys. Rev. D {\bf 99}
  (2019) no.~5, 055015}, \href{http://arxiv.org/abs/1808.07927}{{\tt
  arXiv:1808.07927 [hep-ph]}}.

\bibitem{Brooijmans:2020yij}
G.~Brooijmans {\em et al.}, ``{Les Houches 2019 Physics at TeV Colliders: New
  Physics Working Group Report},'' in {\em {11th Les Houches Workshop on
  Physics at TeV Colliders}: {PhysTeV Les Houches}}.
\newblock 2, 2020.
\newblock \href{http://arxiv.org/abs/2002.12220}{{\tt arXiv:2002.12220
  [hep-ph]}}.

\bibitem{Zyla:2020zbs}
{\bf Particle Data Group} Collaboration, P.~Zyla {\em et al.}, {\em {Review of
  Particle Physics}}. \href{http://dx.doi.org/10.1093/ptep/ptaa104}{PTEP {\bf
  2020} (2020) no.~8, 083C01}.

\bibitem{Endo:2015ifa}
K.~Endo and Y.~Sumino, {\em {A Scale-invariant Higgs Sector and Structure of
  the Vacuum}}. \href{http://dx.doi.org/10.1007/JHEP05(2015)030}{JHEP {\bf 05}
  (2015)  030}, \href{http://arxiv.org/abs/1503.02819}{{\tt arXiv:1503.02819
  [hep-ph]}}.

\bibitem{Endo:2015nba}
K.~Endo and K.~Ishiwata, {\em {Direct detection of singlet dark matter in
  classically scale-invariant standard model}}.
  \href{http://dx.doi.org/10.1016/j.physletb.2015.08.059}{Phys. Lett. B {\bf
  749} (2015)  583--588}, \href{http://arxiv.org/abs/1507.01739}{{\tt
  arXiv:1507.01739 [hep-ph]}}.

\bibitem{Endo:2016koi}
K.~Endo, K.~Ishiwata, and Y.~Sumino, {\em {$WW$ scattering in a radiative
  electroweak symmetry breaking scenario}}.
  \href{http://dx.doi.org/10.1103/PhysRevD.94.075007}{Phys. Rev. D {\bf 94}
  (2016) no.~7, 075007}, \href{http://arxiv.org/abs/1601.00696}{{\tt
  arXiv:1601.00696 [hep-ph]}}.

\bibitem{Helmboldt:2016mpi}
A.~J. Helmboldt, P.~Humbert, M.~Lindner, and J.~Smirnov, {\em {Minimal
  conformal extensions of the Higgs sector}}.
  \href{http://dx.doi.org/10.1007/JHEP07(2017)113}{JHEP {\bf 07} (2017)  113},
  \href{http://arxiv.org/abs/1603.03603}{{\tt arXiv:1603.03603 [hep-ph]}}.

\bibitem{Fujitani:2017gma}
Y.~Fujitani and Y.~Sumino, {\em {Probing Higgs self-coupling of a classically
  scale invariant model in $e^+e^- \to Zhh$: Evaluation at physical point}}.
  \href{http://dx.doi.org/10.1016/j.physletb.2018.01.067}{Phys. Lett. B {\bf
  779} (2018)  46--51}, \href{http://arxiv.org/abs/1710.08096}{{\tt
  arXiv:1710.08096 [hep-ph]}}.

\bibitem{Passarino:1978jh}
G.~Passarino and M.~Veltman, {\em {One Loop Corrections for e+ e- Annihilation
  Into mu+ mu- in the Weinberg Model}}.
  \href{http://dx.doi.org/10.1016/0550-3213(79)90234-7}{Nucl. Phys. B {\bf 160}
  (1979)  151--207}.

\bibitem{Lee:1977eg}
B.~W. Lee, C.~Quigg, and H.~B. Thacker, {\em {Weak Interactions at Very
  High-Energies: The Role of the Higgs Boson Mass}}.
\href{http://dx.doi.org/10.1103/PhysRevD.16.1519}{Phys. Rev. {\bf D16} (1977)
  1519}.

\bibitem{Politzer:1978ic}
H.~Politzer and S.~Wolfram, {\em {Bounds on Particle Masses in the
  Weinberg-Salam Model}}.
  \href{http://dx.doi.org/10.1016/0370-2693(79)90746-9}{Phys. Lett. B {\bf 82}
  (1979)  242--246}. [Erratum: Phys.Lett.B 83, 421 (1979)].

\bibitem{Bechtle:2008jh}
P.~Bechtle, O.~Brein, S.~Heinemeyer, G.~Weiglein, and K.~E. Williams, {\em
  {HiggsBounds: Confronting Arbitrary Higgs Sectors with Exclusion Bounds from
  LEP and the Tevatron}}.
  \href{http://dx.doi.org/10.1016/j.cpc.2009.09.003}{Comput. Phys. Commun. {\bf
  181} (2010)  138--167},
\href{http://arxiv.org/abs/0811.4169}{{\tt arXiv:0811.4169 [hep-ph]}}.

\bibitem{Bechtle:2011sb}
P.~Bechtle, O.~Brein, S.~Heinemeyer, G.~Weiglein, and K.~E. Williams, {\em
  {HiggsBounds 2.0.0: Confronting Neutral and Charged Higgs Sector Predictions
  with Exclusion Bounds from LEP and the Tevatron}}.
  \href{http://dx.doi.org/10.1016/j.cpc.2011.07.015}{Comput. Phys. Commun. {\bf
  182} (2011)  2605--2631},
\href{http://arxiv.org/abs/1102.1898}{{\tt arXiv:1102.1898 [hep-ph]}}.

\bibitem{Bechtle:2013wla}
P.~Bechtle, O.~Brein, S.~Heinemeyer, O.~Stål, T.~Stefaniak, G.~Weiglein, and
  K.~E. Williams, {\em {$\mathsf{HiggsBounds}-4$: Improved Tests of Extended
  Higgs Sectors against Exclusion Bounds from LEP, the Tevatron and the LHC}}.
  \href{http://dx.doi.org/10.1140/epjc/s10052-013-2693-2}{Eur. Phys. J. {\bf
  C74} (2014) no.~3, 2693},
\href{http://arxiv.org/abs/1311.0055}{{\tt arXiv:1311.0055 [hep-ph]}}.

\bibitem{Bechtle:2015pma}
P.~Bechtle, S.~Heinemeyer, O.~Stal, T.~Stefaniak, and G.~Weiglein, {\em
  {Applying Exclusion Likelihoods from LHC Searches to Extended Higgs
  Sectors}}. \href{http://dx.doi.org/10.1140/epjc/s10052-015-3650-z}{Eur. Phys.
  J. {\bf C75} (2015) no.~9, 421},
\href{http://arxiv.org/abs/1507.06706}{{\tt arXiv:1507.06706 [hep-ph]}}.

\bibitem{Bechtle:2020pkv}
P.~Bechtle, D.~Dercks, S.~Heinemeyer, T.~Klingl, T.~Stefaniak, G.~Weiglein, and
  J.~Wittbrodt, {\em {HiggsBounds-5: Testing Higgs Sectors in the LHC 13 TeV
  Era}}. \href{http://arxiv.org/abs/2006.06007}{{\tt arXiv:2006.06007
  [hep-ph]}}.

\bibitem{Staub:2008uz}
F.~Staub, {\em {SARAH}}.
\href{http://arxiv.org/abs/0806.0538}{{\tt arXiv:0806.0538 [hep-ph]}}.

\bibitem{Staub:2009bi}
F.~Staub, {\em {From Superpotential to Model Files for FeynArts and
  CalcHep/CompHep}}. \href{http://dx.doi.org/10.1016/j.cpc.2010.01.011}{Comput.
  Phys. Commun. {\bf 181} (2010)  1077--1086},
\href{http://arxiv.org/abs/0909.2863}{{\tt arXiv:0909.2863 [hep-ph]}}.

\bibitem{Staub:2010jh}
F.~Staub, {\em {Automatic Calculation of supersymmetric Renormalization Group
  Equations and Self Energies}}.
  \href{http://dx.doi.org/10.1016/j.cpc.2010.11.030}{Comput. Phys. Commun. {\bf
  182} (2011)  808--833},
\href{http://arxiv.org/abs/1002.0840}{{\tt arXiv:1002.0840 [hep-ph]}}.

\bibitem{Staub:2012pb}
F.~Staub, {\em {SARAH 3.2: Dirac Gauginos, UFO output, and more}}.
  \href{http://dx.doi.org/10.1016/j.cpc.2013.02.019}{Comput. Phys. Commun. {\bf
  184} (2013)  1792--1809},
\href{http://arxiv.org/abs/1207.0906}{{\tt arXiv:1207.0906 [hep-ph]}}.

\bibitem{Staub:2013tta}
F.~Staub, {\em {SARAH 4 : A tool for (not only SUSY) model builders}}.
  \href{http://dx.doi.org/10.1016/j.cpc.2014.02.018}{Comput. Phys. Commun. {\bf
  185} (2014)  1773--1790},
\href{http://arxiv.org/abs/1309.7223}{{\tt arXiv:1309.7223 [hep-ph]}}.

\bibitem{Porod:2003um}
W.~Porod, {\em {SPheno, a program for calculating supersymmetric spectra, SUSY
  particle decays and SUSY particle production at e+ e- colliders}}.
  \href{http://dx.doi.org/10.1016/S0010-4655(03)00222-4}{Comput. Phys. Commun.
  {\bf 153} (2003)  275--315},
\href{http://arxiv.org/abs/hep-ph/0301101}{{\tt arXiv:hep-ph/0301101
  [hep-ph]}}.

\bibitem{Porod:2011nf}
W.~Porod and F.~Staub, {\em {SPheno 3.1: Extensions including flavour,
  CP-phases and models beyond the MSSM}}.
  \href{http://dx.doi.org/10.1016/j.cpc.2012.05.021}{Comput. Phys. Commun. {\bf
  183} (2012)  2458--2469},
\href{http://arxiv.org/abs/1104.1573}{{\tt arXiv:1104.1573 [hep-ph]}}.

\bibitem{Funakubo:1993jg}
K.~Funakubo, A.~Kakuto, and K.~Takenaga, {\em {The Effective potential of
  electroweak theory with two massless Higgs doublets at finite temperature}}.
  \href{http://dx.doi.org/10.1143/PTP.91.341}{Prog. Theor. Phys. {\bf 91}
  (1994)  341--352}, \href{http://arxiv.org/abs/hep-ph/9310267}{{\tt
  arXiv:hep-ph/9310267}}.

\bibitem{Glashow:1976nt}
S.~L. Glashow and S.~Weinberg, {\em {Natural Conservation Laws for Neutral
  Currents}}.
\href{http://dx.doi.org/10.1103/PhysRevD.15.1958}{Phys. Rev. {\bf D15} (1977)
  1958}.

\bibitem{Paschos:1976ay}
E.~A. Paschos, {\em {Diagonal Neutral Currents}}.
\href{http://dx.doi.org/10.1103/PhysRevD.15.1966}{Phys. Rev. {\bf D15} (1977)
  1966}.

\bibitem{Braathen:2017izn}
J.~Braathen, M.~D. Goodsell, and F.~Staub, {\em {Supersymmetric and
  non-supersymmetric models without catastrophic Goldstone bosons}}.
  \href{http://dx.doi.org/10.1140/epjc/s10052-017-5303-x}{Eur. Phys. J. C {\bf
  77} (2017) no.~11, 757}, \href{http://arxiv.org/abs/1706.05372}{{\tt
  arXiv:1706.05372 [hep-ph]}}.

\bibitem{Barger:1989fj}
V.~D. Barger, J.~L. Hewett, and R.~J.~N. Phillips, {\em {New Constraints on the
  Charged Higgs Sector in Two Higgs Doublet Models}}.
\href{http://dx.doi.org/10.1103/PhysRevD.41.3421}{Phys. Rev. {\bf D41} (1990)
  3421--3441}.

\bibitem{Grossman:1994jb}
Y.~Grossman, {\em {Phenomenology of models with more than two Higgs doublets}}.
  \href{http://dx.doi.org/10.1016/0550-3213(94)90316-6}{Nucl. Phys. {\bf B426}
  (1994)  355--384},
\href{http://arxiv.org/abs/hep-ph/9401311}{{\tt arXiv:hep-ph/9401311
  [hep-ph]}}.

\bibitem{Aoki:2009ha}
M.~Aoki, S.~Kanemura, K.~Tsumura, and K.~Yagyu, {\em {Models of Yukawa
  interaction in the two Higgs doublet model, and their collider
  phenomenology}}. \href{http://dx.doi.org/10.1103/PhysRevD.80.015017}{Phys.
  Rev. {\bf D80} (2009)  015017},
\href{http://arxiv.org/abs/0902.4665}{{\tt arXiv:0902.4665 [hep-ph]}}.

\bibitem{Misiak:2017bgg}
M.~Misiak and M.~Steinhauser, {\em {Weak radiative decays of the B meson and
  bounds on $M_{H^\pm }$ in the Two-Higgs-Doublet Model}}.
  \href{http://dx.doi.org/10.1140/epjc/s10052-017-4776-y}{Eur. Phys. J. {\bf
  C77} (2017) no.~3, 201},
\href{http://arxiv.org/abs/1702.04571}{{\tt arXiv:1702.04571 [hep-ph]}}.

\bibitem{Kanemura:1993hm}
S.~Kanemura, T.~Kubota, and E.~Takasugi, {\em {Lee-Quigg-Thacker bounds for
  Higgs boson masses in a two doublet model}}.
  \href{http://dx.doi.org/10.1016/0370-2693(93)91205-2}{Phys. Lett. B {\bf 313}
  (1993)  155--160}, \href{http://arxiv.org/abs/hep-ph/9303263}{{\tt
  arXiv:hep-ph/9303263}}.

\bibitem{Akeroyd:2000wc}
A.~G. Akeroyd, A.~Arhrib, and E.-M. Naimi, {\em {Note on tree level unitarity
  in the general two Higgs doublet model}}.
  \href{http://dx.doi.org/10.1016/S0370-2693(00)00962-X}{Phys. Lett. {\bf B490}
  (2000)  119--124},
\href{http://arxiv.org/abs/hep-ph/0006035}{{\tt arXiv:hep-ph/0006035
  [hep-ph]}}.

\bibitem{Staub:2011dp}
F.~Staub, T.~Ohl, W.~Porod, and C.~Speckner, {\em {A Tool Box for Implementing
  Supersymmetric Models}}.
  \href{http://dx.doi.org/10.1016/j.cpc.2012.04.013}{Comput. Phys. Commun. {\bf
  183} (2012)  2165--2206}, \href{http://arxiv.org/abs/1109.5147}{{\tt
  arXiv:1109.5147 [hep-ph]}}.

\bibitem{Fujii:2017vwa}
K.~Fujii {\em et al.}, {\em {Physics Case for the 250 GeV Stage of the
  International Linear Collider}}.
\href{http://arxiv.org/abs/1710.07621}{{\tt arXiv:1710.07621 [hep-ex]}}.

\bibitem{Kakizaki:2015wua}
M.~Kakizaki, S.~Kanemura, and T.~Matsui, {\em {Gravitational waves as a probe
  of extended scalar sectors with the first order electroweak phase
  transition}}. \href{http://dx.doi.org/10.1103/PhysRevD.92.115007}{Phys. Rev.
  {\bf D92} (2015) no.~11, 115007},
\href{http://arxiv.org/abs/1509.08394}{{\tt arXiv:1509.08394 [hep-ph]}}.

\bibitem{Hashino:2016xoj}
K.~Hashino, M.~Kakizaki, S.~Kanemura, P.~Ko, and T.~Matsui, {\em {Gravitational
  waves and Higgs boson couplings for exploring first order phase transition in
  the model with a singlet scalar field}}.
  \href{http://dx.doi.org/10.1016/j.physletb.2016.12.052}{Phys. Lett. {\bf
  B766} (2017)  49--54},
\href{http://arxiv.org/abs/1609.00297}{{\tt arXiv:1609.00297 [hep-ph]}}.

\bibitem{Hashino:2018wee}
K.~Hashino, R.~Jinno, M.~Kakizaki, S.~Kanemura, T.~Takahashi, and M.~Takimoto,
  {\em {Selecting models of first-order phase transitions using the synergy
  between collider and gravitational-wave experiments}}.
  \href{http://dx.doi.org/10.1103/PhysRevD.99.075011}{Phys. Rev. {\bf D99}
  (2019) no.~7, 075011},
\href{http://arxiv.org/abs/1809.04994}{{\tt arXiv:1809.04994 [hep-ph]}}.

\bibitem{Martin:2003qz}
S.~P. Martin, {\em {Evaluation of two loop selfenergy basis integrals using
  differential equations}}.
  \href{http://dx.doi.org/10.1103/PhysRevD.68.075002}{Phys. Rev. D {\bf 68}
  (2003)  075002}, \href{http://arxiv.org/abs/hep-ph/0307101}{{\tt
  arXiv:hep-ph/0307101}}.

\bibitem{Degrassi:2009yq}
G.~Degrassi and P.~Slavich, {\em {On the radiative corrections to the neutral
  Higgs boson masses in the NMSSM}}.
  \href{http://dx.doi.org/10.1016/j.nuclphysb.2009.09.018}{Nucl. Phys. B {\bf
  825} (2010)  119--150}, \href{http://arxiv.org/abs/0907.4682}{{\tt
  arXiv:0907.4682 [hep-ph]}}.

\bibitem{Braathen:2016cqe}
J.~Braathen and M.~D. Goodsell, {\em {Avoiding the Goldstone Boson Catastrophe
  in general renormalisable field theories at two loops}}.
  \href{http://dx.doi.org/10.1007/JHEP12(2016)056}{JHEP {\bf 12} (2016)  056},
\href{http://arxiv.org/abs/1609.06977}{{\tt arXiv:1609.06977 [hep-ph]}}.

\end{thebibliography}\endgroup

\end{document}